\documentclass[twocolumn,trackchanges]{aastex701}

\received{TBD}
\revised{TBD}
\accepted{TBD}
\submitjournal{The Astrophysical Journal}
\shorttitle{Dust Reservoirs in SN 1983V with JWST/HST}
\shortauthors{Kressy et al.}

\usepackage{color}
\usepackage{xcolor}
\usepackage{natbib}
\usepackage{graphicx}
\usepackage{epstopdf}
\usepackage{amsmath}
\usepackage{times}
\usepackage{textcomp}
\usepackage{wrapfig}

\newcommand{\JWST}{{\it JWST}\ }
\newcommand{\HST}{{\it HST}\ }

\newcommand{\lsolar}{L$_{\odot}$}
\newcommand{\msolar}{M$_{\odot}$}

\usepackage{soul}
\newcommand\N[1]{{\color{blue} \bf #1}}

\begin{document}



\title{A Significant Dust Reservoir Uncovered with JWST in the Type Ic SN 1983V More Than 40 Years Post-Explosion}

\author[0000-0002-3469-5774]{Sophia S. Kressy}
\affiliation{Department of Physics and Astronomy, University of North Carolina Chapel Hill, Chapel Hill, NC 27599, USA}
\email[show]{skressy@email.unc.edu}

\author[orcid=0000-0001-8385-3727,sname=Moore]{Thomas Moore}
\affiliation{Space Telescope Science Institute, 3700 San Martin Drive, Baltimore, MD 21218, USA}
\email{tmoore@stsci.edu}

\author[0000-0003-2238-1572]{Ori D. Fox}
\affiliation{Space Telescope Science Institute, 3700 San Martin Drive, Baltimore, MD 21218, USA}
\email{ofox@stsci.edu}

\author[0000-0001-5510-2424]{Nathan Smith}
\affiliation{Steward Observatory, University of Arizona, 933 N. Cherry Avenue, Tucson, AZ 85721, USA}
\email[show]{nathans@as.arizona.edu}

\author[0009-0004-7268-7283]{Raphael Baer-Way}
\affiliation{Department of Astronomy, University of Virginia, Charlottesville, VA 22904-4325, USA}
\email{bek5cw@virginia.edu}

\author[0000-0003-0599-8407]{Luc Dessart}
\affiliation{French-Chilean Laboratory for Astronomy, IRL 3386, CNRS and Instituto de Astrof\'isica, Pontificia Universidad Cat\'olica de Chile, Casilla 306, Santiago, Chile}
\affiliation{Institut d'Astrophysique de Paris, CNRS-Sorbonne Universit\'e, 98 bis boulevard Arago, F-75014 Paris, France}
\email{dessart@iap.fr}

\author[0009-0003-0227-1445]{Michael J. Dulude}
\affiliation{Space Telescope Science Institute, 3700 San Martin Drive, Baltimore, MD 21218, USA}
\email{dulude@stsci.edu}

\author[0000-0002-3934-2644]{W.~V.~Jacobson-Gal\'{a}n}
\altaffiliation{NASA Hubble Fellow}
\affiliation{Cahill Center for Astrophysics, California Institute of Technology, MC 249-17, 1216 E California Boulevard, Pasadena, CA 91125, USA}
\email{wynnjg@caltech.edu}

\author[0000-0002-9301-5302]{Melissa Shahbandeh}
\affiliation{Space Telescope Science Institute, 3700 San Martin Drive, Baltimore, MD 21218, USA}
\email{mshahbandeh@stsci.edu}

\author[0000-0001-7380-3144]{Tea Temim}
\affiliation{Department of Astrophysical Sciences, Princeton University, 4 Ivy Lane, Princeton, NJ 08544, USA}
\email{temim@astro.princeton.edu}

\author[0000-0002-5221-7557]{Chris Ashall}
\affiliation{Institute for Astronomy, University of Hawai{'i}, 2680 Woodlawn Dr, Honolulu, HI 96822, USA}
\email{cashall@hawaii.edu}

\author[0000-0003-4263-2228]{David A. Coulter}
\affiliation{Space Telescope Science Institute, 3700 San Martin Drive, Baltimore, MD 21218, USA}
\affiliation{Johns Hopkins University, 3400 N. Charles Street, Baltimore, MD 21218, USA}
\email{dcoulter@stsci.edu}

\author[orcid=0000-0003-3460-0103,sname=Filippenko]{Alexei V. Filippenko}
\affiliation{Department of Astronomy, University of California, Berkeley, CA 94720-3411, USA}
\email{afilippenko@berkeley.edu}

\author[0000-0001-6395-6702]{Sebastian Gomez}
\affiliation{Department of Astronomy, The University of Texas at Austin, 2515 Speedway, Stop C1400, Austin, TX 78712, USA}
\email{sebastian.gomez@austin.utexas.edu}

\author[0000-0001-5754-4007]{Jacob Jencson}
\affiliation{IPAC, Mail Code 100-22, Caltech, 1200 E. California Blvd., Pasadena, CA 91125}
\email{jjencson@ipac.caltech.edu}

\author[0000-0001-7186-105X]{Kyle Medler}
\affiliation{Institute for Astronomy, University of Hawai{'i}, 2680 Woodlawn Dr, Honolulu, HI 96822, USA}
\email{kmedler@hawaii.edu}

\author[0000-0002-7507-8115]{Daniel Patnaude}
\affiliation{Center for Astrophysics | Harvard \& Smithsonian, 60 Garden Street, Cambridge, MA 02138, USA}
\email{dpatnaude@cfa.harvard.edu}

\author[0000-0003-4610-1117]{Tam\'as Szalai}
\affiliation{Department of Experimental Physics, Institute of Physics, University of Szeged, D{\'o}m t{\'e}r 9, 6720 Szeged, Hungary}
\email{szaszi@titan.physx.u-szeged.hu}

\author[0000-0001-9038-9950]{Schuyler D. Van Dyk}
\affiliation{Caltech/IPAC, Mailcode 100-22, Pasadena, CA 91125, USA}
\email{vandyk@ipac.caltech.edu}

\author[0009-0004-5244-1884]{Aiden Agostinelli}
\affiliation{Department of Physics \& Astronomy, University of Montana, Missoula, MT 59801, USA}
\email{aiden.agostinelli@umontana.edu}

\author[]{Sarah Healy}
\affiliation{Department of Physics, Virginia Tech, Blacksburg, VA 24061, USA}
\email{healys@vt.edu}

\author[0009-0003-8380-4003]{Zachary G. Lane}
\affiliation{Space Telescope Science Institute, 3700 San Martin Drive, Baltimore, MD 21218, USA}
\email[show]{zachary.lane@pg.canterbury.ac.nz}

\author[0000-0002-0763-3885]{Dan Milisavljevic}
\affiliation{Department of Physics and Astronomy, Purdue University, 525 Northwestern Ave., West Lafayette, IN 47907, USA}
\email{dmilisav@purdue.edu}

\author[]{Shazrene Mohamed}
\affiliation{Department of Astronomy, University of Virginia, Charlottesville, VA 22904, USA}
\affiliation{Virginia Institute for Theoretical Astronomy, University of Virginia, Charlottesville, VA 22904, USA}
\affiliation{Department of Astronomy, University of Cape Town, Private Bag X3, Rondebosch 7701, South Africa}
\affiliation{South African Astronomical Observatory, P.O. Box 9, Observatory, 7935, Cape Town, South Africa}
\email{shazrene@virginia.edu}

\author[0009-0000-3565-8134]{Benjamin Radmore}
\affiliation{Maria Mitchell Association, 4 Vestal St., Nantucket, MA 02254, USA}
\affiliation{Department of Physics and Astronomy, University of Waterloo, 200 University Avenue West, Waterloo, ON N2L 3G1, Canada}
\email{bradmore@uwaterloo.ca}

\author[0000-0002-4410-5387]{Armin Rest}
\affiliation{Space Telescope Science Institute, 3700 San Martin Drive, Baltimore, MD 21218, USA}
\affiliation{Johns Hopkins University, 3400 N. Charles Street, Baltimore, MD 21218, USA}
\email{arest@stsci.edu}

\author[0000-0002-9820-679X]{Arkaprabha Sarangi}
\affiliation{Indian Institute of Astrophysics, 100 Feet Rd, Koramangala, Bengaluru 560034, Karnataka, India}
\email{arkaprabha.sarangi@iiap.res.in}

\author[0000-0002-0632-8897]{Yossef Zenati}
\affiliation{Astrophysics Research Center of the Open University (ARCO), Department of Natural Sciences, Ra’anana 4353701, Israel}
\affiliation{William H. Miller III Department of Physics \& Astronomy, Johns Hopkins University, 3400 N Charles St, Baltimore, MD 21218, USA}
\affiliation{The Observatories of the Carnegie Institution for Science, Pasadena, CA 91101, USA}
\email{oyossefzm@gmail.com}

\author[0009-0006-5127-8290]{Noah Zimmer}
\affiliation{Department of Physics and Astronomy, Purdue University, 525 Northwestern Avenue, West Lafayette, IN 47907-2036, USA}
\email{zimmer74@purdue.edu}

\keywords{\uat{Type Ic supernovae}{1730} --- \uat{Core-collapse supernovae}{304} --- \uat{Supernova dynamics}{1664} --- \uat{James Webb Space Telescope}{2291} --- \uat{Circumstellar dust}{236}}

\begin{abstract}

Searching for late-time emission from supernovae (SNe) is an active field. Infrared (IR) wavelengths are sensitive to thermal emission from dust, which can be used to probe SN contributions to the cosmic dust budget and progenitor mass-loss histories. The older an SN, the more likely it is for any existing dust to have cooled below the detection threshold of most observatories, even {\it JWST}. Decades-old IR observations of SNe are therefore exceedingly rare. Here we present fortuitous and serendipitous {\it JWST} IR observations that detect a point source at the position of the Type Ic SN 1983V more than 40 yr post-explosion. We demonstrate that the source is unlike nearby H\,\textsc{ii} regions and likely to be the dusty SN. We further show evidence from archival {\it HST} data of a plausible H$\alpha$~counterpart associated with ongoing SN shock interaction that collisionally heats the dust. In this scenario, the dust is distributed in a torus, more consistent with mass loss from binary interaction than a spherical wind. While not the oldest SN detected by {\it JWST} (SN 1980K), SN 1983V is a close second. Moreover, it has a relatively large dust mass ($\sim 7.7 \times 10^{-3}$ \msolar), particularly for a stripped-envelope SN. Although the dust is not likely newly formed, it does suggest such systems may contribute to dust production, particularly in the early Universe where massive stars and binary systems were more common. Spectroscopic observations can ultimately confirm the SN nature of this source. 
\end{abstract}

\section{Introduction}
\label{sec:intro}
Dust has emerged as a powerful tool for studying supernovae (SNe). A number of {\it Spitzer Space Telescope} ({\it Spitzer}) surveys revealed that dust can remain bright in a variety of different SN subclasses for many years and even decades post-explosion, contradicting conventional models that limited which progenitors should be able to form a dusty circumstellar medium (CSM) or dusty ejecta \citep{fox11,szalai19a,szalai21}, including even some thermonuclear Type Ia SNe \citep{fox13b,fox16}. The existence of dust is particularly intriguing for Type Ib/c stripped-envelope SNe (SESNe), which are expected to have neither dense CSM nor the necessarily decelerated ejecta velocities required to form significant amounts of dust.

Despite {\it Spitzer}'s valuable contributions, most of its observations were obtained during the ``Warm Mission,'' when only 3.6 and 4.5~\micron\ imaging was possible (dust temperatures $\gtrsim$500~K). These wavelengths are inadequate for constraining details about the dust and detecting the likely large reservoirs of dust at cooler temperatures ($100 < T_d < 500$~K). The MIRI~(5~$\mu$m~to 28~$\mu$m) and NIRCam~(0.6--5~$\mu$m) instruments on the {\it James Webb Space Telescope (JWST)} have the sensitivity and wavelength coverage necessary to span a wider dust temperature range, and identify previously undetectable dust reservoirs, although even \JWST\ is still mostly limited to dust at $>100$~K.

Already, \JWST has uncovered some of the largest-ever SN dust masses. Recent MIRI imaging of the Type IIP SN 2004et detected one of the largest newly formed ejecta dust masses in an extragalactic SN besides SN 1987A; \citet{Shahbandeh+2023} measure $\gtrsim10^{-2}$~\msolar~dust at $\sim 150$ K more than 6500 days post-explosion. The dust was likely formed in the ejecta, but was warmer and easier to detect (i.e., more luminous) than originally predicted. The reason for this is radiative heating arising from a forward shock (FS) that was stronger than previously expected for interaction with CSM around an SN IIP. The presence of such CSM in SNe IIP is now being modeled and interpreted to arise from the slow buildup of the pre-SN stellar wind \citep{dessart22}. This result offers a new hope that the cooler dust reservoirs in SNe IIP that went undetected by {\it Spitzer} can, in fact, be detected with {\it JWST}.

In addition, recent MIRI MRS spectroscopy of the Type IIn SN 2005ip revealed the presence of nearly $\sim$0.1~\msolar~of dust at $>$6000 days post-explosion \citep{Shahbandeh+2025}. Unlike SN 2004et, the dust in SN 2005ip condensed in a cold, dense shell (CDS) that formed behind the FS. A similar result was found for the Type IIn SN 2010jl \citep{smith26}. Given their extraordinarily dense CSM, SNe IIn are known to have a prominent CDS (as opposed to SNe IIP), but these two cases are the largest amount of dust observed to form in a CDS to date. 

Beyond Type II SNe, the stripped-envelope Type Ib SN 2014C formed $\gtrsim$ 0.04~\msolar~of dust in a CDS that developed nearly a year post-explosion \citep{tinyanont25}. SN 2014C is unique because it was classified at explosion as a typical hydrogen-poor SN~Ib, but at $> 100$ days started showing increasing narrow- and intermediate-width hydrogen emission, as well as strong radio and X-ray emission, indicative of shock interaction with extensive pre-existing CSM formed by the progenitor star \citep{milisavljevic15,margutti17,zhai25}. \citet{tinyanont25} report only seven other SESNe with such late-time ($>$100 day) rebrightening and delayed shock interaction, including SNe 2001em \citep{chugai06,chandra20}, 2004dk \citep{mauerhan18,pooley19,balasubramanian21}, 2018ijp \citep{tartaglia21}, 2019oys
\citep{sollerman20}, and 2019yvr \citep{kilpatrick21,ferrari24}. It is also interesting to note that some SESNe, like SN 2019tsf \citep{sollerman20,zenati22} and SN 2022xxf \citep{kuncarayakti23}, have also shown late-time ($>$100 day) rebrightening and radio emission, but without hydrogen or helium lines; this has been attributed to interactions with hydrogen-free and helium-poor CSM. While most of these SESNe do not have mid-IR observations, they present the intriguing possibility that SESNe may be capable of forming and/or heating newly formed dust. Along these lines, there are several SESNe with interactions inferred from archival IR photometry from the Wide-field Infrared Survey Explorer \citep[{\it WISE};][]{myers24,xiao26}. 

The number of late-time ($> 100$ day) mid-IR observations with {\it JWST} is slowly growing, with several new data points added in just the past 1--2 yr \citep{Zsiros+24,Szalai+2025,Pearson+2025,clayton25,sarangi25,subrayan26}. Many of these studies, including those previously mentioned, all have one thing in common: shock interaction is a central aspect of the physical scenario, whether the catalyst or simply the heating mechanism. Furthermore, when taken together with SN 1987A, they suggest the possibility that there may be an empirical trend in the dust growth with time. While most theoretical models have focused on dust formation in the ejecta, these results offer shocks as a notable alternative formation and/or heating mechanism. Although interacting SNe (i.e., SNe IIn, Ibn) make up $<10$\% of the core-collapse SN population in the low-redshift Universe \citep{li11}, they may be more common at high redshift. A top-heavy initial mass function (IMF) with more high-mass stars, similar to luminous blue variable (LBV) progenitors \citep{smith17_lbv}, and a higher fraction of binary stars that undergo stripping \citep{doughty21}, could lead to more SN~IIn-like environments with dense CSM, resulting in significant shock interaction and a CDS.

Nonetheless, the number of data points remains sparse at best, and questions remain regarding the dominant subclass, origin, heating mechanism, and timescale of dust formation. Very late-time (i.e., $\gtrsim$30 yr) observations of historic SNe are of particular interest given that the SN has entered its remnant phase, the FS has reached a radius corresponding to material lost thousands of years pre-explosion, and a substantial amount of any new dust is expected to have condensed. To better identify and interpret any trends, it is necessary to build a sample of SNe of different subclasses over a range of epochs. 

As part of \JWST Program ID 6356 (PI O. Fox), we searched the {\it JWST} data archive for serendipitous observations of SNe of any type at any epoch. In this paper, we present \JWST MIRI and NIRCam observations which happened to cover the field containing the Type Ic SN 1983V, which shows increasingly strong emission out to $\sim$20 \micron. SN 1983V is an interesting historic SESN that was discovered by R. O. Evans on 25 November 1983 UTC in the nearby galaxy NGC 1365, which has a reported redshift of $z = 0.0054$  \citep{Modjaz+2016}. The SN was observed by \citet{Wheeler+1987} 13 days post-discovery; they found a spectrum similar to that of Type Ib SNe, but with less He and more C and O in the ejecta.
Later, \citet{Clocchiatti+1997} published observations of SN 1983V with CCD photometry and low-resolution spectra taken in 1983--1984.
They found a lack of evidence of H~{\sc i}, He~{\sc i}, or Si~{\sc ii} near maximum light, and thus categorized SN 1983V as a Type Ic SN. \citet{Clocchiatti+1997} suggested that SN 1983V could be an extreme case of a stripped H/He envelope.

In this paper, we present a photometric analysis of SN 1983V.
In Section~\ref{sec:obs}, \JWST MIRI and NIRCam images are aligned and reprocessed, and photometry is then performed to extract fluxes for SN 1983V. We fit a two-component dust model to our spectral energy distribution (SED) and discuss possible geometric implications in Section~\ref{sec:analysis}. Hot and cold dust components are fitted to the SED for silicate and carbon compositions.
We discuss the geometric and evolutionary implications of these findings in Section~\ref{sec:heating_mechanism}. Our results are summarized in Section~\ref{sec:conclusions}.

Throughout this paper, we assume that the distance to NGC 1365 is 18.2\,$\pm$\,1.6\;(statistical)\,$\pm$\,1.0\;(systematic)\;Mpc based on Cepheid observations taken with the {\it Hubble Space Telescope (HST)} \citep{freedman97}. Following \citet{Clocchiatti+1997}, we adopt $E(B-V)=0.4 \pm 0.07$ mag as the reddening toward SN 1983V. Unless otherwise noted, all SEDs and spectra have been corrected for this extinction by assuming $R_{\rm V}\approx3.1$. All listed dates are UTC.

\begin{figure*}[!ht]
\centering
\includegraphics[scale = 0.34, trim= 0in 0in 0in 0in]{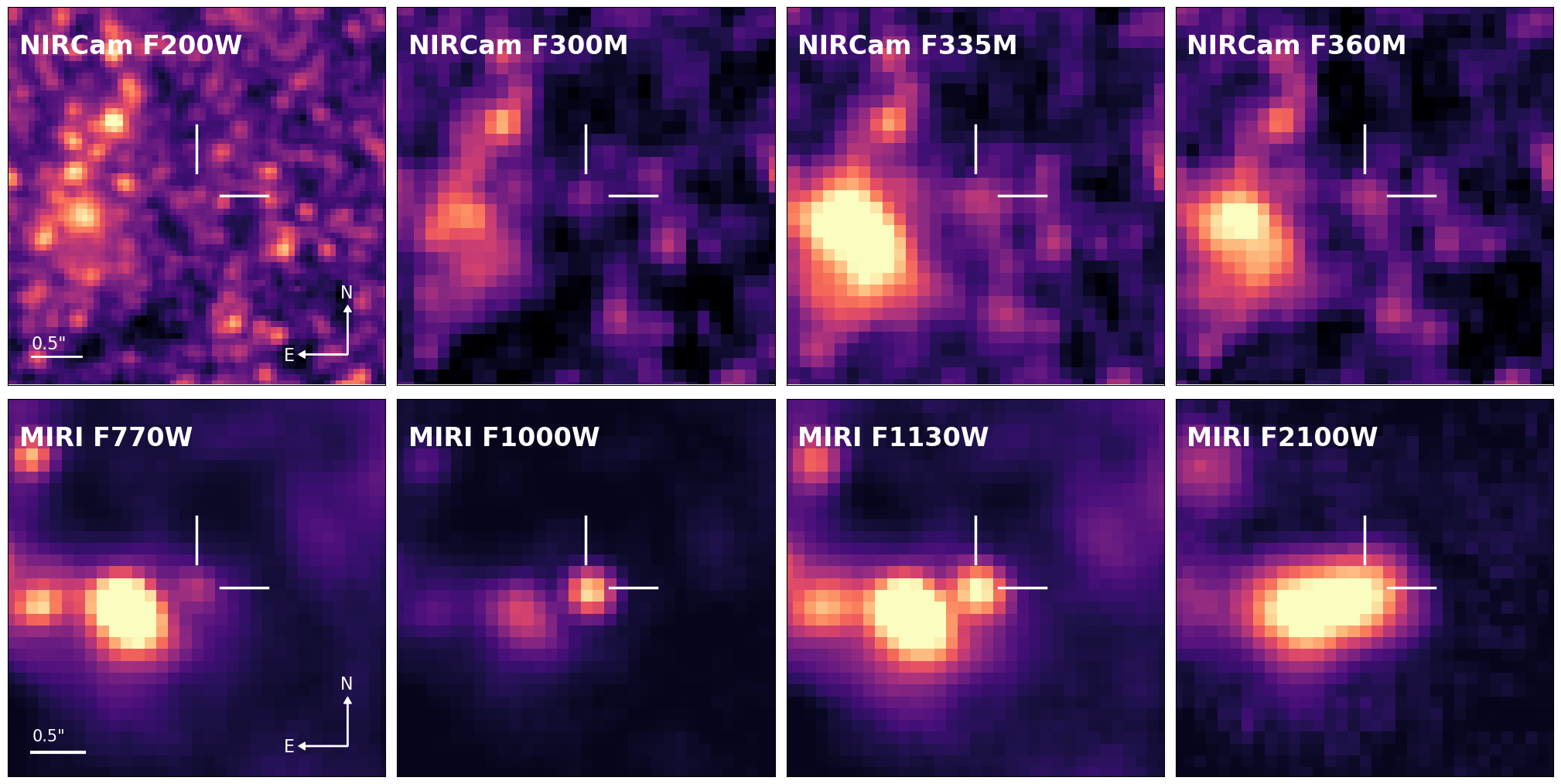}
\caption{MIRI/NIRCam observations of SN 1983V for all available filters. The probable surviving remnant of SN 1983V is indicated with tick marks. A scale bar is shown at bottom left for both NIRCam and MIRI images.}
\label{fig:all_filters}
\end{figure*}

\begin{deluxetable*}{lccccc}
\tablecaption{{\it JWST} Photometry of SN 1983V\label{tab:jwst_obs}}
\tablehead{
\colhead{Instrument} & \colhead{Filter} & \colhead{Effective Exp. Time (s)} & \colhead{Mag (eMag 1$\sigma$)} & \colhead{Flux (eFlux 1$\sigma$) [Jy]} 
}
\startdata
NIRCam & F200W  & 2405 & 26.76 (0.35) & 7.17$\times 10^{-8}$ (2.42$\times 10^{-8}$) \\
NIRCam & F300M  & 773 & 26.03 (0.48) & 1.39$\times 10^{-7}$ (6.24$\times 10^{-8}$) \\
NIRCam & F335M  & 773 & 25.07 (0.20) & 3.39$\times 10^{-7}$ (6.34$\times 10^{-8}$) \\
NIRCam & F360M  & 859 & 24.96 (0.27) & 3.75$\times 10^{-7}$ (9.56$\times 10^{-8}$) \\
MIRI   & F770W  & 533 & 21.36 (0.05) & 1.02$\times 10^{-5}$ (5.27$\times 10^{-7}$) \\
MIRI   & F1000W & 533 & 19.85 (0.05) & 4.15$\times 10^{-5}$ (6.48$\times 10^{-7}$) \\
MIRI   & F1130W & 533 & 19.59 (0.05) & 5.25$\times 10^{-5}$ (7.58$\times 10^{-7}$) \\
MIRI   & F2100W & 533 & 18.96 (0.06) & 9.40$\times 10^{-5}$ (1.90$\times 10^{-6}$) \\
\enddata
\tablenotetext{}{All observations were obtained on 2022-08-13 (MJD 59,805) under \JWST PID GO-2107.}

\end{deluxetable*}

\section{Observations}
\label{sec:obs}

\subsection{JWST Imaging}
\label{sec:jwst_obs}
SN 1983V was observed as part of the Cycle 1 General Observers (GO) 2107 program (PI J. Lee).
Images of SN 1983V were taken with the \JWST Mid-Infrared Instrument \citep[MIRI;][]{Bouchet+2015, Ressler+2015, Rieke+2015, Rieke+2022} and Near-Infrared Camera \citep[NIRCam;][]{Rieke+2023, Rieke+2005} on 13 Aug. 2022, 14,141 days post-explosion. 
Observations in MIRI covered bands F770W, F1000W, F1130W, and F2100W; NIRCam bands included F200W, F300M, F335M, F360M (Table \ref{tab:jwst_obs}).
The FASTR1 readout pattern in the FULL array mode and a 4-point extended-source dither pattern were used. 

\begin{figure}[t]
\centering
\includegraphics[width=0.35\paperwidth]{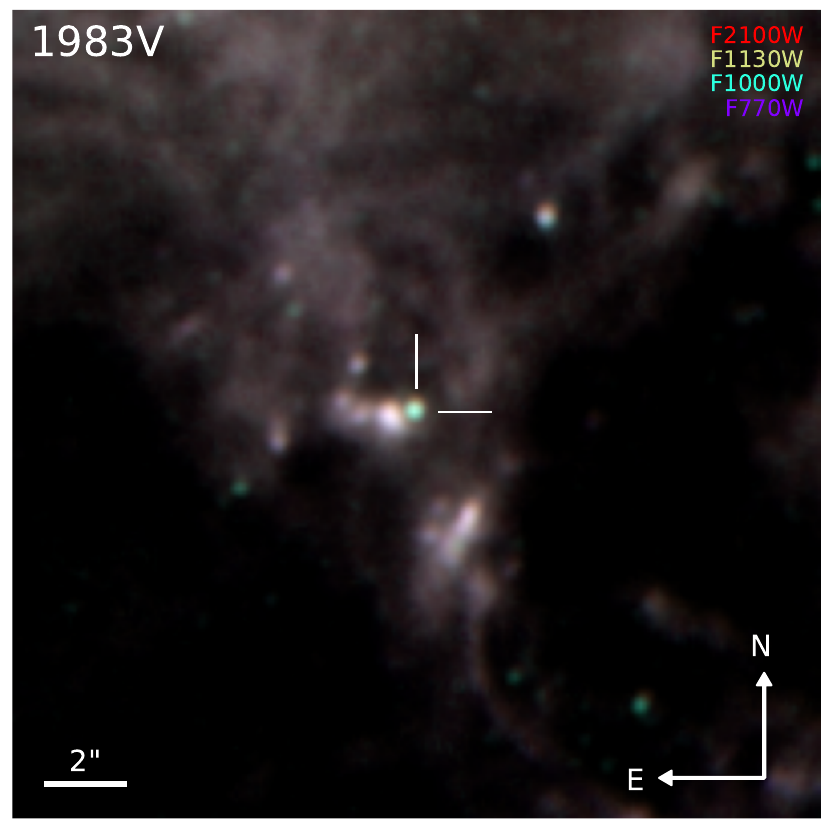}
\caption{
SN 1983V color composite made from archival MIRI images (F770W, F1000W, F1130W, and F2100W) taken on 13 August 2022 under PID GO-2107.}
\label{fig:color_image}
\end{figure}

As part of \JWST PID AR-6356 (PI O. Fox), we downloaded all Level 1\footnote{https://jwst-pipeline.readthedocs.io/en/latest/jwst/data\_products/stages.html} ramp data from the Mikulski Archive for Space Telescopes (MAST)\footnote{https://mast.stsci.edu/portal/Mashup/Clients/Mast/Portal.html} at \url{http://dx.doi.org/10.17909/c55c-rc43} and reprocessed the ramps with the \JWST Calibration Pipeline version 1.7.2, using the Calibration Reference Data System version 11.16.9 \citep{Bushouse+2022}. 
To address cosmic rays, calibrated images for each filter were cosmic-ray flagged. A ``background image'' was constructed for each filter by taking a sigma-clipped average of the individual dithers, and this background was then subtracted from the Level 2 calibrated individual dither images in the corresponding filter. 
Finally, the background-subtracted Level 2 images were mosaicked into a Level 3 final product. 
Figure~\ref{fig:all_filters} shows the cutouts of the Level 3 final product of NIRCam and MIRI filters at the reported position of the SN. 
Figure~\ref{fig:color_image} displays the combined color image of all the MIRI filters. 

\begin{figure*}[t!]
\centering
\includegraphics[height=0.3\textheight]{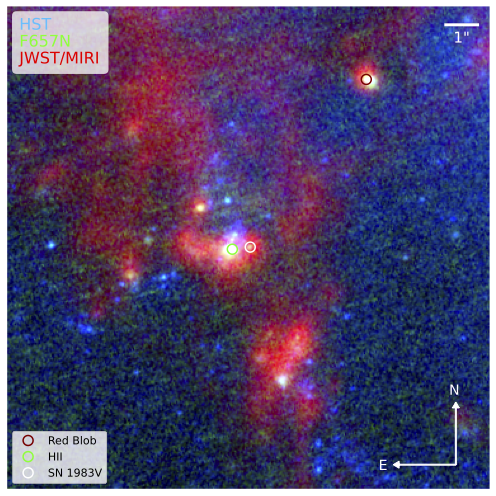}
\includegraphics[height=0.3\textheight]{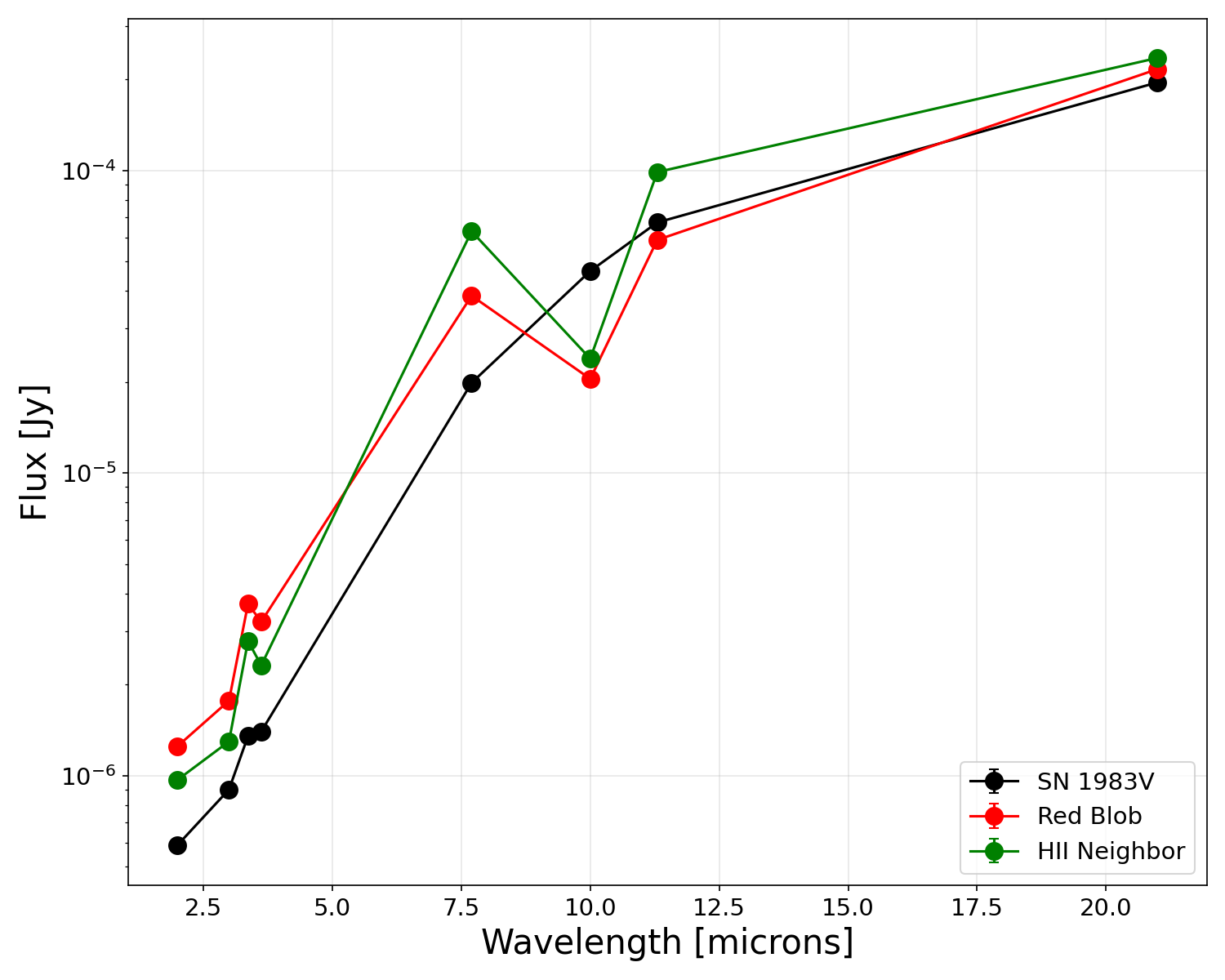}

\caption{Comparison of putative SN identification to nearby sources that are both known clusters (see Section \ref{sec:hst_counterpart} and Figure \ref{fig:hst}) with likely H\,\textsc{ii} regions. ({\it Left:}) SN 1983V color composite made from archival {\it HST}/WFC3 multifilter (blue), {\it HST}/WFC3 F765N (green), and {\it JWST}/MIRI multifilter (red) aligned images, as described in Sections \ref{sec:jwst_obs} and \ref{sec:hst_counterpart}, with circle identification for the putative SN (white) and nearby clusters (green and red). ({\it Right:)} Aperture photometry for the three sources. Photometry was performed using the \texttt{PHOTUTILS.aperture} package, with apertures set at $0.5 \times$ FWHM in MIRI filters\footnote{https://jwst-docs.stsci.edu/jwst-mid-infrared-instrument/miri-performance/miri-point-spread-functions\#gsc.tab=0} to minimize contributions from the underlying galaxy in the wings of the PSF.}
\label{fig:sed_comp}
\end{figure*}

\subsection{JWST Source Identification}
\label{sec:alignment}
Decades after the explosions of historical SNe, it can be challenging to confidently associate a faint late-time source with the residual emission of the SN itself, particularly when the SN position was not well-constrained owing to the limited astrometric precision at the time of explosion. 
For this case of SN 1983V, we do not have high-resolution or deep images of the early-time SN with which to align our \JWST images for comparison. 
In the combined color image (Figure \ref{fig:color_image}), a cyan/turquoise point source stands out as a promising candidate for the surviving remnant of SN 1983V.
The color and SED of the candidate SN are unlike anything else in the vicinity, and the source is unresolved, as opposed to extended. 
Nearby, there is a white-blue, bright, extended source to the east, which is a known cluster (see Section \ref{sec:hst_counterpart}), and is presumably associated with a giant H\,\textsc{ii} region. 

While it is not possible to conclusively characterize this point source as the SN without spectra, we highlight some additional supporting evidence. 
Figure \ref{fig:sed_comp} shows that the IR SED for the putative SN differs from the neighboring extended source.
Both SEDs are very faint in the near-IR and rise steadily to longer wavelengths, but a few key differences exist. 
The SN continuum is smoother, with likely significant silicate emission in the F1000W filter. 
The H\,\textsc{ii} region has a very different and wildly fluctuating SED. Specifically, it brightens in F335M and F770W, dips at F1000W, and then brightens again in F1130W and F2100W. This may be explained as very bright PAH emission at 3.3 \micron~(F335M), 7.7 \micron~+8.6 \micron~ (F770W), and 11.3 \micron~(F1130W), and then emission related to dust within the H\,\textsc{ii} region at F2100W. Alternatively, the dip at 10 \micron~may be associated with silicate absorption from optically thick dust \citep{gordon03,karambelkar26}. For comparison, we also plot the IR SED of the mid-IR bright source in the upper northwest, which is another known cluster and likely H\,\textsc{ii} region. The SEDs of the two likely H\,\textsc{ii} regions are almost identical. The candidate SN does not have jumps in flux associated with these strong PAH bands (or silicate absorption), making it unlikely that the mid-IR emission detected by \JWST arises from a comparable H\,\textsc{ii} region coincident with SN~1983V. We find that the mid-IR color and SED shape are unlike any other source in the field. Combined with a location that is consistent with SN~1983V to within the localization precision, we consider this compelling evidence that this mid-IR source is the late-time counterpart of SN 1983V, and we proceed with this assumption for our analysis in this paper. 

\subsection{HST Optical Counterpart?}
\label{sec:hst_counterpart}

Given the putative detection of SN 1983V in the {\it JWST} imaging, we also explore the possibility that there is an optical counterpart in archival {\it HST} data. SN 1983V was observed by {\it HST} WFC3/UVIS as part of GO-15654 and GO-17126 (PIs J.C. Lee and R. Chandar, respectively; see Table \ref{tab:hst_obs}). Wide filters (F275W, F336W, F438W, F555W, F814W) were observed on 20 July 2019 and narrow (F657N) on 20 July 2023. The individual UVIS {\tt flc} frames in all bands were obtained from the Barbara A. Mikulski Archive for Space Telescopes (MAST) at \url{http://dx.doi.org/10.17909/c55c-rc43}. The data followed standard pipeline processing. The frames in each band then had cosmic-ray hits masked and were combined into mosaics by running them through \texttt{AstroDrizzle} in \texttt{PyRAF}.

\begin{figure*}[t!]
\centering
\includegraphics[width=0.27\paperwidth]{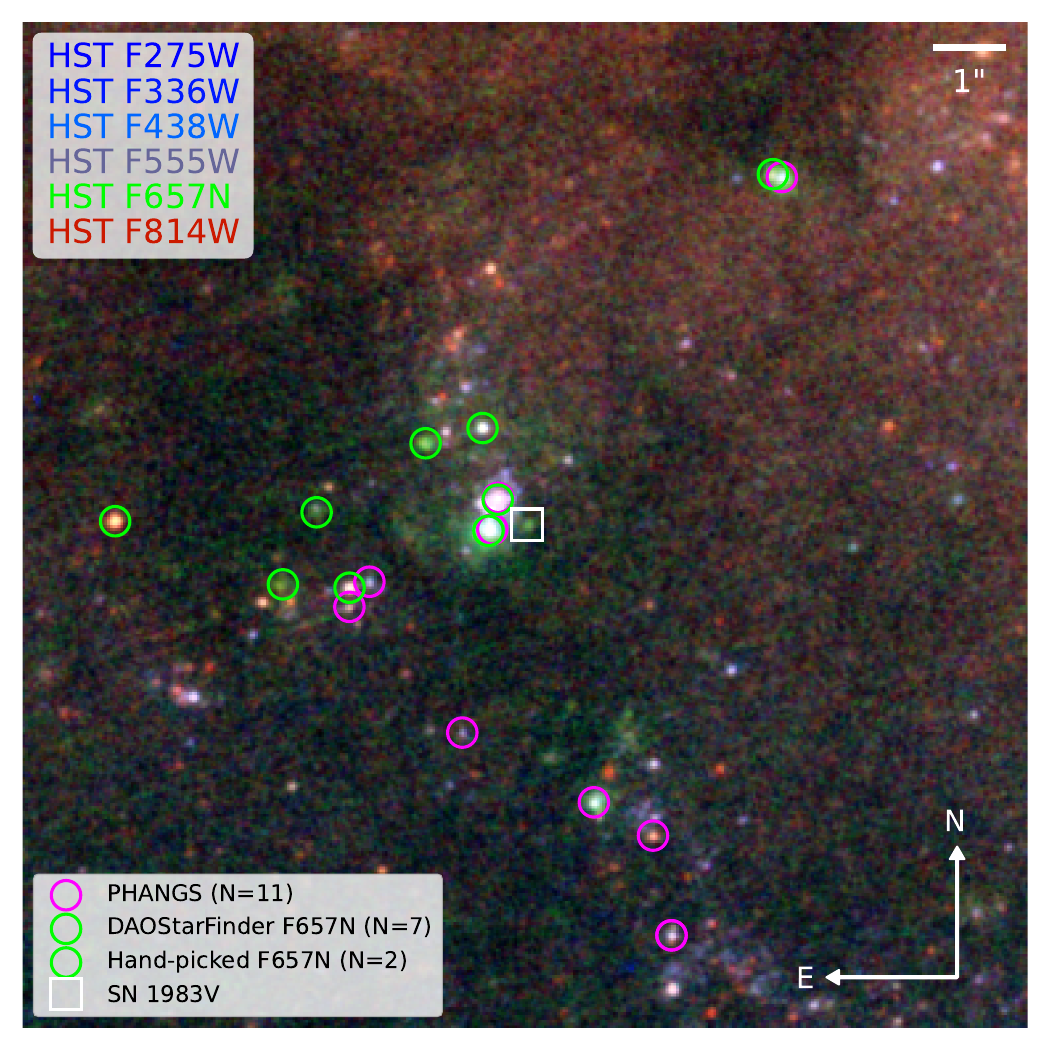}
\includegraphics[width=0.27\paperwidth]{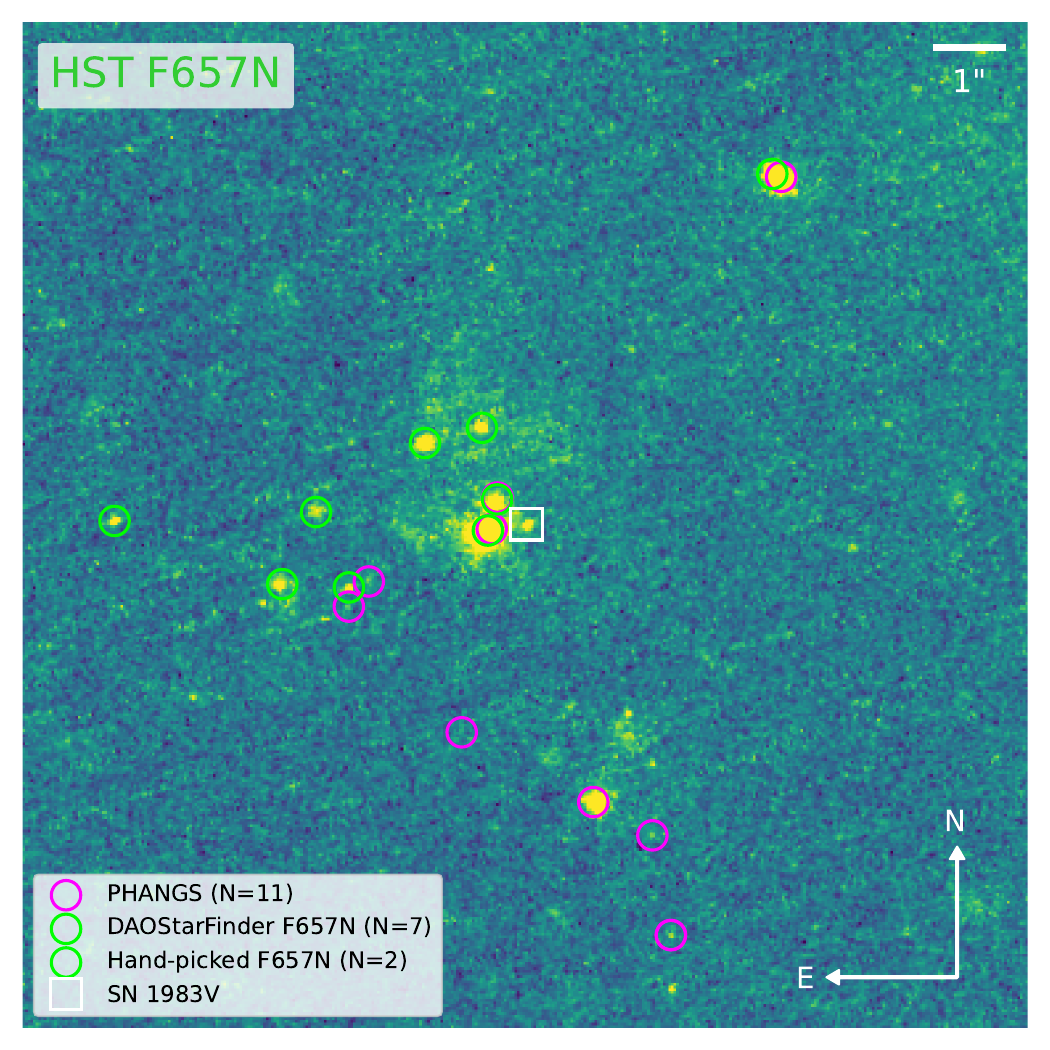}
\includegraphics[width=0.27\paperwidth]{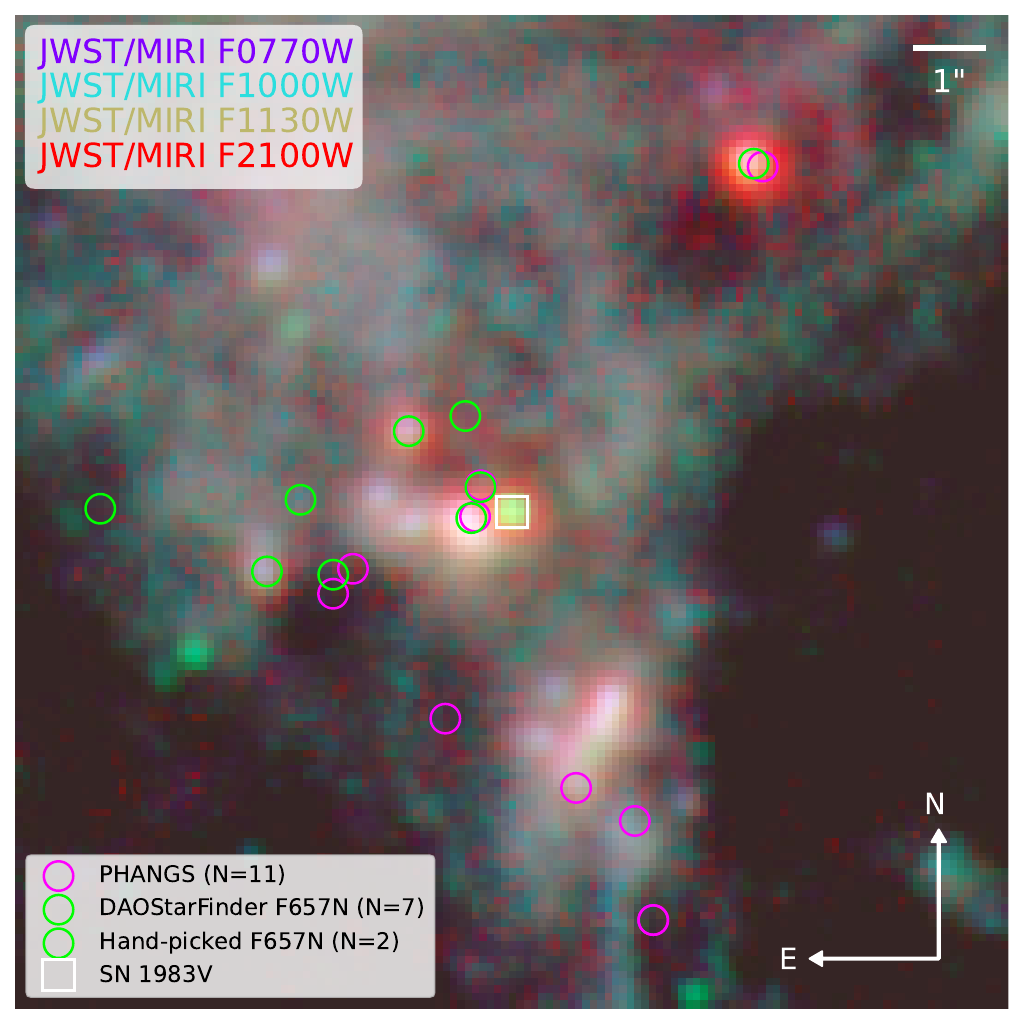}
\caption{We show {\it (left)} \textit{HST}/WFC3 composite, {\it (center)} \textit{HST}/WFC3 F657N, and {\it (right)} \textit{JWST}/MIRI composite of SN 1983V. Details of the observations are listed in Tables \ref{tab:jwst_obs} and \ref{tab:hst_obs}. All images have been aligned, as described in the text. The position of SN 1983V (determined from the \JWST images) is indicated with a white box. Overplotted for comparison are catalogs, as described in the text. The source at the position of SN 1983V stands out as the only source not identified by PHANGS to have both H$\alpha$~optical emission {\it and} colder, redder mid-IR emission.}
\label{fig:hst}
\end{figure*}

We align the \JWST Level 3 mosaicked images to the {\it HST}/WFC3 reference images (Table \ref{tab:hst_obs}) with the {\it JWST}--\HST Alignment Tool (\texttt{JHAT})\footnote{https://jhat.readthedocs.io/en/latest/}. Figure \ref{fig:hst} shows that an optical point source (white box) coincident with the \JWST\ mid-IR source is present in the {\it HST}/F657N narrow-band filter, but none of the other broad-band filters (Table \ref{tab:hst_obs}). We note that the {\it HST} broad-band filters span the continuum, while F657N is specifically sensitive to H$\alpha$.

The bright narrow-band H$\alpha$ point source is a promising optical counterpart. Since the mid-IR SED shape is inconsistent with expectations for an H\,\textsc{ii} region at the position (see Section \ref{sec:alignment}), the H$\alpha$ source at SN~1983V's position may arise from late-time shock interaction.  It is difficult to definitively rule out the hypothesis that some or all of the narrow H$\alpha$ emission arises from a more evolved H\,\textsc{ii} region that is coincident with the dusty SN, but the absence of any detectable blue continuum source detected in {\it HST} broad-band images is not consistent with a stellar cluster at this position. 

For completeness, we also overplot known clusters (pink circles) identified in catalogs generated by a $V$-band-selected source list in the Physics at High Angular resolution in Nearby GalaxieS (PHANGS) {\it HST} dataset \citep{lee22,maschmann24}. No known cluster is at the position of the SN. The cluster identification criteria used by PHANGS optimally track young, luminous clusters, but clusters may be smaller, fainter, or older and, ultimately, below any such catalog detection limit. We therefore also generate our own catalog of F657N sources of visually identified sources (green circles) for comparison to the source at the position of SN 1983V.

By qualitatively comparing these catalogs across the {\it HST} and {\it JWST} images in Figure \ref{fig:hst}, SN 1983V stands out. It is the only source not identified by PHANGS to have both H$\alpha$~optical emission {\it and} colder, redder mid-IR emission.
While most of the F657N sources do not have a mid-IR counterpart, there is still the possibility that this is simply a highly reddened H\,\textsc{ii} region. However, the PHANGS identified clusters that do have associated mid-IR detections show PAH features (see Section \ref{sec:alignment}). Given this source is unique from every other F657N and mid-IR source, it is a plausible conjecture that the H$\alpha$ flux we detect is associated with ongoing shock interaction from a decades-old SN. We consider this possibility throughout the rest of this article.

\subsection{Photometry}
\label{sec:photometry}

With the SN source convincingly identified, we conduct photometry using \texttt{space\_phot} \citep{Pierel24}, a package optimized for forced point-spread-function (PSF) photometry on mosaicked images for \HST and \JWST (i.e., DRZ and Level 3 data products, respectively). We build a drizzled PSF by taking a model PSF at each position in the individual exposures (i.e., FLT and Level 2 data products, respectively), and drizzling them together using the same pipeline implementation as the data. For {\it JWST}, Level 2 PSF models are taken from {\tt webbpsf} v. 1.2.1 \citep{perrin12,perrin14}\footnote{\url{https://webbpsf.readthedocs.io}}. For {\it HST}, individual PSF models are obtained from the STScI PSF website.\footnote{https://www.stsci.edu/hst/instrumentation/wfc3/data-analysis/psf} 

\begin{figure*}[!t]
\centering
\includegraphics[scale = 1, trim= 0in 0in 0in 0in]{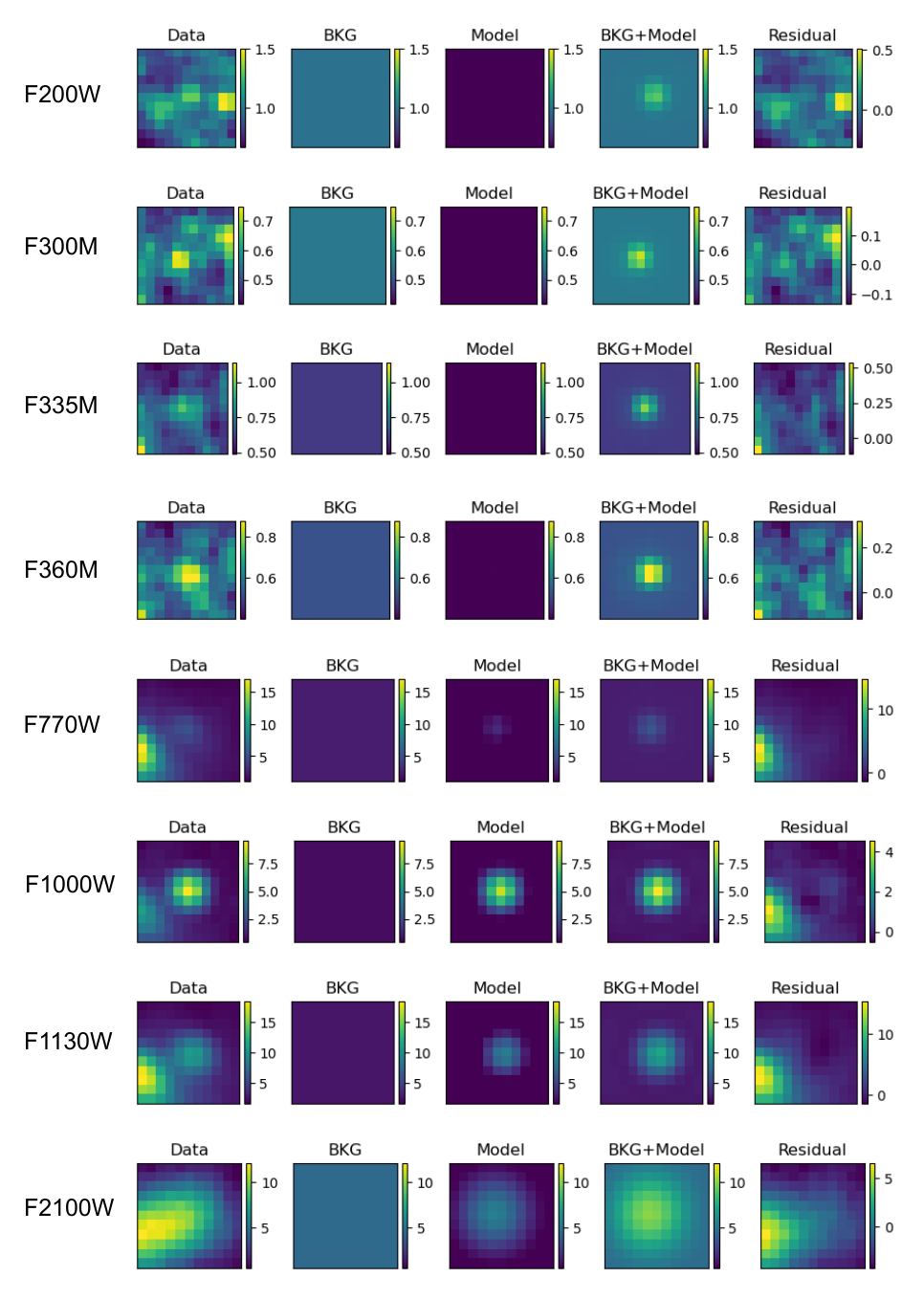}
\caption{Photometry for NIRCam and MIRI filters. 
From left to right, panels show the data, the background value (constant), the PSF  model (derived from \texttt{space\_phot}), the combined background and PSF model, and the residual between the data and the background+model.}
\label{fig:opt_photo}
\end{figure*}

We perform PSF photometry with the \texttt{space\_phot} drizzled PSF fitting routine using $5\times5$ pixel cutouts (Figure~\ref{fig:opt_photo}). Forced photometry was run at the centroid position obtained from the F1000W image. The measured fluxes, which are in units of mJy~sr$^{-1}$, are converted to AB magnitudes \citep{Oke+1983} using the pixel scale of each image ($0\arcsec03$ pix$^{-1}$ for short wavelengths (SW), $0\arcsec06$ pix$^{-1}$ for long wavelengths (LW)). Given the complicated nature of the underlying galaxy background and the nearby H\,\textsc{ii} in the MIRI bands (Figure \ref{fig:all_filters}), we identify an additional systematic uncertainty in our photometric values of $\sim$0.03-0.04 mags in the F1000W, F1130W, and F2100W bands. A final source of photometric uncertainty is a systematic uncertainty on the zero-points, which is $\lesssim0.01$ mag for all filters \citep{boyer22} and negligible compared to the uncertainties derived by the \texttt{space\_phot} PSF-fitting algorithm. Tables \ref{tab:jwst_obs} and \ref{tab:hst_obs} list the resulting photometry. 

\begin{deluxetable*}{lcccccc}
\tablecaption{\textit{HST}/WFC3 Archival Observations of SN 1983V\label{tab:hst_obs}}
\tablehead{
\multicolumn{7}{c}{}  \\
\cline{1-7} 
\colhead{Filter}&\colhead{Date}&\colhead{MJD}&\colhead{Program ID}&\colhead{Effective Exp. Time (s)}&\colhead{Mag (eMag 1$\sigma$)}&\colhead{Flux (eFlux 1$\sigma$) [Jy]}}
\startdata
F275W & 2019-07-20 & 58684 & GO 15654 & 2190 & $<$22.4 & $<$4.0$\times 10^{-6}$ \\
F336W & 2019-07-20 & 58684 & GO 15654 & 1110 & $<$23.1 & $<$2.1$\times 10^{-6}$  \\
F438W & 2019-07-20 & 58684 & GO 15654 & 1050 & $<$22.7 & $<$3.2$\times 10^{-6}$  \\
F555W & 2019-07-20 & 58684 & GO 15654 & 664 & $<$23.9  & $<$1.0$\times 10^{-6}$  \\
F657N & 2023-07-20 & 60145 & GO 17126 & 2392 & 23.19 (0.01)& 1.92$\times 10^{-6}$ (1.78$\times 10^{-8}$)\\
F814W & 2019-07-20 & 58684 & GO 15654 & 836 & $<$23.3 & $<$1.8$\times 10^{-6}$  
\enddata
\tablenotetext{}{All \HST wide filters are non-detections; we report their $3\sigma$ upper limits here.}
\end{deluxetable*}

We note that no PSF model exists for the F657N filter\footnote{https://www.stsci.edu/hst/instrumentation/wfc3/data-analysis/psf}. In this case, we instead perform photometry using a nonvarying effective PSF (ePSF) following \citet{2000PASP..112.1360A} and \citet{2016wfc..rept...12A}. An F657N ePSF was built using standard ePSF fitting routines in \texttt{photutils} for the 10 brightest stars in the image. Using \texttt{photutils} PSF-fitting routines, a magnitude of $m_{\rm 657N}^{} = 23.19 \pm 0.01$ was found for the SN. If we assume the F657N flux arises predominantly from CSM interaction associated with the SN, then the H$\alpha$~flux corresponds to a luminosity $L_{{\rm H}\alpha} \approx 1.6 \times 10^3$~\lsolar. 

\subsection{Optical Spectra}
\label{sec:spectra}

\subsubsection{MUSE}

The location of SN 1983V was serendipitously observed with the Multi Unit Spectroscopic Explorer \citep[MUSE;][]{2010SPIE.7735E..08B} integral field spectrograph on 5 November 2018. The instrument was configured in wide-field mode, with a resolution of $R \approx 3000$ across a wavelength range of 4750--9350~\AA. A spectrum of the explosion site of SN 1983V was obtained using two dithers as part of a larger mosaic with an exposure time of 2580~s each and seeing of 0.64--0.78\arcsec. The data were obtained from ESO Science Portal\footnote{https://archive.eso.org/cms.html} and were reduced by the PHANGS collaboration. 

\begin{figure*}[t!]
\centering
\includegraphics[height=2.25in]{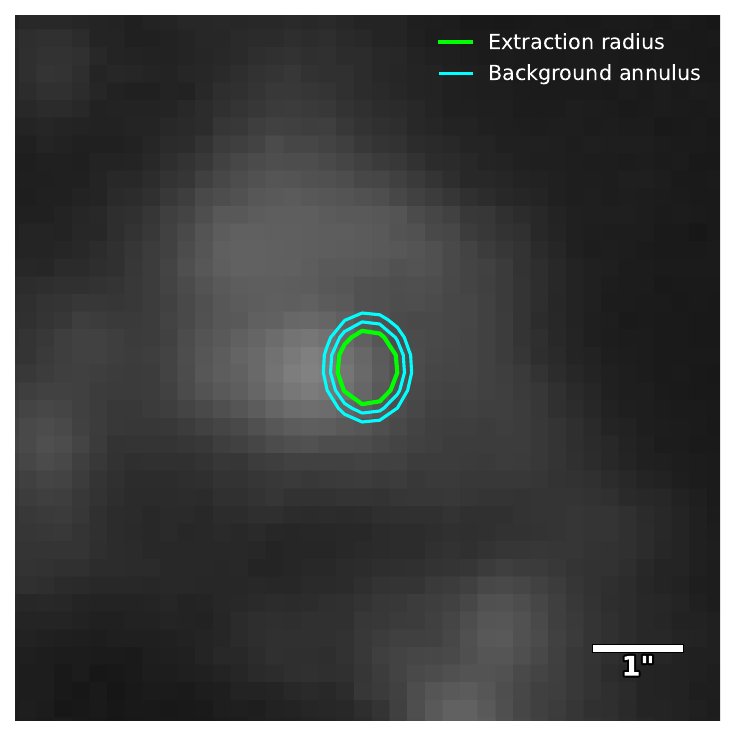}
\includegraphics[height=2.25in]{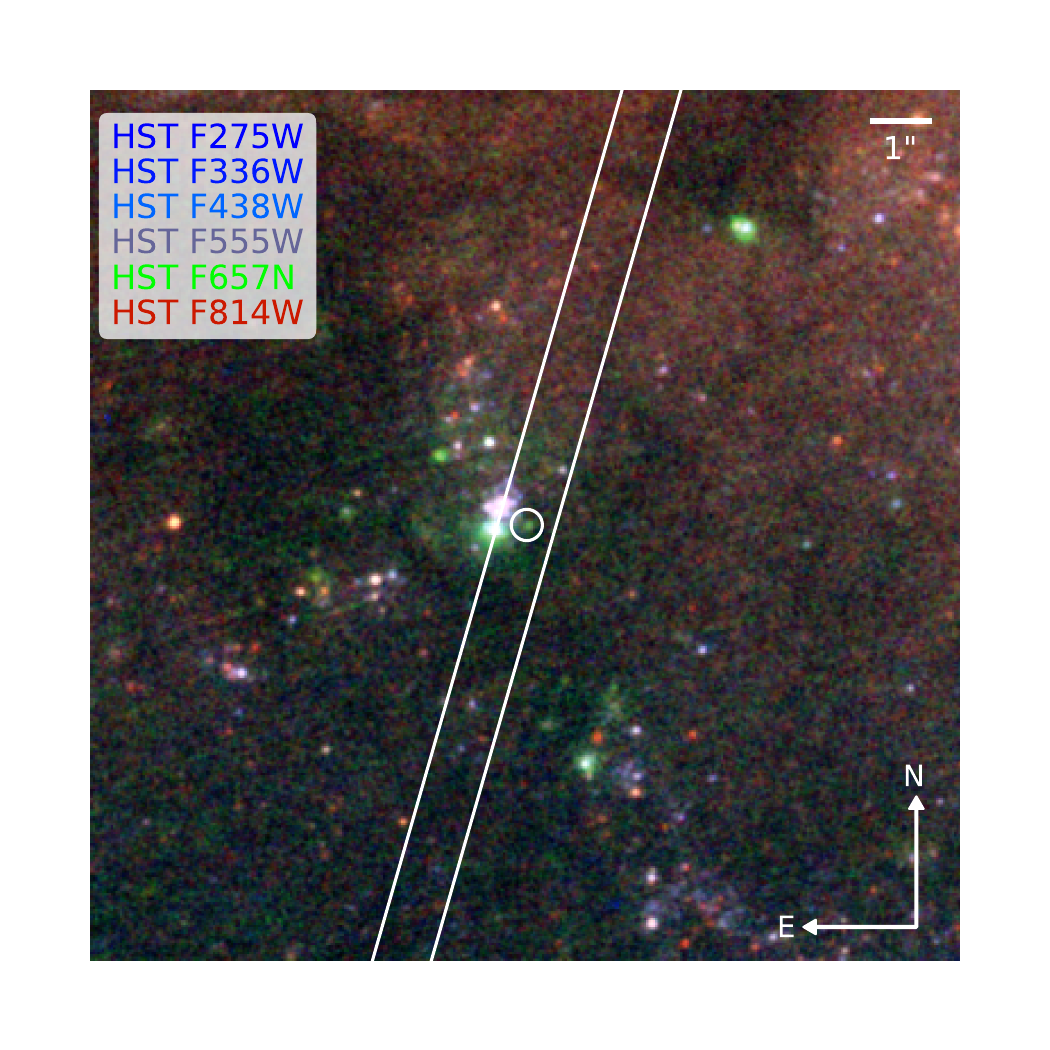}
\includegraphics[height=2.25in]{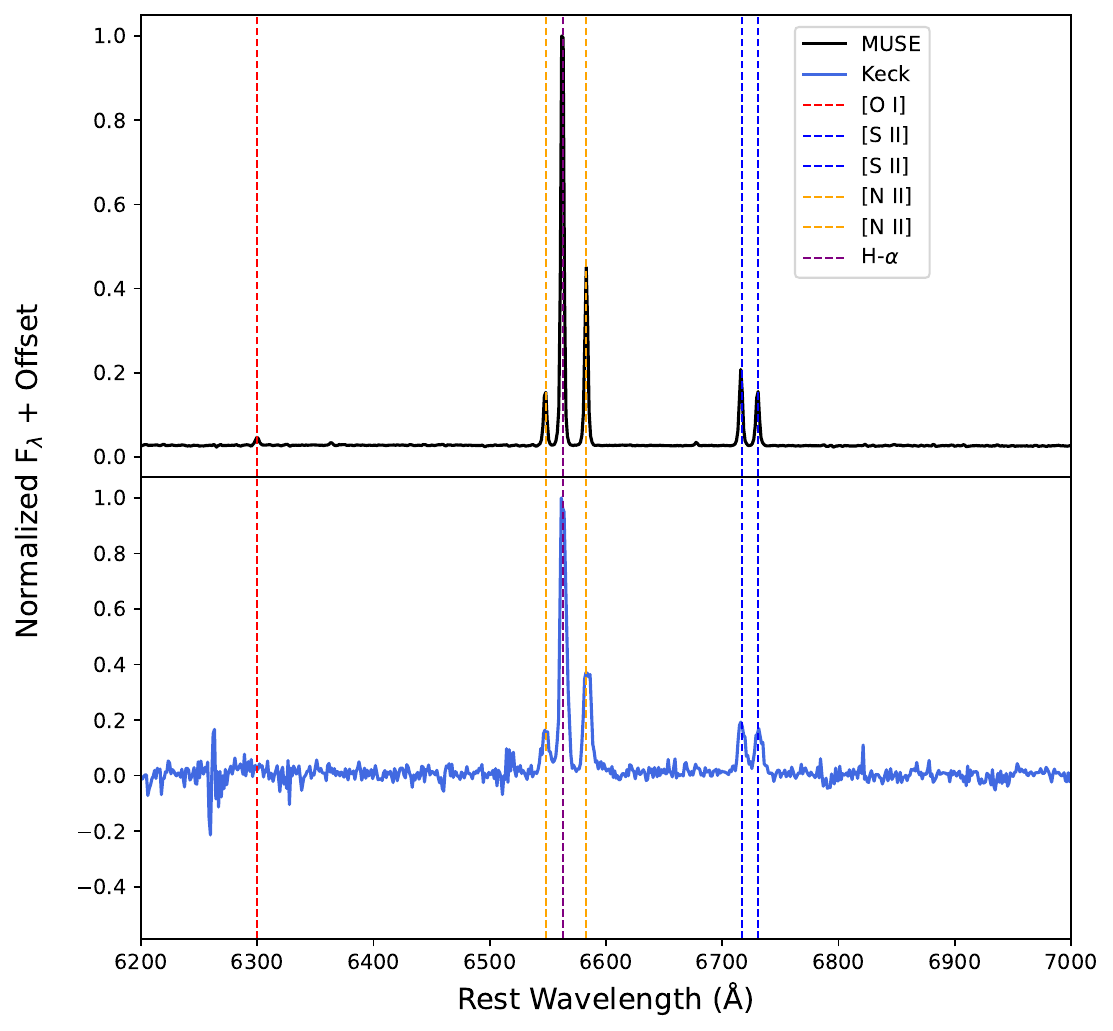}
\caption{({\it left}) Visualization of MUSE spectral data cube summed over the H$\alpha$~line width. The primary extraction aperture and corresponding background annulus are illustrated by the green and blue circles, respectively. ({\it center}) Visualization of the Keck/LRIS slit position overlaid on the HST multi-color image from Figure \ref{fig:hst}. ({\it right}) The resulting ground-based spectra extracted by Keck (top) and MUSE (bottom). Dashed vertical lines indicate emission lines: [O\,\textsc{i}] (red), [S\,\textsc{ii}] (blue), [N\,\textsc{ii}] (yellow), and H$\alpha$ (purple). We note the H\,\textsc{ii} region likely contaminates a significant fraction of the resulting spectrum in both cases.}
    \label{fig:opt_spec}
\end{figure*}

Following the MUSE reduction methods presented by \citet{2025A&A...700A.223K}, to find broad spectral features in SN remnants we extracted spaxels within an aperture around the location of SN 1983V with a radius equal to the seeing at the time of observations (0\farcs695). This aperture includes the nearby bright H\,\textsc{ii} region. We attempted to account for this by subtracting a summed spectrum of spaxels within an annulus around this location with an inner and outer radius larger than the seeing by 0.2\arcsec and 0.4\arcsec, respectively. Figure \ref{fig:opt_spec} visualizes these apertures on the MUSE data cube and plots the resulting MUSE spectrum.

The spectrum is dominated by narrow line emission, including H$\alpha$, [O\,\textsc{i}], [S\,\textsc{ii}], and [N\,\textsc{ii}]. We do not observe a broad boxy H$\alpha$ profile in the spectrum, which would be indicative of CSM interaction \citep[e.g.,][]{Shahbandeh+2023,Shahbandeh+2025}. However, we again note that our extracted spectrum is likely dominated by the nearby H\,\textsc{ii} region to the east. The lack of broad emission does not necessarily prove that that there is no CSM interaction. We calculate an upper limit to the possible contribution from an SN by injecting a broad H$\alpha$ line into the spectra with a width of 4000 km~s$^{-1}$ until the injected profile is detected with a significance of 3$\sigma$ above the noise. We derive a limit of $L_{\rm H{\alpha}} < 1.7 \times 10^4$~L$_{\odot}$. This limit is consistent with our previously found luminosity estimate from \HST in Section~\ref{sec:photometry}.

\subsubsection{Keck/LRIS}


We also attempted to obtain an optical spectrum at the location of SN 1983V using the Low Resolution Imaging Spectrometer \citep[LRIS;][]{Oke_1995} in long-slit mode mounted on the Keck-I 10\,m telescope on Maunakea on 16 December 2025. These observations were conducted with the 600/4000 grism and the 400/8500 grating and a $1.0^{\prime\prime}$ slit width. Target acquisition of SN 1983V was done using an offset star, with the RA/Dec offsets to SN 1983V calculated from the \HST images. The six exposures of SN 1983V were obtained using the blue and red arm of LRIS with 1200~s per exposure, with a position angle of -106 degrees on the sky (Figure \ref{fig:opt_spec}). Data reduction was done with the automated reduction code {\tt Pypeit} \citep{Prochaska_2020} version 1.18.0 \footnote{https://pypeit.readthedocs.io/en/stable/} setup in the {\tt keck\_lris\_blue} and {\tt keck\_lris\_red\_mark4} routines. The exposures were corrected for flat fielding and wavelength calibrated; then the trace was extracted by manually fitting the strongest emission feature in the image. Afterward, flux calibration and coaddition of the images was done to construct the finalized spectrum. Further details on the full reduction procedure that we followed are given by \citet{Medler_2025}. Figure \ref{fig:opt_spec} plots the resulting Keck spectrum.

The LRIS observation is consistent with the earlier MUSE spectrum, showing only emission features associated with the nearby H\,\textsc{ii} region. The main difference is the lack of the [O\,\textsc{i}] emission located at 6300~\AA, which is hidden in the noise of the continuum. The H$\alpha$~line is narrow ($280 \pm 110~\mathrm{km~s^{-1}}$) and looks similar to the MUSE spectrum. Again, however, with a $1.0^{\prime\prime}$ slit, this spectrum also contains significant contamination from the neighboring H\,\textsc{ii} region and does not rule out CSM interaction.

\subsection{Radio Observations with the VLA}
\label{sec:vla}

SN 1983V has been serendipitously observed at radio wavelengths more than 10 times with the Very Large Array (VLA) over the last 40 yr. Of all the VLA observations, we opt to analyze only data at 6 GHz or lower, where any synchrotron emission from ejecta-CSM interaction should remain bright for the longest time \citep{Chevalier_1998}. All data were reduced in the standard manner using CASA \citep{CASA_cite}. Data taken before the upgrade of the VLA were reduced manually, and other data were reduced using the upgraded VLA pipeline. We end up with four C-band (6 GHz) high-resolution observations (VLA PIDs 23A-271 (PI E. Behrens), AS314 (PI D. Saikia), 18A-418 (PI J. Jencson), 11B-177 (PI A. Soderberg)) covering a range  1500--14,000 days post-discovery. The SN is not detected at any epoch, with 3$\sigma$ limits from the root-mean-square (RMS) flux-density values in a $10''$ region surrounding the SN of 0.01--0.1~mJy. Specifically, at 1553, 10,592, 12,585, and 14,572 days post-explosion, we find 3$\sigma$ upper limits on the radio flux density of 0.409, 0.023, 0.021, and 0.010 mJy, respectively.

\subsection{High-Energy Observations}
\label{sec:high_energy}

\begin{deluxetable}{lcccC}
\tablecaption{Best Fitting Dust Model Components \label{tab:photo_output}}
\tablehead{
\multicolumn{5}{c}{}  \\
\cline{1-5} 
\colhead{Dust Component}&\colhead{Composition}&\colhead{Grain Size (\micron)}&\colhead{$M_{\rm d}$ [M$_{\odot}$]}&\colhead{$T_{\rm d}$ [K]}}
\startdata
Hot  & Silicate & 0.1 & 1.8$\pm0.43\times10^{-6}$  & 629$\pm$32\\
Cold & Carbon   & 0.1 & 7.7$\pm 0.48\times 10^{-3}$ & 173$\pm$2\\
\enddata
\end{deluxetable}

NGC 1365 was also observed extensively with \textit{Chandra} and \textit{XMM-Newton} between 1999 and 2021, though many observations involved the use of the \textit{Chandra} HETG, with SN~1983V positioned away from the aim point. Bare ACIS imaging observations were obtained in 2002 and 2006 (\textit{Chandra} sequence numbers 700616, 701286--701291), and were used for analysis. The data were reprocessed with standard processing tools in \textsc{Ciao 4.18}. Data were extracted from 10\arcsec\ diameter circular regions centered on the SN position, employing a local background subtraction. We measure rates of $< 10^{-3} \pm 10^{-4}$ counts~s$^{-1}$ in all observations. Using either a model for power-law emission or a 1~keV thermal plasma, we estimate
luminosities $< 8.7 \times 10^{37}$ erg~ s$^{-1}$ ($2.2 \times 10^4$ \lsolar), consistent with upper limits quoted by \cite{Perna+2008}.

Beyond the 0.3--10 keV energy band of \textit{Chandra} and \textit{XMM-Newton}, SN 1983V was also serendipitously observed multiple times with \textit{NuSTAR} at $\sim 10,500$ and $\sim 14,000$ days post-explosion. \textit{NuSTAR} has an effective energy range of 3--79 keV, and thus could potentially capture high-temperature X-ray emission not seen with \textit{Chandra}. Very high-energy X-ray emission could occur in certain interaction scenarios with special conditions in the shocks \citep{chevalier03}. However, owing to the proximity to the host galaxy and the lower resolution of \textit{NuSTAR}, the SN could not be isolated, and any potential flux variability at the SN location could simply be due to the active galactic nucleus (AGN) variability (\textit{NuSTAR} did not exist before SN 1983V exploded, so no pre- vs. post-explosion comparison is possible). Additionally, while it is true that \textit{NuSTAR} goes to higher energies than \textit{Chandra}, it would be quite unusual for \textit{NuSTAR} to catch emission that was fully missed by \textit{Chandra} owing to instrument specifics \citep[see, e.g.,][]{brethauer22}. We thus conclude that there is no clear evidence for any soft or hard X-ray emission from SN 1983V.

Finally, we note that {\it Swift}/UVOT UV observations exist, but the resolution is too poor to distinguish the SN from the nearby H\,\textsc{ii} region. We determine the total UV luminosity to be $<10^4$~\lsolar.

\begin{figure}[t]
\centering
\includegraphics[scale = .425, trim= 0in 0in 0in 0in]{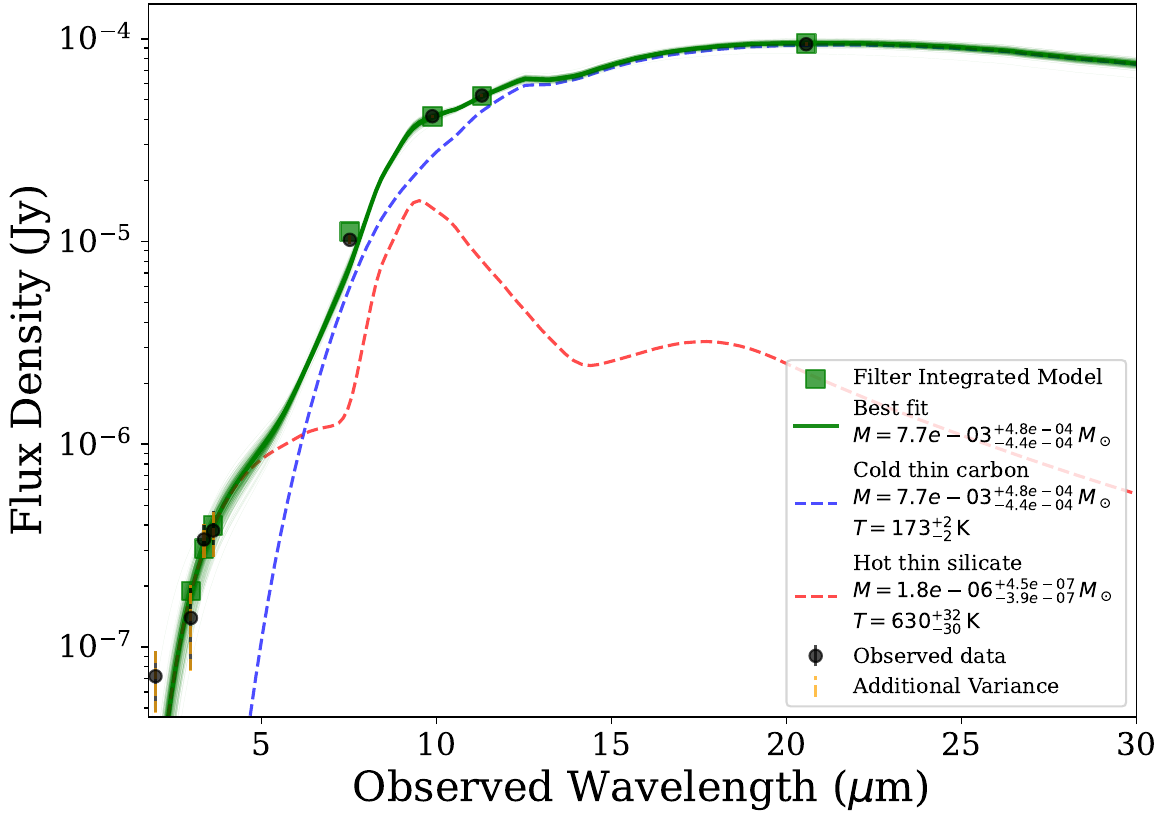}
\caption{A two-component dust model was fitted to the SED of SN 1983V. The green points show the integrated dust model, containing both hot (red, dashed) and cold (blue, dashed) dust components. The black points indicate the observed flux at each filter, with error bars. 
We find a total dust mass of $7.7 \times 10^{-3}$~\msolar\ best fits the SED.}
\label{fig:dust_model}
\end{figure}

\begin{deluxetable}{lcccc}
\tablecaption{Geometric Properties of Modeled Dust Components  \label{tab:param_fit}}
\tablehead{
\multicolumn{5}{c}{}  \\
\cline{1-5} 
\colhead{Dust Component}&\colhead{Composition}&\colhead{$L_{\rm tot}$ [erg s$^{-1}$]}&\colhead{$r_{\rm bb}$ [cm]}&\colhead{$\nu_{\rm exp}$ [km~s$^{-1}$]}}
\startdata
Hot  & Silicate & 8.62$\times 10^{37}$ & 8.78$\times 10^{14}$ & 7.13\\
Cold & Carbon   & 6.05$\times 10^{38}$ & 3.07$\times 10^{16}$ & 249.75\\
\enddata
\end{deluxetable}

\section{Analysis}
\label{sec:analysis}
\subsection{Dust Modeling}
\label{sec:dust_model}

We fit the IR SED (Table \ref{tab:jwst_obs})  as a function of dust mass ($M_{\rm d}$) and dust temperature ($T_{\rm d}$) by assuming that the fluxes are dominated by thermal emission from dust and can be modeled by a modified blackbody emission spectrum, following similar equations, derivations, and models outlined in previous work \citep[e.g.,][]{fox10, dwek19,Shahbandeh+2023}. In this analysis, we use the \texttt{dustysn}\footnote{https://dustysn.readthedocs.io/en/latest/index.html} implementation of these methods \citep{gomez25}. \texttt{dustysn} is a Python package that fits dust models with the {\tt emcee} implementation of Markov chain Monte Carlo \citep[MCMC;][]{Foreman-Mackey+2013}. The dust modeling in \texttt{dustysn} relies on the carbon and silicate mass absorption coefficients from \citet{Sarangi+2022}, which can be used to fit one-, two-, or three-component dust models with a range of fixed grain sizes. At the current time, \texttt{dustysn} only supports optically thin models. However, for this particular target, we only have four mid-IR photometry points, which limits our fits to two optically thin components (each with two parameters). 

We attempt both one- and two-component (hot and cold) dust models, considering a range of different combinations of dust compositions and grain sizes. We find the best-fitting model (lowest $\chi^2$) is a combination of hot silicate and cold carbon components for a grain size of 0.1~\micron\ (Figure \ref{fig:dust_model}; Table \ref{tab:photo_output}). The cold carbon component is found to have a temperature of 173~K and a mass of $7.7 \times 10^{-3}$~\msolar, and the hot silicate component has a temperature of 629~K and a mass of $1.8 \times 10^{-6}$~\msolar\ (see Appendix~\ref{appendix} for posterior).
The hot component contributes very little to the overall dust mass, with the cold component dominating the mid-IR flux at $>$10~\micron.
Dust models with different component and compositions yielded similar fits, temperatures and dust masses.


\subsection{Geometric Dust Properties}
\label{sec:geometry}
The overall geometry of the dust in SN 1983V cannot be derived directly from the dust models in Section \ref{sec:dust_model}. By assuming an optically thin and spherically symmetric shell, however, we can calculate a reasonable estimate of the dust position with the blackbody radius, $r_{\rm bb}^{}$. The blackbody radius corresponds to the minimum radius at which all of the observed dust can reside while still remaining optically thin. In other words, the dust may reside in a spherical shell with an even larger radius or may not even reside in a shell at all. However, if the radius of the shell were any smaller, the amount of dust observed would be distributed in an optically thick geometry.

The radius can be found by first calculating the total integrated IR luminosity, $L_{\rm IR}$, from the photometric flux, $F(\lambda)$,
\begin{align}
    L_{\rm IR} &= \int 4\pi r^2 \,F(\lambda)d\lambda 
    \, .
\end{align}
\noindent
The blackbody radius is then,
\begin{align}
    r_{\rm bb}^{} &= \sqrt{\frac{L_{\rm IR}^{}}{4\pi \sigma T^4}}\, , \\
     r_{\rm bb}^{} &=  \sqrt{\frac{\int 4\pi d^2\, F(\lambda)d\lambda}{4\pi \sigma T^4}}\, .
\end{align}

\begin{figure}[t]
\centering
\includegraphics[scale = .41, trim= 0in 0in 0in 0in]{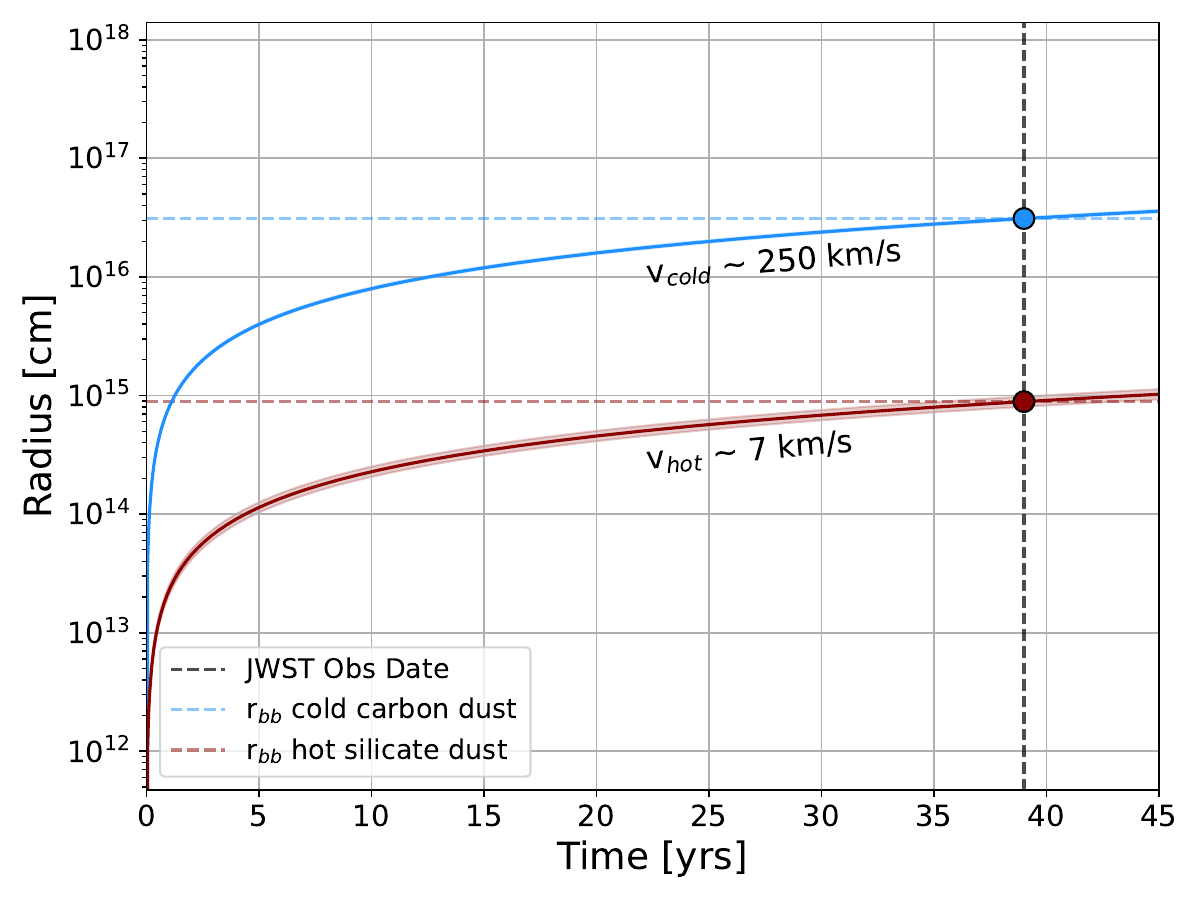}
\caption{Time since explosion versus the blackbody radius. 
The blackbody radius for each dust component is displayed as circled points. 
The derived expansion velocity is mapped across time since explosion.}
\label{fig:time_radius}
\end{figure}

It is a worthwhile exercise to consider the corresponding perceived expansion velocity required to reach the blackbody radius, as it serves as a useful reference point when considering dust-origin and heating-mechanism scenarios in Section \ref{sec:heating_mechanism}. For illustrative purposes, Figure~\ref{fig:time_radius} plots curves of constant velocity to highlight the expansion velocity that would correspond to the derived blackbody radii assuming a relatively constant expansion velocity over the past $\sim$40 yr since explosion. The expansion velocities are $\sim$7~km~s$^{-1}$\ and 250~km~s$^{-1}$\ for the hot and cold components, respectively. Table~\ref{tab:param_fit} reports the total IR integrated luminosity, blackbody radius, and perceived expansion velocity for each dust component in Table \ref{tab:photo_output}. 

\subsection{CSM Geometry and Pre-SN Mass-Loss}
\label{sec:csm_radio}

The radio limits from Section \ref{sec:vla} can be interpreted in the context of ejecta-CSM interaction to place limits on the progenitor mass-loss rate, as radio emission can only arise from the shocks generated from ejecta-CSM interaction in the absence of a central engine \citep{Chevalier_1998}. 
Using Equations 1-5 from \citet{Sfaradi_2025}, we assume a radius and magnetic field of the shock given some shock speed and CSM density, and assume no deceleration of the shock but rather some average shock velocity.

We derive limits for the cases where the upper limit is due to either a lack of synchrotron power or due to free-free absorption, without assuming we are observing the radio spectral peak, under standard radio SESN assumptions: $\epsilon_{\rm e}=\epsilon_{\rm B}=0.1$, electron energy index $p=3$, and CSM density gradient $s=2$, corresponding to a stellar wind density profile. Given the progenitor systems of SNe Ib/c are ambiguous (binary or single-star systems), and we have no direct constraints on the mechanism of mass loss \citep[e.g.,][]{fox22,zapartas26}, we assume a wind speed of 500 km~s$^{-1}$, as this is both a lower limit from what is expected for a Wolf-Rayet SN~Ic progenitor wind \citep{Soderberg_2012} and closer to the 100~km~s$^{-1}$ wind speeds expected from binary interaction \citep{Smith_2017}.

At 14,000 days, the density limits obtained from potential free-free absorption are nonconstraining at $>10~\mathrm{M_{\odot}~yr^{-1}}$ and incongruent with the data at other wavelengths, so we ignore this limit. 
For the synchrotron-dominated scenario, we obtain a limit of $\dot{M}<10^{-4}~\mathrm{M_{\odot}~yr^{-1}}$ for the earliest epoch, and similar limits of $<3\times 10^{-4}~\mathrm{M_{\odot}~yr^{-1}}$ at the later epochs.
These limits assume a shock speed of 10,000 km~s$^{-1}$ at 1500 days, decelerating to 4000 km~s$^{-1}$ at $>10,000$ days, consistent with the deceleration typically seen in radio SESNe \citep{Soderberg_2012}. 
Our derived mass-loss limits are sensitive to the choice of wind speed, namely by the linear relation $v_{\rm wind}/500\, \rm km\,s^{-1}$. 

\begin{figure}[!t]
\centering
\includegraphics[width=8 cm, height= 6 cm]{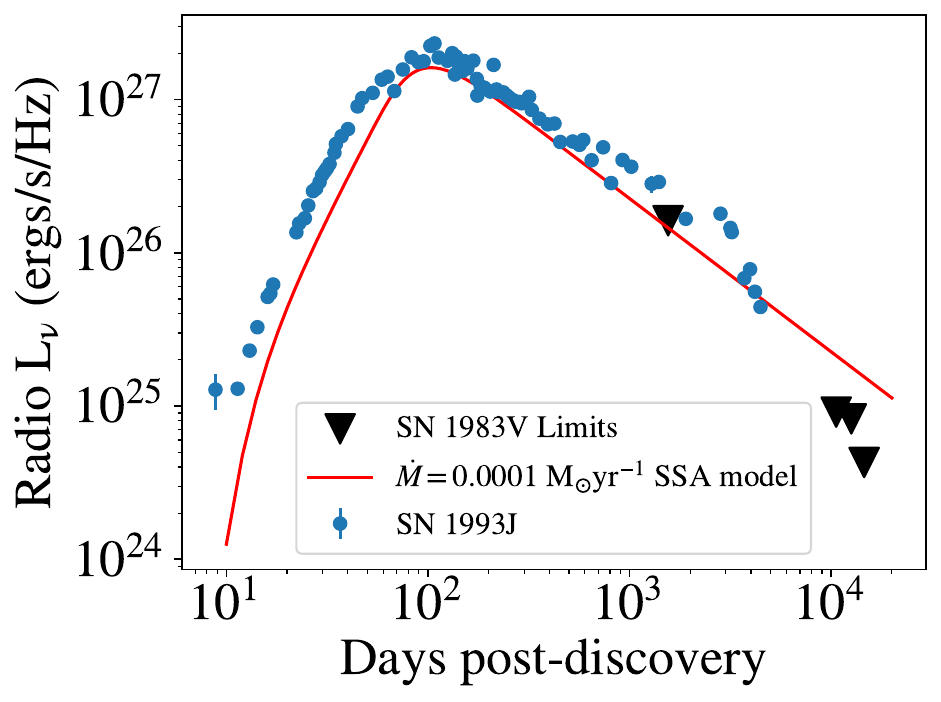}
\caption{A view of the radio upper limits of SN 1983V compared to SN 1993J (a standard radio SESN) with a model light curve for $\dot{M}=0.0001~\mathrm{M_{\odot}~yr^{-1}}$ and $v_{\rm sh}=10,000$~km~s$^{-1}$. Note that the plotted model assumes a constant shock speed, while the mass-loss limits at later epochs assume a decelerating shock; the late-epoch limits are therefore somewhat higher than they appear here. Data for SN 1993J are from \citet{Weiler_2007}.}
\label{fig:Radio_fig}
\end{figure}

Figure \ref{fig:Radio_fig} shows the radio light curve with a model for a $10^{-4}~\mathrm{M_{\odot}~yr^{-1}}$ progenitor mass-loss rate, alongside the radio C-band light curve of SN~1993J for comparison as a prototypical radio-bright SESN \citep{Weiler_2007}. This radio limit on mass-loss rate implies a limit on total shock luminosity of $<4\times 10^{39}$ erg~s$^{-1}$, assuming shock kinetic energy is fully converted to shock power, which may not be the case.
This follows \citet{Chugai_1991}: $L_{\rm sh}=(1/2)\dot{M}v_{\rm sh}^3/v_{\rm w}$. 

\section{Dust Origin and Heating Mechanism}
\label{sec:heating_mechanism}
The dust, CSM, and SN properties presented in Section \ref{sec:analysis} allow us to constrain the possible dust geometry and distribution. The derived blackbody radii, along with the implied expansion velocities, of both the cold and warm components place any corresponding dust shell well within the radius of the FW (assuming a shock velocity of $\sim 2000$--3000~km~s$^{-1}$). These shell properties are consistent with the radii and velocities of innermost ejecta. While it is tempting to conclude that the dust is newly formed in the ejecta, the dust geometry must tell a self-consistent story with any potential origin and heating mechanism. The various possibilities are outlined in detail by \citet{fox10}, including an IR light echo, radiative heating, and collisional heating. We explore these here.

\subsection{IR Light Echo}
\label{sec:ir_echo}
For an IR light-echo scenario, the peak SN luminosity heats an outer ring or shell of dust at a radius $r_{\rm d}^{}$, to a peak temperature $T_{\rm d}$. Light-travel-time effects cause the thermal radiation from the dust grains to reach the observer over an extended period, thereby forming an ``IR echo'' \citep{dwek83}. The IR luminosity plateau occurs on year-long timescales, corresponding to the light-travel time across the inner edge of the dust shell. As dust cools from the peak temperature, it will contribute flux at longer wavelengths. The fact that we observe IR emission from SN 1983V at $>40$~yr would imply a cavity of $>20$~ly ($>6$ pc) across in this scenario. This radius is not consistent with the blackbody radii derived above. Of course, the blackbody radii are only lower limits. If the dust was in fact distributed at $>20$~ly, Figure 8 of \citet{fox10} shows that the peak SN luminosity would be insufficient to heat the dust to observed temperatures. We therefore rule out the IR light-echo scenario for SN 1983V.

\subsection{Radiative Heating from the Forward Shock}
\label{sec:rad_heat}

While radioactive decay may be a viable source at early times ($< 1500$ days), it is not sufficient to heat dust to the temperatures derived from the \JWST data at $>10,000$~days. In a growing fraction of SNe with $>1000$ day dust emission, radiative emission (i.e., free-free bremsstrahlung, synchrotron, inverse Compton, and line emission) may be generated by either the FS or the reverse shock (RS), emitted in the X-rays, UV, and optical \citep[e.g.,][]{fox11,Shahbandeh+2023,Shahbandeh+2025,smith26}. We explore the multiwavelength data here.

Observationally, the H$\alpha$~luminosity (Section \ref{sec:hst_counterpart}) is nearly two orders of magnitude smaller than the total integrated dust luminosity and insufficient to singularly heat the dust to the observed luminosity. Not all emission from the FS necessarily escapes in H$\alpha$. In fact, observations of other interacting SNe reveal that only a fraction ($\sim 1$--10\%) of the energy emerges in the H$\alpha$ line \citep[e.g., SN 1978K;][]{ryder93}. This is particularly relevant for SN 1983V given it is an SN~Ib/c and likely devoid of a hydrogen envelope. More recent simulations predict that, in certain physical scenarios, a majority of shock power is likely to come out as X-rays, while the thermalized part of the power will emerge in the UV, especially in Ly$\alpha$~1215.67~\AA\ and Mg II~2800~\AA\ \citep[see][]{dessart22}. This has been observed in SN~2023ixf, where panchromatic UV-to-IR observations revealed that late-time X-ray emission arises from thermalization of the RS, with reprocessed shock power emerging across UV, optical, and IR wavelengths \citep{JacobsonGalan+2025}. The upper limits on the luminosities associated with {\it Chandra}~and {\it XMM-Newton} (Section \ref{sec:high_energy}), however, are an order of magnitude smaller than the total integrated dust luminosity. Furthermore, the upper limits on the luminosities associated with the radio data (Section \ref{sec:csm_radio}) are not particularly constraining. 

\begin{figure}[t!]
\centering
\includegraphics[width=\linewidth]{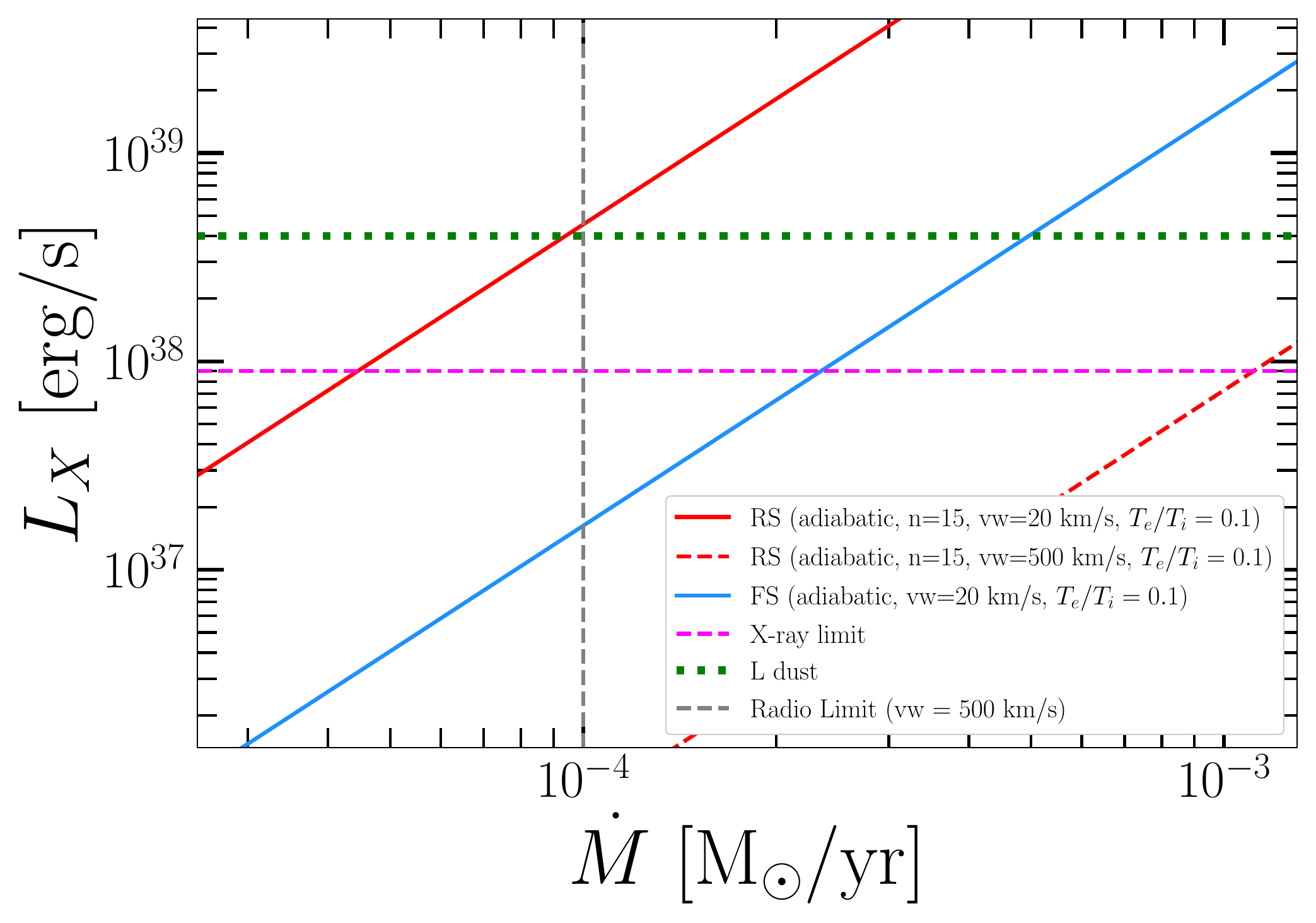}
\caption{X-ray luminosity versus mass-loss rate for different physical scenarios.}
\label{fig:mass_loss}
\end{figure}

Given these constraints on the radiated luminosity at each wavelength, it is increasingly difficult to invoke a radiative heating scenario as the dominant heating mechanism in SN 1983V. However, it could be the case that the dust is optically thick and the mid-IR emission corresponds to completely reprocessed luminosity. In this scenario, we would not expect to see any X-ray, UV, or even optical photons from this shock because they all get dumped into the ejecta/CDS and emerge as IR photons. Figure \ref{fig:mass_loss} plots the theoretical X-ray luminosity generated as a function of CSM pre-SN mass-loss rate for various RS and FS formalisms \citep{Fransson96, chevalier03, Chevalier17}. The limits for the H$\alpha$, X-ray, and radio suggest a nonradiative shock. We therefore only model adiabatic shocks. Overplotted horizontal lines indicate the integrated IR luminosity from the dust and the observational limits on the X-ray flux. The vertical line indicates the constraint on the pre-SN mass-loss rate from the radio observations ($10^{-4}~\mathrm{M_{\odot}~yr^{-1}}$; Section \ref{sec:csm_radio}), assuming a wind speed of 500 km~s$^{-1}$. 

For a 500 km~s$^{-1}$ wind, $10^{-4}~\mathrm{M_{\odot}~yr^{-1}}$, and free-free RS emission (dotted red line in Figure \ref{fig:mass_loss}), the luminosity is not close enough to heat the dust to the observed IR luminosity, even with 100\% reprocessing. To successfully power the dust with radiative emission, we require a scenario with a pre-SN wind speed of only 20 km~s$^{-1}$, while maintaining a mass-loss rate of $10^{-4}~\mathrm{M_{\odot}~yr^{-1}}$ (solid red line in Figure \ref{fig:mass_loss}). Such a low wind speed may be physically possible in some binary stripping scenarios, but with the radio limits, a wind speed of only 20 km~s$^{-1}$~would result in an adjusted limit on the mass-loss rate of $4\times 10^{-6}~\mathrm{M_{\odot}~yr^{-1}}$ (i.e., dividing by 25).

To circumvent the radio limits, we would need to consider that the radio observations were not taken at contemporaneous epochs to the \JWST\ data (last radio data taken in 2023, $\sim 1$ yr after the \JWST\ data) and thus missed the CSM interaction with a very dense shell, akin to SN 2014C \citep{milisavljevic15,tinyanont25}. Even then, the necessary combination of wind speed and mass-loss rate would require quite extraordinary circumstances for an SN Ibc. While a typical WR SN~Ibc progenitor star has a wind of 2000--4000 km~s$^{-1}$, the extreme mass loss invoked for the radiative-heating scenario would suggest something different from the traditional spherical tenuous winds. Instead, it would imply a nonspherical and circumbinary wind, most likely due to binary interaction.

Given all of the above constraints, it is increasingly difficult to invoke a scenario requiring radiative heating from the FS to heat the dust. It may be, after all, that the F657N source coincident with the SN is simply a nearby H\,\textsc{ii} region. Or, the F657N source may be associated with CSM interaction from the SN, but an alternative heating mechanism is present. 

\subsection{Ejecta Dust and the Reverse Shock} 
Another scenario invokes the RS as the heating source. In this case, the newly formed dust may reside in the inner unshocked SN ejecta. Heating could occur either by collisional heating of ejecta grains if they can reach high enough expansion velocities to pass through the RS at this epoch, or radiative heating that propagates inward to the unshocked SN ejecta dust.

In normal core-collapse SN ejecta, one typically expects most of the dust to form within a velocity coordinate of $\sim 2650$ km~s$^{-1}$ \citep{Sarangi+2022}.  The RS radius is a fraction ($\sim 75$\%) of the FS radius since the RS location is defined by the contact discontinuity between the ejecta and FS \citep[e.g.,][]{chevalier82,chevalier03}. We do not have an observational estimate of the RS location or corresponding velocity, but we assume an upper limit of $\sim$5000~km~s$^{-1}$.  Even if most of the newly formed dust has not reached this velocity, we cannot rule out the possibility that a very small mass of some dust has reached these velocities in denser portions of the ejecta. After all, the {\it HST}/F657N imaging supports the possibility of some H$\alpha$~emission. If so, this collisional heating may explain the hot ($\sim 630$~K) low-mass component (Figure \ref{fig:dust_model}). 

Given the expected velocity coordinate below which most dust forms, however, it seems more likely that the bulk of the mid-IR emission in this scenario (i.e., the cooler $\sim$170~K component) must arise from inner ejecta dust that is radiatively heated by inward-propagating RS radiation. If the SN ejecta dust is optically thick, the bulk of the dust we are seeing is an outer layer, and an even larger fraction of cooler ejecta dust is yet to be revealed by observations at longer far-IR or sub-mm wavelengths.  For this reason, the dust mass of $8 \times 10^{-3}$~M$_\odot$ estimated from the 170~K component is likely to be a lower limit to the true dust mass formed by SN 1983V.

This scenario, however, faces the same challenges as that invoking radiative heating from the FS (Section \ref{sec:rad_heat}). The upper limits on the optical, X-ray, and radio luminosities are all too small. Furthermore, unlike the pre-existing CSM, because the RS is beyond the ejecta, there is no obvious geometry that would allow for complete absorption and reprocessing of the radiative emission.

\subsection{CSM Dust and Collisional Heating by the FS}
\label{sec:col_heat}

While the radiative emission described in Section \ref{sec:rad_heat} is insufficient to heat the dust to the observed luminosity, the fact that there likely exists any shock interaction (Section \ref{sec:hst_counterpart}) at all comes as a surprise. CSM interaction is common in many SNe, particularly SNe~IIn, but quite rare in SESNe. SN 1983V was initially classified as an SESN and early-time observations did not show evidence for significant CSM interaction. This late-time detection of CSM interaction is most akin to the Type Ib SN 2014C \citep{milisavljevic15,margutti17,zhai25,tinyanont25}. 

An alternative energy source from the shock may be direct collisional heating. In such a scenario, hot electrons in the post-shock environment collisionally heat pre-existing dust grains \citep{dwek87,dwek08,fox10}. For collisional heating, however, the dust must be located at the position of the shock, which would imply the dust was most likely pre-existing and not newly formed. 

For SN 1983V, we already showed that the blackbody radius is much smaller than the shock radius. Of course, the blackbody radius is only a lower limit and spherical symmetry is assumed. It is highly probable that the CSM is distributed asymmetrically or in a torus, as it is with SN 2014C, likely owing to binary stripping \citep{mauerhan18,bietenholz18,thomas22,brethauer22,bietenholz21,orlando24}. If the dust shells are not spherically symmetric, and instead in clumps or a torus, the actual radius (i.e., distance from the explosion) will be larger (if we maintain it must be optically thin). To be at or beyond the CDS, assuming an FS velocity of $>2000$~km~s$^{-1}$, the cold dust component would have to be located a factor of 10 beyond its current blackbody radius ($\sim 10^{16}$--$10^{17}$ cm), which is quite feasible. 
\citet{orlando24} modeled the early interaction of SN 2014C with its CSM and found the torodial shell located at $10^{16}$--$10^{17}$ cm, in agreement with our findings above if we assume a toroidal geometry.

Following Equation 9 of \citet{fox10}, the upper limit of the total mass of gas heated by the shock (assuming spherical symmetry) at any given time is
\begin{equation}
\label{eqn:mass}
    M_{\rm g} (M_{\odot}) \approx 8.3\times10^{-5}~\bigg(\frac{v_{\rm sh}}{1000~{\rm km~s}^{-1}}\bigg)^3~\bigg(\frac{t}{{\rm yr}}\bigg)^2~\bigg(\frac{a}{\mu{\rm m}}\bigg)\, .
\end{equation}
\noindent
For $v_{\rm sh}\approx 4000$~km~s$^{-1}$, $t=42$~yr, and $a=0.1$~\micron, we find $M_{\rm g}\approx1$~\msolar. Assuming a typical gas-to-dust mass ratio of $\sim100$, this corresponds to a dust mass of $M_{\rm d}\approx0.01$~\msolar, close to what we derive in Table \ref{tab:photo_output}. Equation \ref{eqn:mass} assumes spherical symmetry, so the total amount of dust collisionally heated is only a fraction of the number derived above. On the other hand, the derivation assumes a post-shock gas temperature of $10^6$~K, which could easily be off by an order of magnitude. 

While we do not have strong constraints on the exact values that go into the final estimate of the dust mass, the calculation illustrates the feasibility of the collisional-heating scenario. Along with the fact that most other heating mechanisms are ruled out, we find that shock heating of a pre-existing ring of material is a viable explanation.

\subsection{Central Engine}
A final scenario invokes a central engine within a very dense, dusty shell. Here, the emission from the central source is nearly all reprocessed by the dust and emitted in the IR. Central engines include a central cooling neutron star (CNS) or pulsar wind nebula (PWN), akin to SN 1987A \citep{fransson24} or the Crab Nebula \citep{temim24}, or possibly a neighboring binary companion star, which are to be expected in a large fraction of SNe Ib/c \citep{fox22,zapartas26}. 

The observed dust luminosity of $\sim10^{5}$~\lsolar~(Table \ref{tab:photo_output}) requires a central engine of at least this brightness. Typically, any CNS or PWN is too faint, nearly two orders of magnitude dimmer. However, companion stars to SNe Ib/c are expected to be sufficiently massive and bright, such as a $\sim 35$~\msolar\ main-sequence star or a 25~\msolar\ blue supergiant \citep{zapartas26}.

The advantage of this scenario is that it does not require broad and/or boxy spectroscopic lines associated with the SN and CSM interaction. However, this scenario is difficult to prove since there is no observational test. Furthermore, while a central engine may be present but undetected, it has difficulty explaining the combination of an {\it HST}/F657N detection and {\it HST} broad-band nondetections (Section \ref{sec:hst_counterpart}). We therefore do not prefer this scenario. 

\begin{figure}[t]
\centering
\includegraphics[scale = .42, trim= 0in 0in 0in 0in]{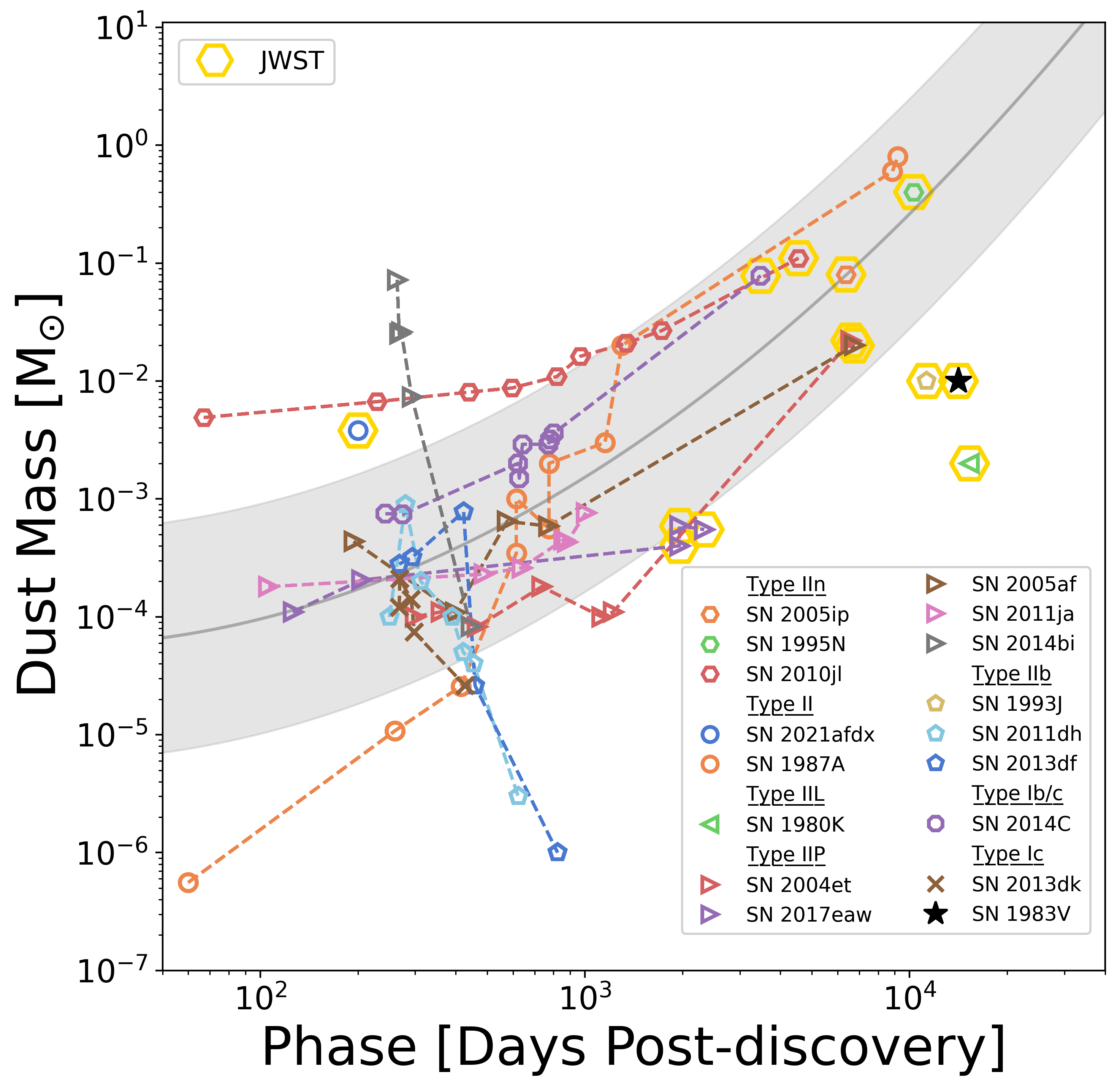}
\caption{The total observed SN dust mass versus time:
a compilation of SNe observed with mid-IR and far-IR observatories and recorded dust masses, denoting their specific SN type.
{\it JWST}-specific observations are indicated with a yellow hexagon \citep{Zsiros+24, Szalai+2025, clayton25, Shahbandeh+2023, Shahbandeh+2025, Sarangi+2025, tinyanont25, Pearson+2025, Hosseinzadeh+2023,smith26}.
All other SN were observed with {\it Spitzer}, some at multiple epochs \citep{szalai19a}.
The point for SN 1983V is shown as a black star. 
The only other Type Ib/c SESN observed and detected at day $>1000$ was SN 2014C.
An empirical trend is evident, but the source of this trend remains undetermined.}
\label{fig:dustmass_time}
\end{figure}

\subsection{Implications on Empirical Trends}
Figure~\ref{fig:dustmass_time} shows the measured dust mass for SN 1983V in context with other SN findings previously reported (see Section \ref{sec:intro}). 
SN 1983V exhibits a relatively large dust mass, and yet it is worthwhile to acknowledge that the dust estimate reported here is a lower limit. 
It is likely that dust exists at wavelengths $> 21$\,\micron\ or that dust may be hidden in optically thick regions. 

Although the compilation of SN dust masses empirically suggests a trend indicating increasing dust mass over time, SN 1983V offers an important reminder that this may not be the case physically.
For SN 1983V, we surmise that the dust is pre-existing.
However, in many other cases (e.g., SN 2017eaw, \citealt{Pearson+2025}; SN 2005ip, \citealt{Shahbandeh+2025}), the dust is newly formed or a combination of the two.
To what degree these dust populations affect or contribute to this apparent trend remains inconclusive with the current limit in sample size and breadth. 
It also may be that \JWST is now revealing colder dust previously inaccessible, and masses are reportedly higher. 
Additionally, dust may be revealing itself at later times owing to optical-depth effects (i.e., becoming optically thin).
This sample of SNe is diverse in SN type, instrument, analysis technique, and predicted dust source.
Additional studies, which are consistent and statistically similar in nature, will ascertain whether this trend is physical.

\section{Conclusions}
\label{sec:conclusions}

We present a photometric analysis of \JWST observations of SN 1983V, a $\sim40$~yr-old, stripped-envelope Type~Ic SN. Although we lack spectroscopic confirmation, we show extensive evidence supporting the likelihood that a source at the SN position is indeed the SN. This marks one of the oldest, dustiest SNe observed with {\it JWST}. Furthermore, we show that archival {\it HST} data reveal a coincident source only in the F657N filter, strongly suggesting the dusty SN has an H$\alpha$ point-source counterpart that may be indicative of ongoing CSM interaction.

PSF fitting was performed in each filter to construct the SED. 
A two-component dust model is fitted to the SED, consisting of a hot silicate component ($\sim1.8\times10^{-6}$~M$_{\odot}$; $\sim$630~K) and a cold carbon component ($\sim7.7\times10^{-3}$~M$_{\odot}$; $\sim$170~K). 
A total dust mass of $\sim 7.7 \times 10^{-3}$~M$_{\odot}$ places SN 1983V among the largest SN dust masses observed to date.
This value is likely a lower limit, as colder dust may exist at even longer wavelengths or may be hidden in optically thick regions.

To explain the dust origin and geometry, we consider several possible scenarios. Any preferred scenario must explain a strong IR source, the dust parameters (temperature, luminosity, and blackbody radius), a possible {\it HST}/F657N H$\alpha$~counterpart that is two orders of magnitude fainter than the mid-IR source, {\it HST} broad-band nondetections, the lack of broad H$\alpha$~or other spectroscopic lines typically associated with an SN, and upper limits in the X-rays, UV, and radio.

We rule out many different scenarios, with collisional heating of a pre-existing CSM by the FS being the only self-consistent and viable possibility remaining. In this scenario, any luminosity associated with the shock need not be comparable to the dust luminosity. Any resulting emission lines would be below the detection limit in the spectra and overwhelmed by the nearby H\,\textsc{ii} region. The dust must be located at the shock radius, which is well beyond the dust blackbody radius. However, if the same dust grains are distributed in clumps or a torus, they can easily be beyond the blackbody radius, which is highly likely in scenarios that invoke binary stripping, such as those of SN 2014C \citep{orlando24}. Late-time shocks are now heating this ring, analogous to SN 1987A \citep{panagia91,plait95,jones23}.

Although the dust is not likely newly formed, surviving dust around SESNe suggests their progenitors may still contribute to dust production, particularly in the early Universe when massive stars were more common. Whether this dust survives and contributes to the overall galactic dust budget remains an open question. Ongoing observations with \JWST will determine whether the dust survives the shocks. Additional filters and spectroscopy are essential to confirm the nature of the SN and to allow for more complex modeling that provides constraints on the dust geometry, optical depth, and composition. Furthermore, additional late-time IR studies of other similar events can clarify progenitor mass-loss and envelope-stripping mechanisms for the broader SESN population.

\section{Acknowledgments}
\label{sec:ack}

This work is based on observations made with the NASA/ESA/CSA {\it James Webb Space Telescope} and {\it Hubble Space Telescope}. Support was provided by NASA/{\it JWST} grants GO-01860, GO-3921, AR-06356, AR-8883, and AR-12264. Data were obtained from the Mikulski Archive for Space Telescopes (MAST) at the Space Telescope Science Institute (STScI), which is operated by the Association of Universities for Research in Astronomy, Inc., under NASA contract NAS 5-03127 for {\it JWST}. The \JWST observations are associated with PID 2107, while the {\it HST} observations are associated with PID 58684 and 60145. The specific observations analyzed can be accessed via \dataset[https://doi.org/10.17909/je9z-mb61]{https://doi.org/10.17909/je9z-mb61}.  Support to MAST for these data is provided by the NASA Office of Space Science via grant NAG5–7584 and by other grants and contracts.  This work originated from a project of the Summer Program in Astrophysics 2025 held at the University of Virginia, and funded by the Center for Global Inquiry and Innovation, the National Science Foundation (grant 2452494), the National Radio Astronomy Observatory (NRAO), the Kavli Foundation, and the Heising-Simons Foundation. We thank Shazrene Mohamed for organizing this year's program at the University of Virginia. This work was supported by the NASA Future Investigators in NASA Earth and Space Science and Technology (FINESST) award, proposal number 23-ASTRO23-0053. We would like to thank Claes Fransson for useful discussions on the forward and reverse shock models. A.V.F. was additionally funded by private donations. Some of the data presented herein were obtained at the W. M. Keck Observatory, which is operated as a scientific partnership among the California Institute of Technology, the University of California, and NASA; the observatory was made possible by the generous financial support of the W. M. Keck Foundation.

\clearpage
\appendix
\label{appendix}
\section{Dust-Model Priors and Posterior}

The priors on dust mass and temperature for the \texttt{dustysn} models presented in \ref{sec:dust_model} are flat priors, chosen over physically motivated ranges.

\begin{figure}[h!]
\centering
\includegraphics[width=0.7\paperwidth]{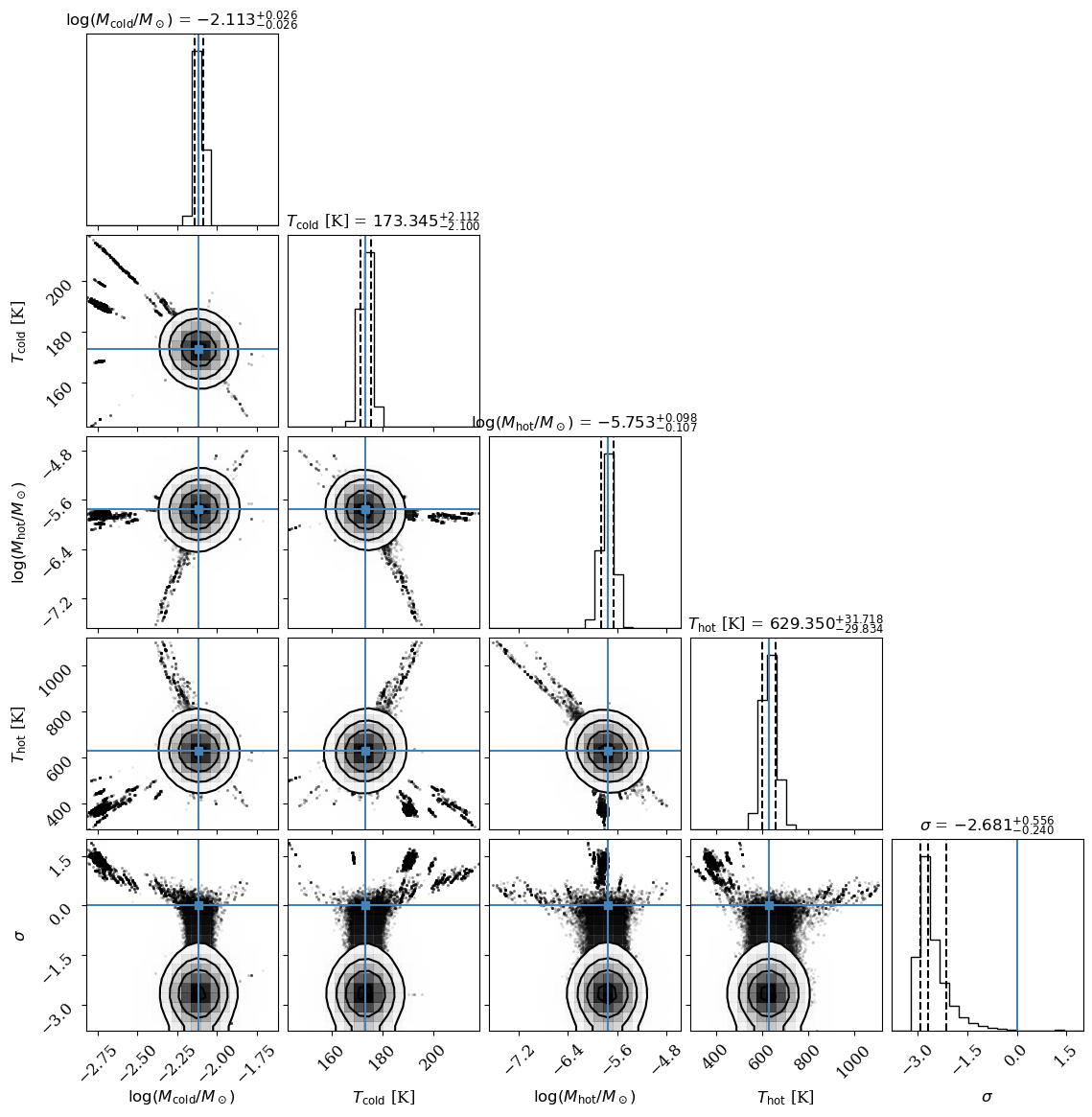}
\caption{Posterior distribution of dust-model parameters for the dust model presented in Figure \ref{fig:dust_model}.}
\label{fig:corner}
\end{figure}

We adopt a two-component model consisting of a cold carbon and a hot silicate dust component, with a grain size of $\rm 0.1\mu m$. For the cold component, we apply a log-uniform prior on dust mass of $\log(M_{\rm dust, cold}/{\rm M}_\odot) \in [-6, 1]$ and a uniform prior on temperature of $T_{\rm cold} \in [20, 2000] \, {\rm K}$. For the hot component, we use a log-uniform prior on dust mass of $\log(M_{\rm dust, hot}/{\rm M}_\odot) \in [-8, 1]$ and a uniform prior on temperature of $T_{\rm hot} \in [20, 3000] \, {\rm K}$. The model uses a redshift and distance of $z=0.0054$ and $18.2$ Mpc, respectively. Finally, $\sigma$ is included to account for potential underestimation of flux uncertainties, with a uniform prior of $\sigma \in [-3,3]$ (dimensionless, in log space). The posterior distribution corner plots for the dust model fit shown in Figure \ref{fig:dust_model} are presented in Figure \ref{fig:corner}.

\pagebreak
\bibliography{refs}{}

@ARTICLE{gordon03,
       author = {{Gordon}, Karl D. and {Clayton}, Geoffrey C. and {Decleir}, Marjorie and {Fitzpatrick}, E.~L. and {Massa}, Derck and {Misselt}, Karl A. and {Tollerud}, Erik J.},
        title = "{One Relation for All Wavelengths: The Far-ultraviolet to Mid-infrared Milky Way Spectroscopic R(V)-dependent Dust Extinction Relationship}",
      journal = {\apj},
         year = 2023,
        month = jun,
       volume = {950},
       number = {2},
          eid = {86},
        pages = {86},
          doi = {10.3847/1538-4357/accb59},
archivePrefix = {arXiv},
       eprint = {2304.01991},
 primaryClass = {astro-ph.GA},
       adsurl = {https://ui.adsabs.harvard.edu/abs/2023ApJ...950...86G}
}

@ARTICLE{karambelkar26,
       author = {{Karambelkar}, Viraj and {Kasliwal}, Mansi M. and {Lau}, Ryan M. and {Jencson}, Jacob E. and {Blagorodnova}, Nadejda and {G{\'o}mez-Mu{\~n}oz}, Marco A. and {Tranin}, Hugo and {Wavasseur}, Maxime and {Shahbandeh}, Melissa and {De}, Kishalay},
        title = "{Hot Springs and Dust Reservoirs: JWST Reveals the Dusty, Molecular Aftermath of Extragalactic Stellar Mergers}",
      journal = {\apj},
         year = 2026,
        month = mar,
       volume = {999},
       number = {1},
          eid = {16},
        pages = {16},
          doi = {10.3847/1538-4357/ae38bf},
archivePrefix = {arXiv},
       eprint = {2508.03932},
 primaryClass = {astro-ph.SR},
       adsurl = {https://ui.adsabs.harvard.edu/abs/2026ApJ...999...16K}
}

@ARTICLE{subrayan26,
       author = {{Subrayan}, Bhagya M. and {Sand}, David J. and {Culbert}, Olivia and {Andrews}, Jennifer E. and {Pearson}, Jeniveve and {Hosseinzadeh}, Griffin and {Jha}, Saurabh W. and {Valenti}, Stefano and {Bostroem}, K. Azalee and {Ransome}, Conor L. and {Ravi}, Aravind P. and {Aamer}, Aysha and {Andrews}, Moira and {Beasor}, Emma R. and {Christy}, Collin and {Dong}, Yize and {Franz}, Noah and {Hoang}, Emily and {Hsu}, Brian and {Jencson}, Jacob and {Kwok}, Lindsey A. and {Lundquist}, M.~J. and {Mehta}, Darshana and {Meza Retamal}, Nicolas and {Shrestha}, Manisha and {Smith}, Nathan and {Vasylyev}, Sergiy},
        title = "{A JWST/MIRI Study of Dust in a Sample of Normal Type IIP Core Collapse Supernovae}",
      journal = {arXiv e-prints},
         year = 2026,
        month = aug,
          eid = {arXiv:2608.16979},
        pages = {arXiv:2608.16979},
          doi = {10.48550/arXiv.2608.16979},
archivePrefix = {arXiv},
       eprint = {2608.16979},
 primaryClass = {astro-ph.HE},
       adsurl = {https://ui.adsabs.harvard.edu/abs/2026arXiv260816979S}
}

@ARTICLE{temim24,
       author = {{Temim}, Tea and {Laming}, J. Martin and {Kavanagh}, P.~J. and {Smith}, Nathan and {Slane}, Patrick and {Blair}, William P. and {De Looze}, Ilse and {Bucciantini}, Niccol{\`o} and {Jerkstrand}, Anders and {Gountanis}, Nicole Marcelina and {Sankrit}, Ravi and {Milisavljevic}, Dan and {Rest}, Armin and {Lyutikov}, Maxim and {DePasquale}, Joseph and {Martin}, Thomas and {Drissen}, Laurent and {Raymond}, John and {Fox}, Ori D. and {Modjaz}, Maryam and {Spitkovsky}, Anatoly and {Strolger}, Louis-Gregory},
        title = "{Dissecting the Crab Nebula with JWST: Pulsar Wind, Dusty Filaments, and Ni/Fe Abundance Constraints on the Explosion Mechanism}",
      journal = {\apjl},
         year = 2024,
        month = jun,
       volume = {968},
       number = {2},
          eid = {L18},
        pages = {L18},
          doi = {10.3847/2041-8213/ad50d1},
archivePrefix = {arXiv},
       eprint = {2406.00172},
 primaryClass = {astro-ph.HE},
       adsurl = {https://ui.adsabs.harvard.edu/abs/2024ApJ...968L..18T}
}

@ARTICLE{fransson24,
       author = {{Fransson}, C. and {Barlow}, M.~J. and {Kavanagh}, P.~J. and {Larsson}, J. and {Jones}, O.~C. and {Sargent}, B. and {Meixner}, M. and {Bouchet}, P. and {Temim}, T. and {Wright}, G.~S. and {Blommaert}, J.~A.~D.~L. and {Habel}, N. and {Hirschauer}, A.~S. and {Hjorth}, J. and {Lenki{\'c}}, L. and {Tikkanen}, T. and {Wesson}, R. and {Coulais}, A. and {Fox}, O.~D. and {Gastaud}, R. and {Glasse}, A. and {Jaspers}, J. and {Krause}, O. and {Lau}, R.~M. and {Nayak}, O. and {Rest}, A. and {Colina}, L. and {van Dishoeck}, E.~F. and {G{\"u}del}, M. and {Henning}, Th. and {Lagage}, P.-O. and {{\"O}stlin}, G. and {Ray}, T.~P. and {Vandenbussche}, B.},
        title = "{Emission lines due to ionizing radiation from a compact object in the remnant of Supernova 1987A}",
      journal = {Science},
         year = 2024,
        month = feb,
       volume = {383},
       number = {6685},
        pages = {898-903},
          doi = {10.1126/science.adj5796},
archivePrefix = {arXiv},
       eprint = {2403.04386},
 primaryClass = {astro-ph.HE},
       adsurl = {https://ui.adsabs.harvard.edu/abs/2024Sci...383..898F}
}

@ARTICLE{zapartas26,
       author = {{Zapartas}, E. and {Fox}, O.~D. and {Su}, J. and {Souropanis}, D. and {Drout}, M.~R. and {Rocha}, K.~A. and {van Dyk}, S.~D. and {Williams}, B.~F. and {Briel}, M. and {Renzo}, M. and {Andrews}, J.~J. and {Fragos}, T. and {Gossage}, S. and {Kruckow}, M.~U. and {Liotine}, C. and {Ryder}, S.~D. and {Srivastava}, P.~M. and {Teng}, E.},
        title = "{The demographics of binary companions to stripped-envelope supernovae: confronting population synthesis models with observations}",
      journal = {\mnras},
         year = 2026,
        month = feb,
       volume = {546},
       number = {2},
          eid = {staf2208},
        pages = {staf2208},
          doi = {10.1093/mnras/staf2208},
archivePrefix = {arXiv},
       eprint = {2508.12677},
 primaryClass = {astro-ph.SR},
       adsurl = {https://ui.adsabs.harvard.edu/abs/2026MNRAS.546f2208Z}
}

@ARTICLE{fox22,
       author = {{Fox}, Ori D. and {Van Dyk}, Schuyler D. and {Williams}, Benjamin F. and {Drout}, Maria and {Zapartas}, Emmanouil and {Smith}, Nathan and {Milisavljevic}, Dan and {Andrews}, Jennifer E. and {Bostroem}, K. Azalee and {Filippenko}, Alexei V. and {Gomez}, Sebastian and {Kelly}, Patrick L. and {de Mink}, S.~E. and {Pierel}, Justin and {Rest}, Armin and {Ryder}, Stuart and {Sravan}, Niharika and {Strolger}, Lou and {Wang}, Qinan and {Weil}, Kathryn E.},
        title = "{The Candidate Progenitor Companion Star of the Type Ib/c SN 2013ge}",
      journal = {\apjl},
         year = 2022,
        month = apr,
       volume = {929},
       number = {1},
          eid = {L15},
        pages = {L15},
          doi = {10.3847/2041-8213/ac5890},
archivePrefix = {arXiv},
       eprint = {2203.01357},
 primaryClass = {astro-ph.HE},
       adsurl = {https://ui.adsabs.harvard.edu/abs/2022ApJ...929L..15F}
}

@ARTICLE{smith26,
       author = {{Smith}, Nathan and {Shahbandeh}, Melissa and {Fox}, Ori D. and {Moore}, Thomas and {Andrews}, Jennifer E. and {Brink}, Thomas G. and {Dwek}, Eli and {Engesser}, Michael and {Filippenko}, Alexei V. and {Nickson}, Bryony and {Temim}, Tea and {Yang}, Yi and {Zheng}, WeiKang and {Ashall}, Chris and {Baer-Way}, Raphael and {Clayton}, Geoffrey C. and {Jencson}, Jacob E. and {Johansson}, Joel and {Lane}, Zachary G. and {Milisavljevic}, Dan and {Rest}, Armin and {Sarangi}, Arkaprabha and {Szalai}, Tam{\'a}s and {Van Dyk}, Schuyler D. and {Williams}, Brian J.},
        title = "{JWST Spectra Indicate a Large Mass of Postshock Dust Formed by SN 2010jl}",
      journal = {\apj},
         year = 2026,
        month = aug,
       volume = {1006},
       number = {2},
          eid = {224},
        pages = {224},
          doi = {10.3847/1538-4357/ae74c5},
       adsurl = {https://ui.adsabs.harvard.edu/abs/2026ApJ..1006..224S}
}

@ARTICLE{Fransson96,
       author = {{Fransson}, Claes and {Lundqvist}, Peter and {Chevalier}, Roger A.},
        title = "{Circumstellar Interaction in SN 1993J}",
      journal = {\apj},
         year = 1996,
        month = apr,
       volume = {461},
        pages = {993},
          doi = {10.1086/177119},
       adsurl = {https://ui.adsabs.harvard.edu/abs/1996ApJ...461..993F}
}

@INCOLLECTION{Chevalier17,
       author = {{Chevalier}, Roger A. and {Fransson}, Claes},
        title = "{Thermal and Non-thermal Emission from Circumstellar Interaction}",
    booktitle = {Handbook of Supernovae},
         year = 2017,
       editor = {{Alsabti}, Athem W. and {Murdin}, Paul},
        pages = {875},
          doi = {10.1007/978-3-319-21846-5_34},
       adsurl = {https://ui.adsabs.harvard.edu/abs/2017hsn..book..875C}
}

@ARTICLE{maschmann24,
       author = {{Maschmann}, Daniel and {Lee}, Janice C. and {Thilker}, David A. and {Whitmore}, Bradley C. and {Deger}, Sinan and {Boquien}, M{\'e}d{\'e}ric and {Chandar}, Rupali and {Dale}, Daniel A. and {Wofford}, Aida and {Hannon}, Stephen and {Larson}, Kirsten L. and {Leroy}, Adam K. and {Schinnerer}, Eva and {Rosolowsky}, Erik and {{\'U}beda}, Leonardo and {Barnes}, Ashley T. and {Emsellem}, Eric and {Grasha}, Kathryn and {Groves}, Brent and {Indebetouw}, R{\'e}my and {Kim}, Hwihyun and {Klessen}, Ralf S. and {Kreckel}, Kathryn and {Levy}, Rebecca C. and {Pinna}, Francesca and {Rodr{\'\i}guez}, M. Jimena and {Tian}, Qiushi and {Williams}, Thomas G.},
        title = "{PHANGS-HST Catalogs for {\ensuremath{\sim}}100,000 Star Clusters and Compact Associations in 38 Galaxies. I. Observed Properties}",
      journal = {\apjs},
         year = 2024,
        month = jul,
       volume = {273},
       number = {1},
          eid = {14},
        pages = {14},
          doi = {10.3847/1538-4365/ad3cd3},
archivePrefix = {arXiv},
       eprint = {2403.04901},
 primaryClass = {astro-ph.GA},
       adsurl = {https://ui.adsabs.harvard.edu/abs/2024ApJS..273...14M}
}

@ARTICLE{lee22,
       author = {{Lee}, Janice C. and {Whitmore}, Bradley C. and {Thilker}, David A. and {Deger}, Sinan and {Larson}, Kirsten L. and {Ubeda}, Leonardo and {Anand}, Gagandeep S. and {Boquien}, M{\'e}d{\'e}ric and {Chandar}, Rupali and {Dale}, Daniel A. and {Emsellem}, Eric and {Leroy}, Adam K. and {Rosolowsky}, Erik and {Schinnerer}, Eva and {Schmidt}, Judy and {Lilly}, James and {Turner}, Jordan and {Van Dyk}, Schuyler and {White}, Richard L. and {Barnes}, Ashley T. and {Belfiore}, Francesco and {Bigiel}, Frank and {Blanc}, Guillermo A. and {Cao}, Yixian and {Chevance}, Melanie and {Congiu}, Enrico and {Egorov}, Oleg V. and {Glover}, Simon C.~O. and {Grasha}, Kathryn and {Groves}, Brent and {Henshaw}, Jonathan D. and {Hughes}, Annie and {Klessen}, Ralf S. and {Koch}, Eric and {Kreckel}, Kathryn and {Kruijssen}, J.~M. Diederik and {Liu}, Daizhong and {Lopez}, Laura A. and {Mayker}, Ness and {Meidt}, Sharon E. and {Murphy}, Eric J. and {Pan}, Hsi-An and {Pety}, J{\'e}r{\^o}me and {Querejeta}, Miguel and {Razza}, Alessandro and {Saito}, Toshiki and {S{\'a}nchez-Bl{\'a}zquez}, Patricia and {Santoro}, Francesco and {Sardone}, Amy and {Scheuermann}, Fabian and {Schruba}, Andreas and {Sun}, Jiayi and {Usero}, Antonio and {Watkins}, E. and {Williams}, Thomas G.},
        title = "{The PHANGS-HST Survey: Physics at High Angular Resolution in Nearby Galaxies with the Hubble Space Telescope}",
      journal = {\apjs},
         year = 2022,
        month = jan,
       volume = {258},
       number = {1},
          eid = {10},
        pages = {10},
          doi = {10.3847/1538-4365/ac1fe5},
archivePrefix = {arXiv},
       eprint = {2101.02855},
 primaryClass = {astro-ph.GA},
       adsurl = {https://ui.adsabs.harvard.edu/abs/2022ApJS..258...10L}
}

@ARTICLE{smith17_lbv,
       author = {{Smith}, Nathan},
        title = "{Luminous blue variables and the fates of very massive stars}",
      journal = {Philosophical Transactions of the Royal Society of London Series A},
         year = 2017,
        month = sep,
       volume = {375},
       number = {2105},
          eid = {20160268},
        pages = {20160268},
          doi = {10.1098/rsta.2016.0268},
       adsurl = {https://ui.adsabs.harvard.edu/abs/2017RSPTA.37560268S}
}

@INCOLLECTION{chevalier03,
       author = {{Chevalier}, R.~A. and {Fransson}, C.},
        title = "{Supernova Interaction with a Circumstellar Medium}",
    booktitle = {Supernovae and Gamma-Ray Bursters},
         year = 2003,
       editor = {{Weiler}, K.},
       volume = {598},
        pages = {171-194},
          doi = {10.1007/3-540-45863-8_10},
       adsurl = {https://ui.adsabs.harvard.edu/abs/2003LNP...598..171C}
}

@ARTICLE{chevalier82,
       author = {{Chevalier}, R.~A.},
        title = "{Self-similar solutions for the interaction of stellar ejecta with an external medium.}",
      journal = {\apj},
         year = 1982,
        month = jul,
       volume = {258},
        pages = {790-797},
          doi = {10.1086/160126},
       adsurl = {https://ui.adsabs.harvard.edu/abs/1982ApJ...258..790C}
}

@ARTICLE{xiao26,
       author = {{Xiao}, Lin and {Peng}, Zeyue and {Galbany}, Lluis and {Szalai}, Tamas and {Fox}, Ori D. and {Hu}, Lei and {Hu}, Maokai and {Pessi}, Thallis and {Yang}, Yi and {Moriya}, Takashi J. and {Han}, Zhanwen and {Wang}, Xiaofeng and {Yan}, Shengyu},
        title = "{The environmental dependence of mid-IR luminous dusty Supernovae}",
      journal = {arXiv e-prints},
         year = 2026,
        month = feb,
          eid = {arXiv:2602.23673},
        pages = {arXiv:2602.23673},
          doi = {10.48550/arXiv.2602.23673},
archivePrefix = {arXiv},
       eprint = {2602.23673},
 primaryClass = {astro-ph.HE},
       adsurl = {https://ui.adsabs.harvard.edu/abs/2026arXiv260223673X}
}

@ARTICLE{plait95,
       author = {{Plait}, Philip C. and {Lundqvist}, Peter and {Chevalier}, Roger A. and {Kirshner}, Robert P.},
        title = "{HST Observations of the Ring around SN 1987A}",
      journal = {\apj},
         year = 1995,
        month = feb,
       volume = {439},
        pages = {730},
          doi = {10.1086/175213},
       adsurl = {https://ui.adsabs.harvard.edu/abs/1995ApJ...439..730P}
}

@ARTICLE{Oke_1995,
       author = {{Oke}, J.~B. and {Cohen}, J.~G. and {Carr}, M. and {Cromer}, J. and {Dingizian}, A. and {Harris}, F.~H. and {Labrecque}, S. and {Lucinio}, R. and {Schaal}, W. and {Epps}, H. and {Miller}, J.},
        title = "{The Keck Low-Resolution Imaging Spectrometer}",
      journal = {\pasp},
         year = 1995,
        month = apr,
       volume = {107},
        pages = {375},
          doi = {10.1086/133562},
       adsurl = {https://ui.adsabs.harvard.edu/abs/1995PASP..107..375O}
}

@ARTICLE{Prochaska_2020,
       author = {{Prochaska}, J. and {Hennawi}, Joseph and {Westfall}, Kyle and {Cooke}, Ryan and {Wang}, Feige and {Hsyu}, Tiffany and {Davies}, Frederick and {Farina}, Emanuele and {Pelliccia}, Debora},
        title = "{PypeIt: The Python Spectroscopic Data Reduction Pipeline}",
      journal = {The Journal of Open Source Software},
         year = 2020,
        month = dec,
       volume = {5},
       number = {56},
          eid = {2308},
        pages = {2308},
          doi = {10.21105/joss.02308},
archivePrefix = {arXiv},
       eprint = {2005.06505},
 primaryClass = {astro-ph.IM},
       adsurl = {https://ui.adsabs.harvard.edu/abs/2020JOSS....5.2308P}
}

@ARTICLE{Medler_2025,
       author = {{Medler}, K. and {Ashall}, C. and {Hoeflich}, P. and {Baron}, E. and {DerKacy}, J.~M. and {Shahbandeh}, M. and {Mera}, T. and {Pfeffer}, C.~M. and {Hoogendam}, W.~B. and {Jones}, D.~O. and {Shiber}, S. and {Fereidouni}, E. and {Fox}, O.~D. and {Jencson}, J. and {Galbany}, L. and {Hinkle}, J.~T. and {Tucker}, M.~A. and {Shappee}, B.~J. and {Huber}, M.~E. and {Auchettl}, K. and {Angus}, C.~R. and {Desai}, D.~D. and {Do}, A. and {Payne}, A.~V. and {Shi}, J. and {Kong}, M.~Y. and {Romagnoli}, S. and {Syncatto}, A. and {Burns}, C.~R. and {Clayton}, G. and {Dulude}, M. and {Engesser}, M. and {Filippenko}, A.~V. and {Gomez}, S. and {Hsiao}, E.~Y. and {de Jaeger}, T. and {Johansson}, J. and {Krisciunas}, K. and {Kumar}, S. and {Lu}, J. and {Matsuura}, M. and {Mazzali}, P.~A. and {Milisavljevic}, D. and {Morrell}, N. and {O'Steen}, R. and {Park}, S. and {Phillips}, M.~M. and {Ravi}, A.~P. and {Rest}, A. and {Rho}, J. and {Suntzeff}, N.~B. and {Sarangi}, A. and {Smith}, N. and {Stritzinger}, M.~D. and {Strolger}, L. and {Szalai}, T. and {Temim}, T. and {Tinyanont}, S. and {Van Dyk}, S.~D. and {Wang}, L. and {Wang}, Q. and {Wesson}, R. and {Yang}, Y. and {Zs{\'\i}ros}, S.},
        title = "{JWST Observations of SN 2023ixf. II. The Panchromatic Evolution between 250 and 720 Days after the Explosion}",
      journal = {\apj},
         year = 2025,
        month = nov,
       volume = {993},
       number = {2},
          eid = {191},
        pages = {191},
          doi = {10.3847/1538-4357/ae0736},
archivePrefix = {arXiv},
       eprint = {2507.19727},
 primaryClass = {astro-ph.SR},
       adsurl = {https://ui.adsabs.harvard.edu/abs/2025ApJ...993..191M}
}

@ARTICLE{JacobsonGalan+2025,
       author = {{Jacobson-Gal{\'a}n}, W.~V. and {Dessart}, L. and {Kilpatrick}, C.~D. and {Patel}, P.~J. and {Auchettl}, K. and {Tinyanont}, S. and {Margutti}, R. and {Dwarkadas}, V.~V. and {Bostroem}, K.~A. and {Chornock}, R. and {Foley}, R.~J. and {Abunemeh}, H. and {Ahumada}, T. and {Arunachalam}, P. and {Bustamante-Rosell}, M.~J. and {Coulter}, D.~A. and {Gall}, C. and {Gao}, H. and {Guo}, X. and {Jones}, D.~O. and {Hjorth}, J. and {Kaewmookda}, M. and {Kasliwal}, M.~M. and {Kaur}, R. and {Larison}, C. and {LeBaron}, N. and {Miao}, H.-Y. and {Narayan}, G. and {Pan}, Y.-C. and {Park}, S.~H. and {Patra}, K.~C. and {Qin}, Y. and {Ransome}, C.~L. and {Rest}, A. and {Rho}, J. and {Rose}, S. and {Sears}, H. and {Swift}, J.~J. and {Taggart}, K. and {Villar}, V.~A. and {Wang}, Q. and {Zenati}, Y. and {Zhou}, H.},
        title = "{A Panchromatic View of Late-time Shock Power in the Type II Supernova 2023ixf}",
      journal = {\apjl},
         year = 2025,
        month = nov,
       volume = {994},
       number = {1},
          eid = {L14},
        pages = {L14},
          doi = {10.3847/2041-8213/ae157a},
archivePrefix = {arXiv},
       eprint = {2508.11747},
 primaryClass = {astro-ph.HE},
       adsurl = {https://ui.adsabs.harvard.edu/abs/2025ApJ...994L..14J}
}

@ARTICLE{2025A&A...700A.223K,
       author = {{Kravtsov}, T. and {Anderson}, J.~P. and {Kuncarayakti}, H. and {Maeda}, K. and {Mattila}, S.},
        title = "{Discovery of young, oxygen-rich supernova remnants in PHANGS-MUSE galaxies}",
      journal = {\aap},
         year = 2025,
        month = aug,
       volume = {700},
          eid = {A223},
        pages = {A223},
          doi = {10.1051/0004-6361/202349083},
archivePrefix = {arXiv},
       eprint = {2409.06504},
 primaryClass = {astro-ph.HE},
       adsurl = {https://ui.adsabs.harvard.edu/abs/2025A&A...700A.223K}
}

@INPROCEEDINGS{2010SPIE.7735E..08B,
       author = {{Bacon}, R. and {Accardo}, M. and {Adjali}, L. and {Anwand}, H. and {Bauer}, S. and {Biswas}, I. and {Blaizot}, J. and {Boudon}, D. and {Brau-Nogue}, S. and {Brinchmann}, J. and {Caillier}, P. and {Capoani}, L. and {Carollo}, C.~M. and {Contini}, T. and {Couderc}, P. and {Daguis{\'e}}, E. and {Deiries}, S. and {Delabre}, B. and {Dreizler}, S. and {Dubois}, J. and {Dupieux}, M. and {Dupuy}, C. and {Emsellem}, E. and {Fechner}, T. and {Fleischmann}, A. and {Fran{\c{c}}ois}, M. and {Gallou}, G. and {Gharsa}, T. and {Glindemann}, A. and {Gojak}, D. and {Guiderdoni}, B. and {Hansali}, G. and {Hahn}, T. and {Jarno}, A. and {Kelz}, A. and {Koehler}, C. and {Kosmalski}, J. and {Laurent}, F. and {Le Floch}, M. and {Lilly}, S.~J. and {Lizon}, J.-L. and {Loupias}, M. and {Manescau}, A. and {Monstein}, C. and {Nicklas}, H. and {Olaya}, J.-C. and {Pares}, L. and {Pasquini}, L. and {P{\'e}contal-Rousset}, A. and {Pell{\'o}}, R. and {Petit}, C. and {Popow}, E. and {Reiss}, R. and {Remillieux}, A. and {Renault}, E. and {Roth}, M. and {Rupprecht}, G. and {Serre}, D. and {Schaye}, J. and {Soucail}, G. and {Steinmetz}, M. and {Streicher}, O. and {Stuik}, R. and {Valentin}, H. and {Vernet}, J. and {Weilbacher}, P. and {Wisotzki}, L. and {Yerle}, N.},
        title = "{The MUSE second-generation VLT instrument}",
    booktitle = {Ground-based and Airborne Instrumentation for Astronomy III},
         year = 2010,
       editor = {{McLean}, Ian S. and {Ramsay}, Suzanne K. and {Takami}, Hideki},
       series = {Society of Photo-Optical Instrumentation Engineers (SPIE) Conference Series},
       volume = {7735},
        month = jul,
          eid = {773508},
        pages = {773508},
          doi = {10.1117/12.856027},
archivePrefix = {arXiv},
       eprint = {2211.16795},
 primaryClass = {astro-ph.IM},
       adsurl = {https://ui.adsabs.harvard.edu/abs/2010SPIE.7735E..08B}
}

@ARTICLE{panagia91,
       author = {{Panagia}, N. and {Gilmozzi}, R. and {Macchetto}, F. and {Adorf}, H.-M. and {Kirshner}, R.~P.},
        title = "{Properties of the SN 1987A Circumstellar Ring and the Distance to the Large Magellanic Cloud}",
      journal = {\apjl},
         year = 1991,
        month = oct,
       volume = {380},
        pages = {L23},
          doi = {10.1086/186164},
       adsurl = {https://ui.adsabs.harvard.edu/abs/1991ApJ...380L..23P}
}

@ARTICLE{jones23,
       author = {{Jones}, O.~C. and {Kavanagh}, P.~J. and {Barlow}, M.~J. and {Temim}, T. and {Fransson}, C. and {Larsson}, J. and {Blommaert}, J.~A.~D.~L. and {Meixner}, M. and {Lau}, R.~M. and {Sargent}, B. and {Bouchet}, P. and {Hjorth}, J. and {Wright}, G.~S. and {Coulais}, A. and {Fox}, O.~D. and {Gastaud}, R. and {Glasse}, A. and {Habel}, N. and {Hirschauer}, A.~S. and {Jaspers}, J. and {Krause}, O. and {Lenki{\'c}}, L. and {Nayak}, O. and {Rest}, A. and {Tikkanen}, T. and {Wesson}, R. and {Colina}, L. and {van Dishoeck}, E.~F. and {G{\"u}del}, M. and {Henning}, Th. and {Lagage}, P.-O. and {{\"O}stlin}, G. and {Ray}, T.~P. and {Vandenbussche}, B.},
        title = "{Ejecta, Rings, and Dust in SN 1987A with JWST MIRI/MRS}",
      journal = {\apj},
         year = 2023,
        month = nov,
       volume = {958},
       number = {1},
          eid = {95},
        pages = {95},
          doi = {10.3847/1538-4357/ad0036},
archivePrefix = {arXiv},
       eprint = {2307.06692},
 primaryClass = {astro-ph.HE},
       adsurl = {https://ui.adsabs.harvard.edu/abs/2023ApJ...958...95J}
}

@ARTICLE{dwek83,
       author = {{Dwek}, E.},
        title = "{The infrared echo of a type II supernova with a circumstellar dust shell : applications to SN 1979c and SN 1980k.}",
      journal = {\apj},
         year = 1983,
        month = nov,
       volume = {274},
        pages = {175-183},
          doi = {10.1086/161435},
       adsurl = {https://ui.adsabs.harvard.edu/abs/1983ApJ...274..175D}
}

@ARTICLE{ryder93,
       author = {{Ryder}, Stuart and {Staveley-Smith}, Lister and {Dopita}, Michael and {Petre}, Robert and {Colbert}, Edward and {Malin}, David and {Schlegel}, Eric},
        title = "{SN 1978K: an Extraordinary Supernova in the Nearby Galaxy NGC 1313}",
      journal = {\apj},
         year = 1993,
        month = oct,
       volume = {416},
        pages = {167},
          doi = {10.1086/173223},
       adsurl = {https://ui.adsabs.harvard.edu/abs/1993ApJ...416..167R}
}

@MISC{2016wfc..rept...12A,
       author = {{Anderson}, Jay},
        title = "{Empirical Models for the WFC3/IR PSF}",
 howpublished = {Instrument Science Report WFC3 2016-12, 42 pages},
         year = 2016,
        month = mar,
        pages = {12},
       adsurl = {https://ui.adsabs.harvard.edu/abs/2016wfc..rept...12A}
}

@ARTICLE{2000PASP..112.1360A,
       author = {{Anderson}, Jay and {King}, Ivan R.},
        title = "{Toward High-Precision Astrometry with WFPC2. I. Deriving an Accurate Point-Spread Function}",
      journal = {\pasp},
         year = 2000,
        month = oct,
       volume = {112},
       number = {776},
        pages = {1360-1382},
          doi = {10.1086/316632},
archivePrefix = {arXiv},
       eprint = {astro-ph/0006325},
 primaryClass = {astro-ph},
       adsurl = {https://ui.adsabs.harvard.edu/abs/2000PASP..112.1360A}
}

@ARTICLE{gomez25,
       author = {{Gomez}, Sebastian and {Temim}, Tea and {Shahbandeh}, Melissa and {Fox}, Ori and {Moore}, Thomas and {Healey}, Sarah},
        title = "{gmzsebastian/dustysn: v0.2 (v0.2)}",
      journal = {Zenodo},
         year = 2025,
        month = "",
       volume = "",
        pages = "",
          doi = {https://doi.org/10.5281/zenodo.17807367},
       adsurl = ""
}

@ARTICLE{Oke+1983,
       author = {{Oke}, J.~B. and {Gunn}, J.~E.},
        title = "{Secondary standard stars for absolute spectrophotometry.}",
      journal = {\apj},
         year = 1983,
        month = mar,
       volume = {266},
        pages = {713-717},
          doi = {10.1086/160817},
       adsurl = {https://ui.adsabs.harvard.edu/abs/1983ApJ...266..713O}
}

@ARTICLE{Pierel24,
       author = {{Pierel}, Justin},
        title = "Space-Phot: Simple Python-Based Photometry for Space Telescopes",
      journal = {Zenodo},
         year = 2024,
        month = "",
       volume = "",
        pages = "",
          doi = {https://doi.org/10.5281/zenodo.12100100},
       adsurl = ""
}

@ARTICLE{dwek08,
       author = {{Dwek}, Eli and {Arendt}, Richard G. and {Bouchet}, Patrice and {Burrows}, David N. and {Challis}, Peter and {Danziger}, I. John and {De Buizer}, James M. and {Gehrz}, Robert D. and {Kirshner}, Robert P. and {McCray}, Richard and {Park}, Sangwook and {Polomski}, Elisha F. and {Woodward}, Charles E.},
        title = "{Infrared and X-Ray Evidence for Circumstellar Grain Destruction by the Blast Wave of Supernova 1987A}",
      journal = {\apj},
         year = 2008,
        month = apr,
       volume = {676},
       number = {2},
        pages = {1029-1039},
          doi = {10.1086/529038},
archivePrefix = {arXiv},
       eprint = {0712.2759},
 primaryClass = {astro-ph},
       adsurl = {https://ui.adsabs.harvard.edu/abs/2008ApJ...676.1029D}
}

@ARTICLE{dwek87,
       author = {{Dwek}, Eli},
        title = "{The Infrared Diagnostic of a Dusty Plasma with Applications to Supernova Remnants}",
      journal = {\apj},
         year = 1987,
        month = nov,
       volume = {322},
        pages = {812},
          doi = {10.1086/165774},
       adsurl = {https://ui.adsabs.harvard.edu/abs/1987ApJ...322..812D}
}

@ARTICLE{orlando24,
       author = {{Orlando}, Salvatore and {Greco}, Emanuele and {Hirai}, Ryosuke and {Matsuoka}, Tomoki and {Miceli}, Marco and {Nagataki}, Shigheiro and {Ono}, Masaomi and {Chen}, Ke-Jung and {Milisavljevic}, Dan and {Patnaude}, Daniel and {Bocchino}, Fabrizio and {Elias-Rosa}, Nancy},
        title = "{Constraining the Circumstellar Medium Structure and Progenitor Mass-loss History of Interacting Supernovae Through 3D Hydrodynamic Modeling: The Case of SN 2014C}",
      journal = {\apj},
         year = 2024,
        month = dec,
       volume = {977},
       number = {1},
          eid = {118},
        pages = {118},
          doi = {10.3847/1538-4357/ad8ac8},
archivePrefix = {arXiv},
       eprint = {2410.17699},
 primaryClass = {astro-ph.HE},
       adsurl = {https://ui.adsabs.harvard.edu/abs/2024ApJ...977..118O}
}

@ARTICLE{bietenholz18,
       author = {{Bietenholz}, Michael F. and {Kamble}, Atish and {Margutti}, Raffaella and {Milisavljevic}, Danny and {Soderberg}, Alicia},
        title = "{SN 2014C: VLBI images of a supernova interacting with a circumstellar shell}",
      journal = {\mnras},
         year = 2018,
        month = apr,
       volume = {475},
       number = {2},
        pages = {1756-1764},
          doi = {10.1093/mnras/stx3194},
archivePrefix = {arXiv},
       eprint = {1707.09935},
 primaryClass = {astro-ph.HE},
       adsurl = {https://ui.adsabs.harvard.edu/abs/2018MNRAS.475.1756B}
}

@ARTICLE{bietenholz21,
       author = {{Bietenholz}, Michael F. and {Bartel}, Norbert and {Kamble}, Atish and {Margutti}, Raffaella and {Matthews}, David Jacob and {Milisavljevic}, Danny},
        title = "{SN 2014C: VLBI image shows a shell structure and decelerated expansion}",
      journal = {\mnras},
         year = 2021,
        month = apr,
       volume = {502},
       number = {2},
        pages = {1694-1701},
          doi = {10.1093/mnras/staa4003},
archivePrefix = {arXiv},
       eprint = {2012.12049},
 primaryClass = {astro-ph.HE},
       adsurl = {https://ui.adsabs.harvard.edu/abs/2021MNRAS.502.1694B}
}

@ARTICLE{brethauer22,
       author = {{Brethauer}, Daniel and {Margutti}, Raffaella and {Milisavljevic}, Dan and {Bietenholz}, Michael F. and {Chornock}, Ryan and {Coppejans}, Deanne L. and {De Colle}, Fabio and {Hajela}, Aprajita and {Terreran}, Giacomo and {Vargas}, Felipe and {DeMarchi}, Lindsay and {Harris}, Chelsea and {Jacobson-Gal{\'a}n}, Wynn V. and {Kamble}, Atish and {Patnaude}, Daniel and {Stroh}, Michael C.},
        title = "{Seven Years of Coordinated Chandra-NuSTAR Observations of SN 2014C Unfold the Extreme Mass-loss History of Its Stellar Progenitor}",
      journal = {\apj},
         year = 2022,
        month = nov,
       volume = {939},
       number = {2},
          eid = {105},
        pages = {105},
          doi = {10.3847/1538-4357/ac8b14},
archivePrefix = {arXiv},
       eprint = {2206.00842},
 primaryClass = {astro-ph.HE},
       adsurl = {https://ui.adsabs.harvard.edu/abs/2022ApJ...939..105B}
}

@ARTICLE{thomas22,
       author = {{Thomas}, Benjamin P. and {Wheeler}, J. Craig and {Dwarkadas}, Vikram V. and {Stockdale}, Christopher and {Vink{\'o}}, Jozsef and {Pooley}, David and {Xu}, Yerong and {Zeimann}, Greg and {MacQueen}, Phillip},
        title = "{Seven Years of SN 2014C: A Multiwavelength Synthesis of an Extraordinary Supernova}",
      journal = {\apj},
         year = 2022,
        month = may,
       volume = {930},
       number = {1},
          eid = {57},
        pages = {57},
          doi = {10.3847/1538-4357/ac5fa6},
archivePrefix = {arXiv},
       eprint = {2203.12747},
 primaryClass = {astro-ph.HE},
       adsurl = {https://ui.adsabs.harvard.edu/abs/2022ApJ...930...57T}
}

@ARTICLE{myers24,
       author = {{Myers}, Charlotte and {De}, Kishalay and {Yan}, Lin and {Jencson}, Jacob E. and {Earley}, Nicholas and {Fremling}, Christoffer and {Hiramatsu}, Daichi and {Kasliwal}, Mansi M. and {Lau}, Ryan M. and {MacLeod}, Morgan and {Masterson}, Megan and {Panagiotou}, Christos and {Simcoe}, Robert and {Tinyanont}, Samaporn},
        title = "{WTP 19aalnxx: Discovery of a Bright Mid-infrared Transient in the Emerging Class of Low-luminosity Supernovae Revealed by Delayed Circumstellar Interaction}",
      journal = {\apj},
         year = 2024,
        month = dec,
       volume = {976},
       number = {2},
          eid = {230},
        pages = {230},
          doi = {10.3847/1538-4357/ad8922},
archivePrefix = {arXiv},
       eprint = {2405.14663},
 primaryClass = {astro-ph.HE},
       adsurl = {https://ui.adsabs.harvard.edu/abs/2024ApJ...976..230M}
}

@ARTICLE{kuncarayakti23,
       author = {{Kuncarayakti}, H. and {Sollerman}, J. and {Izzo}, L. and {Maeda}, K. and {Yang}, S. and {Schulze}, S. and {Angus}, C.~R. and {Aubert}, M. and {Auchettl}, K. and {Della Valle}, M. and {Dessart}, L. and {Hinds}, K. and {Kankare}, E. and {Kawabata}, M. and {Lundqvist}, P. and {Nakaoka}, T. and {Perley}, D. and {Raimundo}, S.~I. and {Strotjohann}, N.~L. and {Taguchi}, K. and {Cai}, Y.-Z. and {Charalampopoulos}, P. and {Fang}, Q. and {Fraser}, M. and {Guti{\'e}rrez}, C.~P. and {Imazawa}, R. and {Kangas}, T. and {Kawabata}, K.~S. and {Kotak}, R. and {Kravtsov}, T. and {Matilainen}, K. and {Mattila}, S. and {Moran}, S. and {Murata}, I. and {Salmaso}, I. and {Anderson}, J.~P. and {Ashall}, C. and {Bellm}, E.~C. and {Benetti}, S. and {Chambers}, K.~C. and {Chen}, T.-W. and {Coughlin}, M. and {De Colle}, F. and {Fremling}, C. and {Galbany}, L. and {Gal-Yam}, A. and {Gromadzki}, M. and {Groom}, S.~L. and {Hajela}, A. and {Inserra}, C. and {Kasliwal}, M.~M. and {Mahabal}, A.~A. and {Martin-Carrillo}, A. and {Moore}, T. and {M{\"u}ller-Bravo}, T.~E. and {Nicholl}, M. and {Ragosta}, F. and {Riddle}, R.~L. and {Sharma}, Y. and {Srivastav}, S. and {Stritzinger}, M.~D. and {Wold}, A. and {Young}, D.~R.},
        title = "{The broad-lined Type-Ic supernova SN 2022xxf and its extraordinary two-humped light curves. I. Signatures of H/He-free interaction in the first four months}",
      journal = {\aap},
         year = 2023,
        month = oct,
       volume = {678},
          eid = {A209},
        pages = {A209},
          doi = {10.1051/0004-6361/202346526},
archivePrefix = {arXiv},
       eprint = {2303.16925},
 primaryClass = {astro-ph.SR},
       adsurl = {https://ui.adsabs.harvard.edu/abs/2023A&A...678A.209K}
}

@ARTICLE{zenati22,
       author = {{Zenati}, Yossef and {Wang}, Qinan and {Bobrick}, Alexey and {DeMarchi}, Lindsay and {Glanz}, Hila and {Rozner}, Mor and {Rest}, Armin and {Metzger}, Brian D. and {Margutti}, Raffaella and {Gomez}, Sebastian and {Smith}, Nathan and {Toonen}, Silvia and {Bright}, Joe S. and {Norman}, Colin and {Foley}, Ryan J. and {Gagliano}, Alexander and {Krolik}, Julian H. and {Smartt}, Stephen J. and {Villar}, Ashley V. and {Narayan}, Gautham and {Fox}, Ori and {Auchettl}, Katie and {Brethauer}, Daniel and {Clocchiatti}, Alejandro and {Coelln}, Sophie V. and {Coppejans}, Deanne L. and {Dimitriadis}, Georgios and {Doroszmai}, Andris and {Drout}, Maria and {Jacobson-Galan}, Wynn and {Gao}, Bore and {Ridden-Harper}, Ryan and {Kilpatrick}, Charles Donald and {Laskar}, Tanmoy and {Matthews}, David and {Rest}, Sofia and {Smith}, Ken W. and {McKenzie Stauffer}, Candice and {Stroh}, Michael C. and {Strolger}, Louis-Gregory and {Terreran}, Giacomo and {Pierel}, Justin D.~R. and {Piro}, Anthony L.},
        title = "{Evidence for Extended Hydrogen-Poor CSM in the Three-Peaked Light Curve of Stripped Envelope Ib Supernova}",
      journal = {arXiv e-prints},
         year = 2022,
        month = jul,
          eid = {arXiv:2207.07146},
        pages = {arXiv:2207.07146},
          doi = {10.48550/arXiv.2207.07146},
archivePrefix = {arXiv},
       eprint = {2207.07146},
 primaryClass = {astro-ph.HE},
       adsurl = {https://ui.adsabs.harvard.edu/abs/2022arXiv220707146Z}
}

@ARTICLE{ferrari24,
       author = {{Ferrari}, Luc{\'\i}a and {Folatelli}, Gast{\'o}n and {Kuncarayakti}, Hanindyo and {Stritzinger}, Maximilian and {Maeda}, Keiichi and {Bersten}, Melina and {Rom{\'a}n Aguilar}, Lili M. and {S{\'a}ez}, M. Manuela and {Dessart}, Luc and {Lundqvist}, Peter and {Mazzali}, Paolo and {Nagao}, Takashi and {Ashall}, Chris and {Bose}, Subhash and {Brennan}, Se{\'a}n J. and {Cai}, Yongzhi and {Handberg}, Rasmus and {Holmbo}, Simon and {Karamehmetoglu}, Emir and {Pastorello}, Andrea and {Reguitti}, Andrea and {Anderson}, Joseph and {Chen}, Ting-Wan and {Galbany}, Llu{\'\i}s and {Gromadzki}, Mariusz and {Guti{\'e}rrez}, Claudia P. and {Inserra}, Cosimo and {Kankare}, Erkki and {M{\"u}ller Bravo}, Tom{\'a}s E. and {Mattila}, Seppo and {Nicholl}, Matt and {Pignata}, Giuliano and {Sollerman}, Jesper and {Srivastav}, Shubham and {Young}, David R.},
        title = "{The metamorphosis of the Type Ib SN 2019yvr: late-time interaction}",
      journal = {\mnras},
         year = 2024,
        month = mar,
       volume = {529},
       number = {1},
        pages = {L33-L40},
          doi = {10.1093/mnrasl/slad195},
archivePrefix = {arXiv},
       eprint = {2401.15052},
 primaryClass = {astro-ph.HE},
       adsurl = {https://ui.adsabs.harvard.edu/abs/2024MNRAS.529L..33F}
}

@ARTICLE{kilpatrick21,
       author = {{Kilpatrick}, Charles D. and {Drout}, Maria R. and {Auchettl}, Katie and {Dimitriadis}, Georgios and {Foley}, Ryan J. and {Jones}, David O. and {DeMarchi}, Lindsay and {French}, K. Decker and {Gall}, Christa and {Hjorth}, Jens and {Jacobson-Gal{\'a}n}, Wynn V. and {Margutti}, Raffaella and {Piro}, Anthony L. and {Ramirez-Ruiz}, Enrico and {Rest}, Armin and {Rojas-Bravo}, C{\'e}sar},
        title = "{A cool and inflated progenitor candidate for the Type Ib supernova 2019yvr at 2.6 yr before explosion}",
      journal = {\mnras},
         year = 2021,
        month = jun,
       volume = {504},
       number = {2},
        pages = {2073-2093},
          doi = {10.1093/mnras/stab838},
archivePrefix = {arXiv},
       eprint = {2101.03206},
 primaryClass = {astro-ph.HE},
       adsurl = {https://ui.adsabs.harvard.edu/abs/2021MNRAS.504.2073K}
}

@ARTICLE{sollerman20,
       author = {{Sollerman}, J. and {Fransson}, C. and {Barbarino}, C. and {Fremling}, C. and {Horesh}, A. and {Kool}, E. and {Schulze}, S. and {Sfaradi}, I. and {Yang}, S. and {Bellm}, E.~C. and {Burruss}, R. and {Cunningham}, V. and {De}, K. and {Drake}, A.~J. and {Golkhou}, V.~Z. and {Green}, D.~A. and {Kasliwal}, M. and {Kulkarni}, S. and {Kupfer}, T. and {Laher}, R.~R. and {Masci}, F.~J. and {Rodriguez}, H. and {Rusholme}, B. and {Williams}, D.~R.~A. and {Yan}, L. and {Zolkower}, J.},
        title = "{Two stripped envelope supernovae with circumstellar interaction. But only one really shows it}",
      journal = {\aap},
         year = 2020,
        month = nov,
       volume = {643},
          eid = {A79},
        pages = {A79},
          doi = {10.1051/0004-6361/202038960},
archivePrefix = {arXiv},
       eprint = {2009.04154},
 primaryClass = {astro-ph.HE},
       adsurl = {https://ui.adsabs.harvard.edu/abs/2020A&A...643A..79S}
}

@ARTICLE{tartaglia21,
       author = {{Tartaglia}, L. and {Sollerman}, J. and {Barbarino}, C. and {Taddia}, F. and {Mason}, E. and {Berton}, M. and {Taggart}, K. and {Bellm}, E.~C. and {De}, K. and {Frederick}, S. and {Fremling}, C. and {Gal-Yam}, A. and {Golkhou}, V.~Z. and {Graham}, M. and {Ho}, A.~Y.~Q. and {Hung}, T. and {Kaye}, S. and {Kim}, Y.-L. and {Laher}, R.~R. and {Masci}, F.~J. and {Perley}, D.~A. and {Porter}, M.~D. and {Reiley}, D.~J. and {Riddle}, R. and {Rusholme}, B. and {Soumagnac}, M.~T. and {Walters}, R.},
        title = "{SN 2018ijp: the explosion of a stripped-envelope star within a dense H-rich shell?}",
      journal = {\aap},
         year = 2021,
        month = jun,
       volume = {650},
          eid = {A174},
        pages = {A174},
          doi = {10.1051/0004-6361/202039068},
archivePrefix = {arXiv},
       eprint = {2009.03331},
 primaryClass = {astro-ph.HE},
       adsurl = {https://ui.adsabs.harvard.edu/abs/2021A&A...650A.174T}
}

@ARTICLE{balasubramanian21,
       author = {{Balasubramanian}, A. and {Corsi}, A. and {Polisensky}, E. and {Clarke}, T.~E. and {Kassim}, N.~E.},
        title = "{Radio Observations of SN2004dk with VLITE Confirm Late-time Rebrightening}",
      journal = {\apj},
         year = 2021,
        month = dec,
       volume = {923},
       number = {1},
          eid = {32},
        pages = {32},
          doi = {10.3847/1538-4357/ac2154},
archivePrefix = {arXiv},
       eprint = {2101.07348},
 primaryClass = {astro-ph.HE},
       adsurl = {https://ui.adsabs.harvard.edu/abs/2021ApJ...923...32B}
}

@ARTICLE{pooley19,
       author = {{Pooley}, David and {Wheeler}, J. Craig and {Vink{\'o}}, Jozsef and {Dwarkadas}, Vikram V. and {Szalai}, Tamas and {Silverman}, Jeffrey M. and {Griesel}, Madelaine and {McCullough}, Molly and {Marion}, G.~H. and {MacQueen}, Phillip},
        title = "{Interaction of SN Ib 2004dk with a Previously Expelled Envelope}",
      journal = {\apj},
         year = 2019,
        month = oct,
       volume = {883},
       number = {2},
          eid = {120},
        pages = {120},
          doi = {10.3847/1538-4357/ab3e36},
archivePrefix = {arXiv},
       eprint = {1910.06395},
 primaryClass = {astro-ph.HE},
       adsurl = {https://ui.adsabs.harvard.edu/abs/2019ApJ...883..120P}
}

@ARTICLE{mauerhan18,
       author = {{Mauerhan}, Jon C. and {Filippenko}, Alexei V. and {Zheng}, WeiKang and {Brink}, Thomas G. and {Graham}, Melissa L. and {Shivvers}, Isaac and {Clubb}, Kelsey I.},
        title = "{Stripped-envelope supernova SN 2004dk is now interacting with hydrogen-rich circumstellar material}",
      journal = {\mnras},
         year = 2018,
        month = aug,
       volume = {478},
       number = {4},
        pages = {5050-5055},
          doi = {10.1093/mnras/sty1307},
archivePrefix = {arXiv},
       eprint = {1803.07051},
 primaryClass = {astro-ph.SR},
       adsurl = {https://ui.adsabs.harvard.edu/abs/2018MNRAS.478.5050M}
}

@ARTICLE{chandra20,
       author = {{Chandra}, Poonam and {Chevalier}, Roger A. and {Chugai}, Nikolai and {Milisavljevic}, Dan and {Fransson}, Claes},
        title = "{Supernova Interaction with a Dense Detached Shell in SN 2001em}",
      journal = {\apj},
         year = 2020,
        month = oct,
       volume = {902},
       number = {1},
          eid = {55},
        pages = {55},
          doi = {10.3847/1538-4357/abb460},
archivePrefix = {arXiv},
       eprint = {2008.13724},
 primaryClass = {astro-ph.HE},
       adsurl = {https://ui.adsabs.harvard.edu/abs/2020ApJ...902...55C}
}

@ARTICLE{chugai06,
       author = {{Chugai}, Nikolai N. and {Chevalier}, Roger A.},
        title = "{Late Emission from the Type Ib/c SN 2001em: Overtaking the Hydrogen Envelope}",
      journal = {\apj},
         year = 2006,
        month = apr,
       volume = {641},
       number = {2},
        pages = {1051-1059},
          doi = {10.1086/500539},
archivePrefix = {arXiv},
       eprint = {astro-ph/0510362},
 primaryClass = {astro-ph},
       adsurl = {https://ui.adsabs.harvard.edu/abs/2006ApJ...641.1051C}
}

@ARTICLE{margutti17,
       author = {{Margutti}, Raffaella and {Kamble}, A. and {Milisavljevic}, D. and {Zapartas}, E. and {de Mink}, S.~E. and {Drout}, M. and {Chornock}, R. and {Risaliti}, G. and {Zauderer}, B.~A. and {Bietenholz}, M. and {Cantiello}, M. and {Chakraborti}, S. and {Chomiuk}, L. and {Fong}, W. and {Grefenstette}, B. and {Guidorzi}, C. and {Kirshner}, R. and {Parrent}, J.~T. and {Patnaude}, D. and {Soderberg}, A.~M. and {Gehrels}, N.~C. and {Harrison}, F.},
        title = "{Ejection of the Massive Hydrogen-rich Envelope Timed with the Collapse of the Stripped SN 2014C}",
      journal = {\apj},
         year = 2017,
        month = feb,
       volume = {835},
       number = {2},
          eid = {140},
        pages = {140},
          doi = {10.3847/1538-4357/835/2/140},
archivePrefix = {arXiv},
       eprint = {1601.06806},
 primaryClass = {astro-ph.HE},
       adsurl = {https://ui.adsabs.harvard.edu/abs/2017ApJ...835..140M}
}

@ARTICLE{zhai25,
       author = {{Zhai}, Qian and {Zhang}, Jujia and {Lin}, Weili and {Mazzali}, Paolo and {Pian}, Elena and {Benetti}, Stefano and {Tomasella}, Lina and {Liu}, Jialian and {Li}, Liping},
        title = "{SN 2014C: A Metamorphic Supernova Exploded in the Intricate and Hydrogen-rich Surroundings}",
      journal = {\apj},
         year = 2025,
        month = jan,
       volume = {978},
       number = {2},
          eid = {163},
        pages = {163},
          doi = {10.3847/1538-4357/ad9c76},
archivePrefix = {arXiv},
       eprint = {2411.17008},
 primaryClass = {astro-ph.HE},
       adsurl = {https://ui.adsabs.harvard.edu/abs/2025ApJ...978..163Z}
}

@ARTICLE{milisavljevic15,
       author = {{Milisavljevic}, D. and {Margutti}, R. and {Kamble}, A. and {Patnaude}, D.~J. and {Raymond}, J.~C. and {Eldridge}, J.~J. and {Fong}, W. and {Bietenholz}, M. and {Challis}, P. and {Chornock}, R. and {Drout}, M.~R. and {Fransson}, C. and {Fesen}, R.~A. and {Grindlay}, J.~E. and {Kirshner}, R.~P. and {Lunnan}, R. and {Mackey}, J. and {Miller}, G.~F. and {Parrent}, J.~T. and {Sanders}, N.~E. and {Soderberg}, A.~M. and {Zauderer}, B.~A.},
        title = "{Metamorphosis of SN 2014C: Delayed Interaction between a Hydrogen Poor Core-collapse Supernova and a Nearby Circumstellar Shell}",
      journal = {\apj},
         year = 2015,
        month = dec,
       volume = {815},
       number = {2},
          eid = {120},
        pages = {120},
          doi = {10.1088/0004-637X/815/2/120},
archivePrefix = {arXiv},
       eprint = {1511.01907},
 primaryClass = {astro-ph.HE},
       adsurl = {https://ui.adsabs.harvard.edu/abs/2015ApJ...815..120M}
}

@ARTICLE{fox16,
       author = {{Fox}, Ori D. and {Johansson}, Joel and {Kasliwal}, Mansi and {Andrews}, Jennifer and {Bally}, John and {Bond}, Howard E. and {Boyer}, Martha L. and {Gehrz}, R.~D. and {Helou}, George and {Hsiao}, E.~Y. and {Masci}, Frank J. and {Parthasarathy}, M. and {Smith}, Nathan and {Tinyanont}, Samaporn and {Van Dyk}, Schuyler D.},
        title = "{An Excess of Mid-infrared Emission from the Type Iax SN 2014dt}",
      journal = {\apjl},
         year = 2016,
        month = jan,
       volume = {816},
       number = {1},
          eid = {L13},
        pages = {L13},
          doi = {10.3847/2041-8205/816/1/L13},
archivePrefix = {arXiv},
       eprint = {1510.08070},
 primaryClass = {astro-ph.HE},
       adsurl = {https://ui.adsabs.harvard.edu/abs/2016ApJ...816L..13F}
}

@ARTICLE{fox13b,
       author = {{Fox}, Ori D. and {Filippenko}, Alexei V.},
        title = "{The Late-time Rebrightening of Type Ia SN 2005gj in the Mid-infrared}",
      journal = {\apjl},
         year = 2013,
        month = jul,
       volume = {772},
       number = {1},
          eid = {L6},
        pages = {L6},
          doi = {10.1088/2041-8205/772/1/L6},
archivePrefix = {arXiv},
       eprint = {1304.4934},
 primaryClass = {astro-ph.HE},
       adsurl = {https://ui.adsabs.harvard.edu/abs/2013ApJ...772L...6F}
}

@ARTICLE{boyer22,
       author = {{Boyer}, Martha L. and {Anderson}, Jay and {Gennaro}, Mario and {Geha}, Marla and {Wingfield McQuinn}, Kristen B. and {Tollerud}, Erik and {Correnti}, Matteo and {Brenner Newman}, Max J. and {Cohen}, Roger E. and {Kallivayalil}, Nitya and {Beaton}, Rachel and {Cole}, Andrew A. and {Dolphin}, Andrew and {Kalirai}, Jason S. and {Sandstrom}, Karin M. and {Savino}, Alessandro and {Skillman}, Evan D. and {Weisz}, Daniel R. and {Williams}, Benjamin F.},
        title = "{The JWST Resolved Stellar Populations Early Release Science Program. I. NIRCam Flux Calibration}",
      journal = {Research Notes of the American Astronomical Society},
         year = 2022,
        month = sep,
       volume = {6},
       number = {9},
          eid = {191},
        pages = {191},
          doi = {10.3847/2515-5172/ac923a},
archivePrefix = {arXiv},
       eprint = {2209.03348},
 primaryClass = {astro-ph.IM},
       adsurl = {https://ui.adsabs.harvard.edu/abs/2022RNAAS...6..191B}
}

@INPROCEEDINGS{perrin14,
       author = {{Perrin}, Marshall D. and {Sivaramakrishnan}, Anand and {Lajoie}, Charles-Philippe and {Elliott}, Erin and {Pueyo}, Laurent and {Ravindranath}, Swara and {Albert}, Lo{\"\i}c.},
        title = "{Updated point spread function simulations for JWST with WebbPSF}",
    booktitle = {Space Telescopes and Instrumentation 2014: Optical, Infrared, and Millimeter Wave},
         year = 2014,
       editor = {{Oschmann}, Jr., Jacobus M. and {Clampin}, Mark and {Fazio}, Giovanni G. and {MacEwen}, Howard A.},
       series = {Society of Photo-Optical Instrumentation Engineers (SPIE) Conference Series},
       volume = {9143},
        month = aug,
          eid = {91433X},
        pages = {91433X},
          doi = {10.1117/12.2056689},
       adsurl = {https://ui.adsabs.harvard.edu/abs/2014SPIE.9143E..3XP}
}

@INPROCEEDINGS{perrin12,
       author = {{Perrin}, Marshall D. and {Soummer}, R{\'e}mi and {Elliott}, Erin M. and {Lallo}, Matthew D. and {Sivaramakrishnan}, Anand},
        title = "{Simulating point spread functions for the James Webb Space Telescope with WebbPSF}",
    booktitle = {Space Telescopes and Instrumentation 2012: Optical, Infrared, and Millimeter Wave},
         year = 2012,
       editor = {{Clampin}, Mark C. and {Fazio}, Giovanni G. and {MacEwen}, Howard A. and {Oschmann}, Jr., Jacobus M.},
       series = {Society of Photo-Optical Instrumentation Engineers (SPIE) Conference Series},
       volume = {8442},
        month = sep,
          eid = {84423D},
        pages = {84423D},
          doi = {10.1117/12.925230},
       adsurl = {https://ui.adsabs.harvard.edu/abs/2012SPIE.8442E..3DP}
}

@ARTICLE{dwek19,
       author = {{Dwek}, Eli and {Sarangi}, Arkaprabha and {Arendt}, Richard G.},
        title = "{The Evolution of Dust Opacity in Core Collapse Supernovae and the Rapid Formation of Dust in Their Ejecta}",
      journal = {\apjl},
         year = 2019,
        month = feb,
       volume = {871},
       number = {2},
          eid = {L33},
        pages = {L33},
          doi = {10.3847/2041-8213/aaf9a8},
archivePrefix = {arXiv},
       eprint = {1812.08234},
 primaryClass = {astro-ph.SR},
       adsurl = {https://ui.adsabs.harvard.edu/abs/2019ApJ...871L..33D}
}

@software{Bushouse+2022,
       author = {{Bushouse}, Howard and {Eisenhamer}, Jonathan and {Dencheva}, Nadia and {Davies}, James and {Greenfield}, Perry and {Morrison}, Jane and {Hodge}, Phil and {Simon}, Bernie and {Grumm}, David and {Droettboom}, Michael and {Slavich}, Edward and {Sosey}, Megan and {Pauly}, Tyler and {Miller}, Todd and {Jedrzejewski}, Robert and {Hack}, Warren and {Davis}, David and {Crawford}, Steven and {Law}, David and {Gordon}, Karl and {Regan}, Michael and {Cara}, Mihai and {MacDonald}, Ken and {Bradley}, Larry and {Shanahan}, Clare and {Jamieson}, William and {Teodoro}, Mairan and {Williams}, Thomas},
        title = "{JWST Calibration Pipeline}",
         year = 2022,
        month = oct,
          eid = {10.5281/zenodo.7325378},
          doi = {10.5281/zenodo.7325378},
      version = {1.8.2},
    publisher = {Zenodo},
       adsurl = {https://ui.adsabs.harvard.edu/abs/2022zndo...7325378B}
}

@ARTICLE{Foreman-Mackey+2013,
       author = {{Foreman-Mackey}, Daniel and {Hogg}, David W. and {Lang}, Dustin and {Goodman}, Jonathan},
        title = "{emcee: The MCMC Hammer}",
      journal = {\pasp},
         year = 2013,
        month = mar,
       volume = {125},
       number = {925},
        pages = {306},
          doi = {10.1086/670067},
archivePrefix = {arXiv},
       eprint = {1202.3665},
 primaryClass = {astro-ph.IM},
       adsurl = {https://ui.adsabs.harvard.edu/abs/2013PASP..125..306F}
}

@ARTICLE{Zsiros+24,
       author = {{Zs{\'\i}ros}, Szanna and {Szalai}, Tam{\'a}s and {De Looze}, Ilse and {Sarangi}, Arkaprabha and {Shahbandeh}, Melissa and {Fox}, Ori D. and {Temim}, Tea and {Milisavljevic}, Dan and {Van Dyk}, Schuyler D. and {Smith}, Nathan and {Filippenko}, Alexei V. and {Brink}, Thomas G. and {Zheng}, WeiKang and {Dessart}, Luc and {Jencson}, Jacob and {Johansson}, Joel and {Pierel}, Justin and {Rest}, Armin and {Tinyanont}, Samaporn and {Niculescu-Duvaz}, Maria and {Barlow}, M.~J. and {Wesson}, Roger and {Andrews}, Jennifer and {Clayton}, Geoff and {De}, Kishalay and {Dwek}, Eli and {Engesser}, Michael and {Foley}, Ryan J. and {Gezari}, Suvi and {Gomez}, Sebastian and {Gonzaga}, Shireen and {Kasliwal}, Mansi and {Lau}, Ryan and {Marston}, Anthony and {O'Steen}, Richard and {Siebert}, Matthew and {Skrutskie}, Michael and {Strolger}, Lou and {Wang}, Qinan and {Williams}, Brian and {Williams}, Robert and {Xiao}, Lin},
        title = "{Serendipitous detection of the dusty Type IIL SN 1980K with JWST/MIRI}",
      journal = {\mnras},
         year = 2024,
        month = mar,
       volume = {529},
       number = {1},
        pages = {155-168},
          doi = {10.1093/mnras/stae507},
archivePrefix = {arXiv},
       eprint = {2310.03448},
 primaryClass = {astro-ph.HE},
       adsurl = {https://ui.adsabs.harvard.edu/abs/2024MNRAS.529..155Z}
}

@ARTICLE{clayton25,
       author = {{Clayton}, Geoffrey C. and {Wesson}, R. and {Fox}, Ori D. and {Shahbandeh}, Melissa and {Filippenko}, Alexei V. and {Nickson}, Bryony and {Engesser}, Michael and {Van Dyk}, Schuyler D. and {Zheng}, WeiKang and {Brink}, Thomas G. and {Yang}, Yi and {Temim}, Tea and {Smith}, Nathan and {Andrews}, Jennifer and {Ashall}, Chris and {De Looze}, Ilse and {Derkacy}, James M. and {Dessart}, Luc and {Dulude}, Michael and {Dwek}, Eli and {Foley}, Ryan J. and {Gezari}, Suvi and {Gomez}, Sebastian and {Gonzaga}, Shireen and {Indukuri}, Siva and {Jencson}, Jacob and {Johansson}, Joel and {Kasliwal}, Mansi and {Lane}, Zachary G. and {Lau}, Ryan and {Law}, David and {Marston}, Anthony and {Milisavljevic}, Dan and {O'Steen}, Richard and {Pierel}, Justin and {Rest}, Armin and {Sarangi}, Arkaprabha and {Siebert}, Matthew and {Skrutskie}, Michael and {Strolger}, Lou and {Szalai}, Tamas and {Tinyanont}, Samaporn and {Wang}, Qinan and {Williams}, Brian and {Xiao}, Lin and {Zsiros}, Szanna},
        title = "{Very Late-Time JWST and Keck Spectra of the Oxygen-Rich Supernova 1995N}",
      journal = {arXiv e-prints},
         year = 2025,
        month = may,
          eid = {arXiv:2505.01574},
        pages = {arXiv:2505.01574},
          doi = {10.48550/arXiv.2505.01574},
archivePrefix = {arXiv},
       eprint = {2505.01574},
 primaryClass = {astro-ph.SR},
       adsurl = {https://ui.adsabs.harvard.edu/abs/2025arXiv250501574C}
}

@ARTICLE{sarangi25,
       author = {{Sarangi}, Arkaprabha and {Zsiros}, Szanna and {Szalai}, Tamas and {Martinez}, Laureano and {Shahbandeh}, Melissa and {Fox}, Ori D. and {Van Dyk}, Schuyler D. and {Filippenko}, Alexei V. and {Bersten}, Melina Cecilia and {De Looze}, Ilse and {Ashall}, Chris and {Temim}, Tea and {Jencson}, Jacob E. and {Rest}, Armin and {Milisavljevic}, Dan and {Dessart}, Luc and {Dwek}, Eli and {Smith}, Nathan and {Tinyanont}, Samaporn and {Brink}, Thomas G. and {Zheng}, WeiKang and {Clayton}, Geoffrey C. and {Andrews}, Jennifer},
        title = "{Two Decades of Dust Evolution in SN 2005af through JWST, Spitzer, and Chemical Modeling}",
      journal = {arXiv e-prints},
         year = 2025,
        month = apr,
          eid = {arXiv:2504.20574},
        pages = {arXiv:2504.20574},
          doi = {10.48550/arXiv.2504.20574},
archivePrefix = {arXiv},
       eprint = {2504.20574},
 primaryClass = {astro-ph.SR},
       adsurl = {https://ui.adsabs.harvard.edu/abs/2025arXiv250420574S}
}

@ARTICLE{fox11,
       author = {{Fox}, Ori D. and {Chevalier}, Roger A. and {Skrutskie}, Michael F. and {Soderberg}, Alicia M. and {Filippenko}, Alexei V. and {Ganeshalingam}, Mohan and {Silverman}, Jeffrey M. and {Smith}, Nathan and {Steele}, Thea N.},
        title = "{A Spitzer Survey for Dust in Type IIn Supernovae}",
      journal = {\apj},
         year = 2011,
        month = nov,
       volume = {741},
       number = {1},
          eid = {7},
        pages = {7},
          doi = {10.1088/0004-637X/741/1/7},
archivePrefix = {arXiv},
       eprint = {1104.5012},
 primaryClass = {astro-ph.SR},
       adsurl = {https://ui.adsabs.harvard.edu/abs/2011ApJ...741....7F}
}

@INPROCEEDINGS{freedman97,
       author = {{Freedman}, Wendy L. and {Madore}, B.~F. and {Kennicutt}, R.~C.},
        title = "{Hubble Space Telescope Key Project on the Extragalactic Distance Scale}",
    booktitle = {The Extragalactic Distance Scale},
         year = 1997,
       editor = {{Livio}, Mario and {Donahue}, Megan and {Panagia}, Nino},
        month = jan,
        pages = {171},
       adsurl = {https://ui.adsabs.harvard.edu/abs/1997eds..proc..171F}
}

@ARTICLE{Sarangi+2022,
       author = {{Sarangi}, Arkaprabha},
        title = "{Formation, distribution, and IR emission of dust in the clumpy ejecta of Type II-P core-collapse supernovae, in isotropic and anisotropic scenarios}",
      journal = {\aap},
         year = 2022,
        month = dec,
       volume = {668},
          eid = {A57},
        pages = {A57},
          doi = {10.1051/0004-6361/202244391},
archivePrefix = {arXiv},
       eprint = {2209.14896},
 primaryClass = {astro-ph.SR},
       adsurl = {https://ui.adsabs.harvard.edu/abs/2022A&A...668A..57S}
}

@ARTICLE{fox10,
       author = {{Fox}, Ori D. and {Chevalier}, Roger A. and {Dwek}, Eli and {Skrutskie}, Michael F. and {Sugerman}, Ben E.~K. and {Leisenring}, Jarron M.},
        title = "{Disentangling the Origin and Heating Mechanism of Supernova Dust: Late-time Spitzer Spectroscopy of the Type IIn SN 2005ip}",
      journal = {\apj},
         year = 2010,
        month = dec,
       volume = {725},
       number = {2},
        pages = {1768-1778},
          doi = {10.1088/0004-637X/725/2/1768},
archivePrefix = {arXiv},
       eprint = {1005.4682},
 primaryClass = {astro-ph.HE},
       adsurl = {https://ui.adsabs.harvard.edu/abs/2010ApJ...725.1768F}
}

@ARTICLE{tinyanont25,
       author = {{Tinyanont}, Samaporn and {Fox}, Ori D. and {Shahbandeh}, Melissa and {Temim}, Tea and {Williams}, Robert and {Wangnok}, Kittipong and {Rest}, Armin and {Lau}, Ryan M. and {Maeda}, Keiichi and {Jencson}, Jacob E. and {Auchettl}, Katie and {Filippenko}, Alexei V. and {Larison}, Conor and {Ashall}, Chris and {Brink}, Thomas G. and {Davis}, Kyle W. and {Dessart}, Luc and {Foley}, Ryan J. and {Galbany}, Llu{\'\i}s and {Grayling}, Matthew and {Johansson}, Joel and {Kasliwal}, Mansi M. and {Lane}, Zachary G. and {LeBaron}, Natalie and {Milisavljevic}, Dan and {Rho}, Jeonghee and {Sakon}, Itsuki and {Sarangi}, Arkaprabha and {Szalai}, Tam{\'a}s and {Taggart}, Kirsty and {Van Dyk}, Schuyler D. and {Wang}, Qinan and {Yang}, Yi and {Zheng}, WeiKang and {Zs{\'\i}ros}, Szanna},
        title = "{Large Cold Dust Reservoir Revealed in Transitional SN Ib 2014C by James Webb Space Telescope Mid-infrared Spectroscopy}",
      journal = {\apj},
         year = 2025,
        month = jun,
       volume = {985},
       number = {2},
          eid = {198},
        pages = {198},
          doi = {10.3847/1538-4357/adccc0},
archivePrefix = {arXiv},
       eprint = {2504.14009},
 primaryClass = {astro-ph.HE},
       adsurl = {https://ui.adsabs.harvard.edu/abs/2025ApJ...985..198T}
}

@ARTICLE{Rieke+2022,
       author = {{Rieke}, George and {Wright}, Gillian},
        title = "{A mid-infrared dream come true}",
      journal = {Nature Astronomy},
         year = 2022,
        month = jul,
       volume = {6},
        pages = {891-891},
          doi = {10.1038/s41550-022-01736-6},
       adsurl = {https://ui.adsabs.harvard.edu/abs/2022NatAs...6..891R}
}

@INPROCEEDINGS{Rieke+2005,
       author = {{Rieke}, Marcia J. and {Kelly}, Douglas and {Horner}, Scott},
        title = "{Overview of James Webb Space Telescope and NIRCam's Role}",
    booktitle = {Cryogenic Optical Systems and Instruments XI},
         year = 2005,
       editor = {{Heaney}, James B. and {Burriesci}, Lawrence G.},
       series = {Society of Photo-Optical Instrumentation Engineers (SPIE) Conference Series},
       volume = {5904},
        month = aug,
        pages = {1-8},
          doi = {10.1117/12.615554},
       adsurl = {https://ui.adsabs.harvard.edu/abs/2005SPIE.5904....1R}
}

@ARTICLE{Rieke+2023,
       author = {{Rieke}, Marcia J. and {Kelly}, Douglas M. and {Misselt}, Karl and {Stansberry}, John and {Boyer}, Martha and {Beatty}, Thomas and {Egami}, Eiichi and {Florian}, Michael and {Greene}, Thomas P. and {Hainline}, Kevin and {Leisenring}, Jarron and {Roellig}, Thomas and {Schlawin}, Everett and {Sun}, Fengwu and {Tinnin}, Lee and {Williams}, Christina C. and {Willmer}, Christopher N.~A. and {Wilson}, Debra and {Clark}, Charles R. and {Rohrbach}, Scott and {Brooks}, Brian and {Canipe}, Alicia and {Correnti}, Matteo and {DiFelice}, Audrey and {Gennaro}, Mario and {Girard}, Julien H. and {Hartig}, George and {Hilbert}, Bryan and {Koekemoer}, Anton M. and {Nikolov}, Nikolay K. and {Pirzkal}, Norbert and {Rest}, Armin and {Robberto}, Massimo and {Sunnquist}, Ben and {Telfer}, Randal and {Wu}, Chi Rai and {Ferry}, Malcolm and {Lewis}, Dan and {Baum}, Stefi and {Beichman}, Charles and {Doyon}, Ren{\'e} and {Dressler}, Alan and {Eisenstein}, Daniel J. and {Ferrarese}, Laura and {Hodapp}, Klaus and {Horner}, Scott and {Jaffe}, Daniel T. and {Johnstone}, Doug and {Krist}, John and {Martin}, Peter and {McCarthy}, Donald W. and {Meyer}, Michael and {Rieke}, George H. and {Trauger}, John and {Young}, Erick T.},
        title = "{Performance of NIRCam on JWST in Flight}",
      journal = {\pasp},
         year = 2023,
        month = feb,
       volume = {135},
       number = {1044},
          eid = {028001},
        pages = {028001},
          doi = {10.1088/1538-3873/acac53},
archivePrefix = {arXiv},
       eprint = {2212.12069},
 primaryClass = {astro-ph.IM},
       adsurl = {https://ui.adsabs.harvard.edu/abs/2023PASP..135b8001R}
}

@ARTICLE{Rieke+2015,
       author = {{Rieke}, G.~H. and {Wright}, G.~S. and {B{\"o}ker}, T. and {Bouwman}, J. and {Colina}, L. and {Glasse}, Alistair and {Gordon}, K.~D. and {Greene}, T.~P. and {G{\"u}del}, Manuel and {Henning}, Th. and {Justtanont}, K. and {Lagage}, P. -O. and {Meixner}, M.~E. and {N{\o}rgaard-Nielsen}, H. -U. and {Ray}, T.~P. and {Ressler}, M.~E. and {van Dishoeck}, E.~F. and {Waelkens}, C.},
        title = "{The Mid-Infrared Instrument for the James Webb Space Telescope, I: Introduction}",
      journal = {\pasp},
         year = 2015,
        month = jul,
       volume = {127},
       number = {953},
        pages = {584},
          doi = {10.1086/682252},
archivePrefix = {arXiv},
       eprint = {1508.02294},
 primaryClass = {astro-ph.IM},
       adsurl = {https://ui.adsabs.harvard.edu/abs/2015PASP..127..584R}
}

@ARTICLE{Ressler+2015,
       author = {{Ressler}, M.~E. and {Sukhatme}, K.~G. and {Franklin}, B.~R. and {Mahoney}, J.~C. and {Thelen}, M.~P. and {Bouchet}, P. and {Colbert}, J.~W. and {Cracraft}, Misty and {Dicken}, D. and {Gastaud}, R. and {Goodson}, G.~B. and {Eccleston}, Paul and {Moreau}, V. and {Rieke}, G.~H. and {Schneider}, Analyn},
        title = "{The Mid-Infrared Instrument for the James Webb Space Telescope, VIII: The MIRI Focal Plane System}",
      journal = {\pasp},
         year = 2015,
        month = jul,
       volume = {127},
       number = {953},
        pages = {675},
          doi = {10.1086/682258},
archivePrefix = {arXiv},
       eprint = {1508.02417},
 primaryClass = {astro-ph.IM},
       adsurl = {https://ui.adsabs.harvard.edu/abs/2015PASP..127..675R}
}

@ARTICLE{Bouchet+2015,
       author = {{Bouchet}, Patrice and {Garc{\'\i}a-Mar{\'\i}n}, Macarena and {Lagage}, P. -O. and {Amiaux}, J{\'e}rome and {Augu{\'e}res}, J. -L. and {Bauwens}, Eva and {Blommaert}, J.~A.~D.~L. and {Chen}, C.~H. and {Detre}, {\"O}. H. and {Dicken}, Dan and {Dubreuil}, D. and {Galdemard}, Ph. and {Gastaud}, R. and {Glasse}, A. and {Gordon}, K.~D. and {Gougnaud}, F. and {Guillard}, Phillippe and {Justtanont}, K. and {Krause}, Oliver and {Leboeuf}, Didier and {Longval}, Yuying and {Martin}, Laurant and {Mazy}, Emmanuel and {Moreau}, Vincent and {Olofsson}, G{\"o}ran and {Ray}, T.~P. and {Rees}, J. -M. and {Renotte}, Etienne and {Ressler}, M.~E. and {Ronayette}, Samuel and {Salasca}, Sophie and {Scheithauer}, Silvia and {Sykes}, Jon and {Thelen}, M.~P. and {Wells}, Martyn and {Wright}, David and {Wright}, G.~S.},
        title = "{The Mid-Infrared Instrument for the James Webb Space Telescope, III: MIRIM, The MIRI Imager}",
      journal = {\pasp},
         year = 2015,
        month = jul,
       volume = {127},
       number = {953},
        pages = {612},
          doi = {10.1086/682254},
archivePrefix = {arXiv},
       eprint = {1508.02488},
 primaryClass = {astro-ph.IM},
       adsurl = {https://ui.adsabs.harvard.edu/abs/2015PASP..127..612B}
}

@ARTICLE{Hosseinzadeh+2023,
       author = {{Hosseinzadeh}, Griffin and {Sand}, David J. and {Jencson}, Jacob E. and {Andrews}, Jennifer E. and {Shivaei}, Irene and {Bostroem}, K. Azalee and {Valenti}, Stefano and {Szalai}, Tam{\'a}s and {Burke}, Jamison and {Howell}, D. Andrew and {McCully}, Curtis and {Newsome}, Megan and {Gonzalez}, Estefania Padilla and {Pellegrino}, Craig and {Terreran}, Giacomo},
        title = "{JWST Imaging of the Cartwheel Galaxy Reveals Dust Associated with SN 2021afdx}",
      journal = {\apjl},
         year = 2023,
        month = jan,
       volume = {942},
       number = {1},
          eid = {L18},
        pages = {L18},
          doi = {10.3847/2041-8213/aca64e},
archivePrefix = {arXiv},
       eprint = {2210.06499},
 primaryClass = {astro-ph.HE},
       adsurl = {https://ui.adsabs.harvard.edu/abs/2023ApJ...942L..18H}
}

@ARTICLE{Sarangi+2025,
       author = {{Sarangi}, Arkaprabha and {Zs{\'\i}ros}, Szanna and {Szalai}, Tam{\'a}s and {Martinez}, Laureano and {Shahbandeh}, Melissa and {Fox}, Ori D. and {Van Dyk}, Schuyler D. and {Filippenko}, Alexei V. and {Bersten}, Melina Cecilia and {De Looze}, Ilse and {Ashall}, Chris and {Temim}, Tea and {Jencson}, Jacob E. and {Rest}, Armin and {Milisavljevic}, Dan and {Dessart}, Luc and {Dwek}, Eli and {Smith}, Nathan and {Tinyanont}, Samaporn and {Brink}, Thomas G. and {Zheng}, WeiKang and {Clayton}, Geoffrey C. and {Andrews}, Jennifer},
        title = "{Two Decades of Dust Evolution in SN 2005af through JWST, Spitzer, and Chemical Modeling}",
      journal = {\apj},
         year = 2025,
        month = nov,
       volume = {993},
       number = {1},
          eid = {94},
        pages = {94},
          doi = {10.3847/1538-4357/ae0645},
archivePrefix = {arXiv},
       eprint = {2504.20574},
 primaryClass = {astro-ph.SR},
       adsurl = {https://ui.adsabs.harvard.edu/abs/2025ApJ...993...94S}
}

@ARTICLE{szalai19a,
       author = {{Szalai}, Tam{\'a}s and {Zs{\'\i}ros}, Szanna and {Fox}, Ori D. and
         {Pejcha}, Ond{\v{r}}ej and {M{\"u}ller}, Tom{\'a}s},
        title = "{A Comprehensive Analysis of Spitzer Supernovae}",
      journal = {ApJS},
         year = 2019,
        month = apr,
       volume = {241},
       number = {2},
          eid = {38},
        pages = {38},
          doi = {10.3847/1538-4365/ab10df},
archivePrefix = {arXiv},
       eprint = {1803.02571},
 primaryClass = {astro-ph.HE},
       adsurl = {https://ui.adsabs.harvard.edu/abs/2019ApJS..241...38S}
}

@ARTICLE{szalai21,
       author = {{Szalai}, Tam{\'a}s and {Fox}, Ori D. and {Arendt}, Richard G. and {Dwek}, Eli and {Andrews}, Jennifer E. and {Clayton}, Geoffrey C. and {Filippenko}, Alexei V. and {Johansson}, Joel and {Kelly}, Patrick L. and {Krafton}, Kelsie and {Marston}, A.~P. and {Mauerhan}, Jon C. and {Van Dyk}, Schuyler D.},
        title = "{Spitzer's Last Look at Extragalactic Explosions: Long-term Evolution of Interacting Supernovae}",
      journal = {\apj},
         year = 2021,
        month = sep,
       volume = {919},
       number = {1},
          eid = {17},
        pages = {17},
          doi = {10.3847/1538-4357/ac0e2b},
archivePrefix = {arXiv},
       eprint = {2106.12427},
 primaryClass = {astro-ph.HE},
       adsurl = {https://ui.adsabs.harvard.edu/abs/2021ApJ...919...17S}
}

@article{li11,
	Affiliation = {AA(Department of Astronomy, University of California, Berkeley, CA 94720-3411, USA), AB(Department of Astronomy, University of California, Berkeley, CA 94720-3411, USA; NASA Ames Research Center, Mountain View, CA 94043, USA), AC(Department of Astronomy, University of California, Berkeley, CA 94720-3411, USA; Harvard-Smithsonian Center for Astrophysics, 60 Garden Street, Cambridge, MA 02138, USA), AD(Department of Astronomy, University of California, Berkeley, CA 94720-3411, USA), AE(Department of Astronomy...},
	Author = {Weidong Li and Jesse Leaman and Ryan Chornock and Alexei V Filippenko and Dovi Poznanski and Mohan Ganeshalingam and Xiaofeng Wang and Maryam Modjaz and Saurabh Jha and Ryan J Foley and Nathan Smith},
	Doi = {10.1111/j.1365-2966.2011.18160.x},
	Journal = {MNRAS},
	Month = {Apr},
	Pages = {1441},
	Pmid = {2011MNRAS.412.1441L},
	Title = {Nearby supernova rates from the Lick Observatory Supernova Search - II. The observed luminosity functions and fractions of supernovae in a complete sample},
	Uri = {papers://4C7415BD-6F56-4EC4-8776-9C82E38B9188/Paper/p12480},
	Url = {http://adsabs.harvard.edu/cgi-bin/nph-data_query?bibcode=2011MNRAS.412.1441L&link_type=ABSTRACT},
	Volume = {412},
	Year = {2011}}

@ARTICLE{doughty21,
       author = {{Doughty}, Caitlin and {Finlator}, Kristian},
        title = "{The effects of binary stars on galaxies and metal-enriched gas during reionization}",
      journal = {\mnras},
         year = 2021,
        month = aug,
       volume = {505},
       number = {2},
        pages = {2207-2223},
          doi = {10.1093/mnras/stab1448},
archivePrefix = {arXiv},
       eprint = {2105.09972},
 primaryClass = {astro-ph.GA},
       adsurl = {https://ui.adsabs.harvard.edu/abs/2021MNRAS.505.2207D}
}

@ARTICLE{dessart22,
       author = {{Dessart}, L. and {Hillier}, D. John},
        title = "{Modeling the signatures of interaction in Type II supernovae: UV emission, high-velocity features, broad-boxy profiles}",
      journal = {\aap},
         year = 2022,
        month = apr,
       volume = {660},
          eid = {L9},
        pages = {L9},
          doi = {10.1051/0004-6361/202243372},
archivePrefix = {arXiv},
       eprint = {2204.00446},
 primaryClass = {astro-ph.SR},
       adsurl = {https://ui.adsabs.harvard.edu/abs/2022A&A...660L...9D}
}

@ARTICLE{Wheeler+1987,
       author = {{Wheeler}, J.~C. and {Harkness}, R.~P. and {Barker}, E.~S. and {Cochran}, A.~L. and {Wills}, D.},
        title = "{Supernovae 1983i and 1983v: Evidence for Abundance Variations in Type Ib Supernovae}",
      journal = {\apjl},
         year = 1987,
        month = feb,
       volume = {313},
        pages = {L69},
          doi = {10.1086/184833},
       adsurl = {https://ui.adsabs.harvard.edu/abs/1987ApJ...313L..69W}
}

@ARTICLE{Clocchiatti+1997,
       author = {{Clocchiatti}, A. and {Wheeler}, J.~C. and {Phillips}, M.~M. and {Suntzeff}, N.~B. and {Cristiani}, S. and {Phillips}, A. and {Harkness}, R.~P. and {Dopita}, M.~A. and {Beuermann}, K. and {Rosa}, M. and {Grosb{\o}l}, P. and {Lindblad}, P.~O. and {Filippenko}, A.~V.},
        title = "{SN 1983V in NGC 1365 and the Nature of Stripped Envelope Core-Collapse Supernovae}",
      journal = {\apj},
         year = 1997,
        month = jul,
       volume = {483},
       number = {2},
        pages = {675-697},
          doi = {10.1086/304268},
       adsurl = {https://ui.adsabs.harvard.edu/abs/1997ApJ...483..675C}
}

@ARTICLE{Modjaz+2016,
       author = {{Modjaz}, Maryam and {Liu}, Yuqian Q. and {Bianco}, Federica B. and {Graur}, Or},
        title = "{The Spectral SN-GRB Connection: Systematic Spectral Comparisons between Type Ic Supernovae and Broad-lined Type Ic Supernovae with and without Gamma-Ray Bursts}",
      journal = {\apj},
         year = 2016,
        month = dec,
       volume = {832},
       number = {2},
          eid = {108},
        pages = {108},
          doi = {10.3847/0004-637X/832/2/108},
archivePrefix = {arXiv},
       eprint = {1509.07124},
 primaryClass = {astro-ph.HE},
       adsurl = {https://ui.adsabs.harvard.edu/abs/2016ApJ...832..108M}
}

@ARTICLE{Szalai+2025,
       author = {{Szalai}, Tam{\'a}s and {Zs{\'\i}ros}, Szanna and {Jencson}, Jacob and {Fox}, Ori D. and {Shahbandeh}, Melissa and {Sarangi}, Arkaprabha and {Temim}, Tea and {De Looze}, Ilse and {Smith}, Nathan and {Filippenko}, Alexei V. and {Van Dyk}, Schuyler D. and {Andrews}, Jennifer and {Ashall}, Chris and {Clayton}, Geoffrey C. and {Dessart}, Luc and {Dulude}, Michael and {Dwek}, Eli and {Gomez}, Sebastian and {Johansson}, Joel and {Milisavljevic}, Dan and {Pierel}, Justin and {Rest}, Armin and {Tinyanont}, Samaporn and {Brink}, Thomas G. and {De}, Kishalay and {Engesser}, Michael and {Foley}, Ryan J. and {Gezari}, Suvi and {Kasliwal}, Mansi and {Lau}, Ryan and {Marston}, Anthony and {O'Steen}, Richard and {Siebert}, Matthew and {Skrutskie}, Michael and {Strolger}, Lou and {Wang}, Qinan and {Williams}, Brian J. and {Williams}, Robert and {Xiao}, Lin and {Zheng}, WeiKang},
        title = "{JWST/MIRI detects the dusty SN1993J about 30 years after explosion}",
      journal = {\aap},
         year = 2025,
        month = may,
       volume = {697},
          eid = {A132},
        pages = {A132},
          doi = {10.1051/0004-6361/202451470},
archivePrefix = {arXiv},
       eprint = {2503.12950},
 primaryClass = {astro-ph.SR},
       adsurl = {https://ui.adsabs.harvard.edu/abs/2025A&A...697A.132S}
}

@ARTICLE{Pearson+2025,
       author = {{Pearson}, Jeniveve and {Subrayan}, Bhagya and {Sand}, David J. and {Andrews}, Jennifer E. and {Beasor}, Emma R. and {Bostroem}, K. Azalee and {Dong}, Yize and {Hoang}, Emily and {Hosseinzadeh}, Griffin and {Hsu}, Brian and {Jacobson-Gal{\'a}n}, Wynn and {Janzen}, Daryl and {Jencson}, Jacob and {Jha}, Saurabh W. and {Kilpatrick}, Charles D. and {Kwok}, Lindsey A. and {Liu}, Chang and {Lundquist}, M.~J. and {Mehta}, Darshana and {Miller}, Adam A. and {Ravi}, Aravind P. and {Rehemtulla}, Nabeel and {Meza Retamal}, Nicol{\'a}s and {Shrestha}, Manisha and {Smith}, Nathan and {Valenti}, Stefano and {Whitler}, Lily},
        title = "{Mid-infrared Dust Evolution and Late-time Circumstellar Medium Interaction in SN 2017eaw}",
      journal = {\apj},
         year = 2025,
        month = nov,
       volume = {993},
       number = {2},
          eid = {213},
        pages = {213},
          doi = {10.3847/1538-4357/ae00ba},
archivePrefix = {arXiv},
       eprint = {2507.00125},
 primaryClass = {astro-ph.HE},
       adsurl = {https://ui.adsabs.harvard.edu/abs/2025ApJ...993..213P}
}

@ARTICLE{Shahbandeh+2023,
       author = {{Shahbandeh}, Melissa and {Sarangi}, Arkaprabha and {Temim}, Tea and {Szalai}, Tam{\'a}s and {Fox}, Ori D. and {Tinyanont}, Samaporn and {Dwek}, Eli and {Dessart}, Luc and {Filippenko}, Alexei V. and {Brink}, Thomas G. and {Foley}, Ryan J. and {Jencson}, Jacob and {Pierel}, Justin and {Zs{\'\i}ros}, Szanna and {Rest}, Armin and {Zheng}, WeiKang and {Andrews}, Jennifer and {Clayton}, Geoffrey C. and {De}, Kishalay and {Engesser}, Michael and {Gezari}, Suvi and {Gomez}, Sebastian and {Gonzaga}, Shireen and {Johansson}, Joel and {Kasliwal}, Mansi and {Lau}, Ryan and {De Looze}, Ilse and {Marston}, Anthony and {Milisavljevic}, Dan and {O'Steen}, Richard and {Siebert}, Matthew and {Skrutskie}, Michael and {Smith}, Nathan and {Strolger}, Lou and {Van Dyk}, Schuyler D. and {Wang}, Qinan and {Williams}, Brian and {Williams}, Robert and {Xiao}, Lin and {Yang}, Yi},
        title = "{JWST observations of dust reservoirs in type IIP supernovae 2004et and 2017eaw}",
      journal = {\mnras},
         year = 2023,
        month = aug,
       volume = {523},
       number = {4},
        pages = {6048-6060},
          doi = {10.1093/mnras/stad1681},
archivePrefix = {arXiv},
       eprint = {2301.10778},
 primaryClass = {astro-ph.HE},
       adsurl = {https://ui.adsabs.harvard.edu/abs/2023MNRAS.523.6048S}
}

@ARTICLE{Perna+2008,
       author = {{Perna}, Rosalba and {Soria}, Roberto and {Pooley}, Dave and {Stella}, Luigi},
        title = "{How rapidly do neutron stars spin at birth? Constraints from archival X-ray observations of extragalactic supernovae}",
      journal = {\mnras},
         year = 2008,
        month = mar,
       volume = {384},
       number = {4},
        pages = {1638-1648},
          doi = {10.1111/j.1365-2966.2007.12821.x},
archivePrefix = {arXiv},
       eprint = {0712.1040},
 primaryClass = {astro-ph},
       adsurl = {https://ui.adsabs.harvard.edu/abs/2008MNRAS.384.1638P}
}
\bibliographystyle{aasjournalv7}

\end{document}